\documentclass[
12pt,                    
a4paper,             
BCOR10mm,     
DIV14,                 
listof=totoc,                    
twoside,
headsepline
]{scrreprt}

\usepackage[english]{babel}
\usepackage[utf8]{inputenc}
\usepackage[T1]{fontenc}
\usepackage{makeidx}
\usepackage{url}
\usepackage{doc}
\usepackage{graphicx}
\usepackage{setspace}
\usepackage{authoraftertitle}
\usepackage[figuresright]{rotating}
\usepackage{adjustbox}
\usepackage{tabularx}
\usepackage{rotating}
\usepackage{ragged2e}
\usepackage{multirow}
\usepackage[bookmarks=true,colorlinks = true,pdfpagelabels,pdfstartview = FitH,bookmarksopen = true,bookmarksnumbered = true,linkcolor = black,plainpages = false,hypertexnames = false,citecolor = black,urlcolor=black]{hyperref} 
\usepackage[round]{natbib}
\usepackage{breqn}
\usepackage{amsfonts}

\usepackage[english]{isodate}

\date{July 30, 2021}

\begin{document}
	
	\pagenumbering{roman} 
	\begin{titlepage}

		\vspace*{1cm}
		\begin{center}
			\vspace*{3cm}
			\textbf{ 
				\Large Ruprecht-Karls-Universität Heidelberg\\
				\smallskip
				\Large Institut für Informatik\\
				\smallskip
				\Large Lehrstuhl für Parallele und Verteilte Systeme\\
				\smallskip
			}
			
			\vspace{3cm}
			
			\textbf{\large Masterarbeit} 
			
			\vspace{0.5\baselineskip}
			{\huge
				Neural Models for Source Code Synthesis and Completion
			}
		\end{center}
		
		\vfill 
		
		{\large
			\begin{tabular}[l]{ll}
				Name: & Mitodru Niyogi\\
				Matrikelnummer: & 3573020\\
  				Datum der Abgabe: & \printdate{2021-07-30}
			\end{tabular}
		}
		
	\end{titlepage}
	
	\onehalfspacing
	
	\thispagestyle{empty}
	
	\vspace*{100pt}
	\noindent
	
I, Mitodru Niyogi, hereby declare that this master thesis is my own work and that I used no sources other than those indicated. \\

Ich, Mitodru Niyogi, erkläre hiermit, dass ich diese Arbeit eigenständig verfasst habe und keine anderen Quellen als die angegebenen verwendet habe. \\

	

	\vspace*{50pt}
	
	\noindent
	Heidelberg, \printdate{2021-07-30}
	\cleardoublepage
	
	
	
	\section*{Abstract}
	

Natural language (NL) to code suggestion systems assist developers in programming Integrated Development Environments (IDEs) by translating NL utterances into compilable code snippet. This facilitates the programmers to answer “how to” questions while using a new language or API, without the need to adequately familiarize themselves with the new programming languages, and the latest commonly used open source APIs. 

The development of these systems poses several challenges because of the lack of large (NL, code) annotated corpora, the misalignment between NL and code tokens, the syntactic and semantic requirements of the target code, and its contextual dependencies.

The current approaches mainly involve hard-coded, rule-based systems based on semantic parsing. These systems make heavy use of hand-crafted rules that map patterns in NL or elements in its syntax parse tree to various query constructs and can only work on a limited subset of NL with a restricted NL syntax. These systems are unable to extract semantic information from the coding intents of the developer, and fail to infer types, names, the context of the source code, and the developer’s intent to get accurate system-level code suggestions. Hence, these systems are non-scalable for real-time deployment. 

In this master thesis, we present novel sequence-to-sequence deep learning models and training paradigms to map NL to general-purpose programming languages that can assist users with suggestions of source code snippets, given a NL intent, and also extend auto-completion functionality of the source code to users while they are writing a code. These developed architectures incorporate contextual awareness into neural models which generate source code tokens directly instead of generating parse trees/abstract meaning representations from the source code and converting them back to source code. 

The proposed pretraining strategy and the data augmentation techniques improve the performance of the proposed architectures. An ablation study of the system was performed to gauge the important components of the developed AI system. The proposed Seq2Seq-BART has been found to exceed the performance of a neural semantic parser, TranX, based on the BLEU-4 metric by 10.82\%. Thereafter, a finer analysis for the parsable code translations from the NL intent for  CoNaLA\footnote{\href{https://conala-corpus.github.io/}{https://conala-corpus.github.io/}}  challenge was introduced. The proposed system is bidirectional as it can be also used to generate NL code documentation given source code. Lastly, a RoBERTa masked language model for Python code was proposed to extend the developed system for code completion.

\section*{Zusammenfassung}
Natural-language-to-Code-Vorschlagssysteme unterstützen Entwickler innerhalb integrierter Entwicklungsumgebungen (IDEs), indem sie passend zu Absichten, ausgedrückt in Natürlicher Sprache (NL), Code-Snippets erzeugen.
Dies ermöglicht Programmierern erleichterten Zugang zu ihnen potenziell unbekannten Programmiersprachen oder (Open-Source) APIs.
Die Entwicklung dieser Systeme stellt eine große Herausforderung dar. Schwierigkeiten sind u.a. fehlende annotierte Korpora (NL bzw. Code), unzureichende Kongruenz von NL und Code-Token, sowie die syntaktischen und semantischen Anforderungen des Ausgabecodes und seine kontextuellen Abhängigkeiten.

Aktuelle Ansätze beschränken sich hauptsächlich auf regelbasierte Systeme, welche auf semantischem Parsing basieren. Diese Systeme machen starken Gebrauch von händisch aufgestellten Regeln, die natürlichsprachliche Muster oder Elemente in ihrem Syntaxbaum auf verschiedene Abfragekonstrukte abbilden. Sie können nur in bestimmten Spezialfällen von NL mit eingeschränkter Syntax funktionieren. Diese Systeme sind nicht in der Lage, semantische Informationen aus Absichten des Programmierers zu extrahieren und können keine Typen, Namen oder den Kontext des Quellcodes ableiten, um genaue Codevorschläge auf Programmierebene zu erhalten. Daher sind sie für den Einsatz in Echtzeit nicht skalierbar.

In dieser Masterarbeit betrachten wir neue Sequence-to-sequence Deep-Learning-Modelle und Trainings-Paradigmen, um NL auf Programmiersprachen abzubilden. Neben dem Vorschlagen von Quellcode ermöglicht dies erweiterte Autovervollständigung. Statt Parse-Bäume bzw. andere abstrakte Bedeutungsdarstellungen aus dem Quellcode zu generieren und zurückzuübersetzen, integrieren unsere Architekturen Kontextbewusstsein in neuronale Modelle, welche Quellcode-Token direkt erzeugen.

Die vorgeschlagene Pretraining-Strategie und die Datenerweiterungstechnik verbessern die Leistungsfähigkeit der Architekturen. Um die wichtigen Komponenten des entwickelten KI-Systems zu messen, wurde eine Ablationsstudie durchgeführt. Wir stellen fest, dass der vorgeschlagene Seq2Seq-BART die Leistung eines neuronalen semantischen Parsers, TranX, basierend auf der BLEU-4-Metrik um 10,82\% übertrifft. Weiterhin wurde eine genauere Analyse der Parser-geeigneten Code-Übersetzungen aus NL für \href{https://conala-corpus.github.io/}{CoNaLa}\footnotemark[1] 
durchgeführt. Unser System ist bidirektional, kann also auch verwendet werden, um eine (natürlichsprachliche) Dokumentation von gegebenem Quellcode zu erzeugen. Schließlich schlagen wir ein RoBERTa Masked Language Model für Python-Code vor, welches das System um Codevervollständigung erweitert.

	\cleardoublepage
	
\section*{Acknowledgements}
Firstly, I would like to express my sincere gratitude to my supervisor, Prof. Dr. Artur Andrejak for introducing me to this domain of neural code synthesis and continuously supporting me with patience, warm encouragement, and guidance. \\ 

Furthermore, I owe my deepest gratitude to my SAP manager, Dr. Dejan Kovachev for providing me with his guidance and dedicated involvement in every step throughout the work. His insightful comments, motivation, and encouragement guided me in every step of this journey. I am also grateful to the SAP family for financially supporting this work and providing me with computing resources and GPU servers. \\

I wish to thank my friends, Antonio, Simon, Pascal, Sebastian, and Ricardo for proofreading the thesis and helping me to translate the abstract into German. Most importantly, none of this could have happened without my family. I want to express my profound gratitude to my beloved parents and my loving sister, Moupiya, for proofreading my thesis and providing me with the opportunity to pursue my dreams with their continuous encouragement, motivation, and support throughout my career. \\

I perceive this opportunity as a big milestone in my career development. I will strive to use gained skills and knowledge in the best possible way, and I will continue to work on their improvement, to attain desired career objectives. I hope to continue cooperation with all of you in the future. \\

Thank you.	

	\clearpage

	\tableofcontents
	\pagenumbering{arabic} 
	
	\listoffigures
	\listoftables
	
	
	\chapter{Introduction}\label{intro}
Mapping instructions in Natural Language (NL) into programs has become a popular task in natural language processing (NLP) with numerous applications for both programmers, and non-programmer users. The non-programmer users can benefit from applications such as home automation devices, digital voice assistants, and NL interfaces to databases (NLIDBs) which take natural language as input and translate them into programs. The programmers can benefit from assistive technological applications such as automatic code generation in Integrated Development Environments (IDEs),  source code retrieval from large codebases, and bug identification systems. These systems are very relevant today when software applications are ubiquitous and the number of people writing source code is growing at a humongous rate. 

A large number of software packages/bundles are being released every month, almost making it impossible for developers to adequately familiarize themselves with the latest and the most used APIs. Therefore, training deep learning models to automatically generate code from natural language can significantly improve developers' efficiency and source code quality besides saving time at work. This also facilitates the non-programmers users to effectively communicate complex intent with WiFi-enabled devices using natural language.
	
In earlier NL to code systems, researchers focused mainly on the task of semantic parsing by mapping NL to special purpose abstract meaning representations such as lambda calculus \citep{Zettlemoyer05learningto}. These hard-coded ruled systems worked to serve the objective but were found to be expensive and inextensible for real-world deployment. These systems made heavy use of hand-crafted rules that mapped patterns in the NL or elements in its syntax parse tree to various query constructs and could only work on a limited subset of natural language with a restricted NL syntax.

In recent years, the use of deep learning in language modeling, text auto-completion, and text generation has made tremendous progress and gained much attention from the research community. Among the different areas related to language processing, one of the most notable is the statistical modeling of the programming languages for source code generation from NL text, and source code prediction. 
	
	\section{Motivation} 
	The use of natural language by a developer to express coding intention and get it translated into code segments, is an interesting problem that, if solved, can reduce the need for developers to search online sources or prevalent documentation for helpful code snippets. The current code completion tools have focused on specific token types or features, and are limited to suggesting only variables or methods, often failing to have a holistic view of the surrounding context. On the other hand, larger code snippet suggestions are also possible, but they are mainly search-based and output generic code that needs to be further adapted by the developer. 
	
	The current approaches don’t consider the coding intent of the developer, which is the most valuable signal about what needs to be inserted in the code next. To correctly suggest a whole line of code, the working system requires to infer types of the target token for method completion, and the correct local variables to be passed as arguments to the method. Furthermore, additional structural and semantic information needs to be extracted from the context to make accurate statement-level suggestions.
	
	In this thesis, we aim to leverage the user input expressed in natural language to translate into source code in order to suggest auto-generated working code from natural language intent that will assist programmers to get code suggestions from an AI system to write code that can be embedded in the working code flow given developer's control to adapt the suggested code for their needs. In the following sections, we discuss the problem definition, goal, approach, challenges, contribution, and the structure of the thesis in detail.
	
	\section{Problem Definition}
	The problem can be further thought of developing an AI system that 1) translates natural language into code on the go, such as when a developer while writing code for a specific program forgets some code recipe to implement, the model will be able to assist the developer by generating source code given natural language text by the developer; the model should understand the context of the intent and the source code of the program while mapping the NL to source code to predict the entire sequence of source code tokens as a valid working line of source code snippet from the natural language, 
	2) to extend its usability for an assistive code completion feature of source code, i.e., the system should be able to predict the next code tokens given the previous tokens written by the developer. E.g., if the user needs to visualize a bar chart of CSV files using Pandas and Matplotlib, then the user would write this intention in-text pointing to the data, the visualization type, and data columns to use, then a valid compilable Python Jupyter code snippet would be generated.
	
	The aim is to create a machine learning system/service which will integrate the above-mentioned subproblems by investing in research of the state-of-the-art deep learning models, evaluation, and improvements over the existing architecture. 
	
	Subproblem 1 defines the model for source code translation from natural language, and subproblem 2 is an extension of the code generation problem from NL intent as an auto code completion feature alike Gmail's Smart-Compose for email writing but for writing code in this case.
	
	\section{Goal}
	\textbf{Natural language to code translation}
	
	This research aims to conduct a set of experiments combining different state-of-the-art deep neural network architectures designed for sequence-to-sequence (Seq2Seq) learning and for developing novel hybrid Seq2Seq architectures with state-of-the-art architectures built for NLP tasks such as abstractive text summarization, machine translation. 
	
	The research also aims to see how using different neural-based tokenization for the input can make the developed architecture learn more about the contextual embeddings and be aware of the context of the input and their respective mapping to the target objective of translating natural language to the source code snippet. 
	
	The research also aims to develop techniques on top of the proposed architectures and to use the concept of transfer learning, which leverages the extra knowledge gained from the pretraining objective of learning the mappings of unstructured text to structured code, to generate more diverse and accurate source code target from the unstructured text input. This also results in achieving better evaluation metrics scores for the generated code snippet. 
	
	The research also delves into developing a versatile Seq2Seq architecture that can be used for both objectives of translating text to code (NL2Code) and also generating comments, docstring, method documentation from structured source code input (Code2NL). The research also aims to develop a data augmentation technique using back-translation from the Code2NL task to make the training and validation dataset bigger for the NL2Code task which would eventually make the developed models learn more from the bigger dataset and generalize the model for unseen data. 
	
	The research also aims at performing an ablation study regarding the developed architectures to gauge the importance of the crucial components of the developed AI system for the NL2Code objective. The research also develops a finer qualitative analysis of the generated source code from the text input and also introduces a new qualitative analysis technique of the generated source code evaluation in terms of the number of valid parsable code generated from the AI system as well as their individual Sentence-BLEU \citep{10.3115/1073083.1073135} metric comparison of the generated code snippet with the ground truth for classification of generated structured source code into finer categories of further qualitative analysis and auto-completion (for example, filling the blanks). 
	
	\textbf{Code completion}
	
	The research aims to develop a novel neural language model based on RoBERTa architecture for source code. This is the first of its kind RoBERTa masked language model trained from scratch on code corpus. The research also aims at investing how well the trained language model on source code and natural language pair works for the fill-in-task where we mask some of the code tokens in the code snippet and allow the language model to predict the masked code tokens and compare the predicted masked tokens with the ground truth.
	
	\section{Approach}
	The natural language to code translation can be perceived as a machine translation problem. Therefore, in this thesis, we plan to model various existing Encoder-Decoder architectures suited for machine translation and fine-tune for natural language to code translation on the CoNaLa dataset. The idea of fine-tuning a model for a downstream task is a way to perform transfer learning from a pretrained model on a large corpus for a specific task and use the pretrained model to train on a different smaller annotated dataset to fine-tune the learned weights for the specific downstream task.
	
	\citep{wang2020fullline}, had found out that syntax-based approaches such as ASDL action sequences/ Abstract Syntax Tree (AST) syntax-based modeling action sequence \citep{10.1145/3387904.3389261}, semantic parsing, etc., do not outperform neural token sequence-based ones, and are even found to perform worse on some datasets such as Py150k\footnote{\href{https://www.sri.inf.ethz.ch/py150}{https://www.sri.inf.ethz.ch/py150}}.  
	
	It has also been seen that BPE-NLM \citep{Karampatsis2020BigC}, a GRU language model for code completion using byte-pair encoding \citep{sennrich-etal-2016-neural} to represent source code tokens, has achieved state-of-the-art results on code completion tasks in multiple programming languages. Thus, we aim to focus more on sequence-based neural source code modeling with different neural tokenization modeling techniques such as Byte-pair encoding, SentencePiece \citep{kudo2018sentencepiece}, WordPiece \citep{6289079}, along with character-based Python tokenizer instead of AST \citep{10.1145/3290353} based representation of source code.

	Firstly, we considered the NL2Code task as a sequence-to-sequence task, i.e.,  encoding the input natural language intent and decoding the generated code snippet from the sequence-to-sequence framework. We formulated our problem by implementing the Seq2Seq method with attention mechanism \citep{bahdanau2016neural} used for machine translation to translate natural language to code (NL2Code). 
	
	Secondly, we implemented Convolutional Seq2Seq learning \citep{pmlr-v70-gehring17a} for Nl2Code which was initially used for text classification.
	
	Thirdly, we considered implementing the Transformer \citep{vaswani2017attention} based architecture and evaluating how the Transformer model exclusively designed for NLP tasks can be used in our use case of the NL2Code task.
	
	Fourthly, we proposed a hybrid Seq2Seq architecture with RoBERTa \citep{liu2019roberta} as encoder and decoder and we also proposed another hybrid Seq2Seq architecture with BART \citep{lewis2019bart}. We used the concept of transfer learning and the data augmentation technique to pretrain the weights of the deep learning NLP models on the mined corpus and then fine-tune the weights for NL2Code on the (NL, code) pairs of the augmented training set.  
	
	We also performed an ablation study to understand the important components of the proposed Seq2Seq architecture and the transformer-based architectures that we have developed in this work. The pretraining process helped the proposed architecture to enrich its knowledge from the natural language, code pairs as the standard deep learning models are exclusively built for NLP tasks such as machine translation, text generation, which are pretrained on unlabeled English corpus. Thus, the architecture we built was fine-tuned for the natural language, source code pairs dataset. 
	
	Lastly, for NL2Code, we performed both quantitative and qualitative studies about the translated prediction from the model and their analysis to see where the model fails and where the model has been able to translate the NL into code correctly. The task was to identify the pattern of failures and try to fix it by adapting possible approaches such as in the preprocessing steps of the input sequence by using different neural-based subword tokenizers to parse the source code and natural language tokens or assisting the proposed model architecture with pretrained knowledge from the (NL, code) dataset learned from the automatically annotated mined corpus, and by augmenting the small available training dataset using back-translation of source code to natural language generation. 
	
	The data augmentation and the pretraining technique of the proposed hybrid Se2Seq architecture on the mined CoNaLa corpus helped us to generate more diverse and accurate code predictions from the natural language intent and achieve better BLEU metric than the state-of-the-art neural semantic parser, TranX \citep{yin-neubig-2018-tranx}. 
	
	For the code completion task, we proposed a RoBERTa \citep{liu2019roberta} based masked language model for source code. We pretrained the RoBERTa mask language model from scratch on the Algorithms' \footnote{\href{https://github.com/TheAlgorithms/}{https://github.com/TheAlgorithms/}} Python code corpus to build a BERT-based language model for source code and further use this source code language model for code completion task.  
	
	\section{Challenges}
	Amongst the many challenges that models have to address for mapping language to general-purpose source code, we explicitly address the main challenges which are as follows:
	
	\textbf{Lack of big (NL,code) pair corpora}
	
	The absence of a proper large (NL, code) pair annotated dataset limited us in exploring the full capacity of our proposed developed deep learning model. As deep learning architectures are data-hungry, the presence of a large annotated training dataset is beneficial for learning and generalization. We have also seen some of the neural architectures were too big to easily overfit on the CoNaLa training dataset that we have used in this work. 
	
	To alleviate this problem, we have proposed a data augmentation technique using back-translation of the hypothesis, i.e., generating new natural language intent from the code snippet to add new training examples in the dataset. We were able to augment the CoNaLa original training and validation sets by $1x$, $3x$, $5x$ of the original size. After training our proposed model on the augmented dataset, the model has given better quantitative results than on the original training set.
	
	\textbf{GPU memory limitations}
	
	The inability to access more than one GPU due to budget and hardware constraints limited us in exploring very intensive computational models on large CoNaLa mined corpus for pretraining purposes. The lack of access to adequate GPU resources didn't allow us to train GPT \citep{Radford2019LanguageMA} model from scratch on a large code database for code completion and source code modeling.
	
	\textbf{Syntax decoding and lack of diverse output} 
	
	The lack of highly featured engineered module structure in our developed architecture to handle structured code generation from a natural language such as parse trees of source code, or incorporation of syntax structures like parent-child links in ASTs, posed a challenge for the deep learning architectures to produce syntactically correct code without any prior knowledge of grammar, syntax, rules of programming languages. The generative models developed especially for text generation problems often suffer from the lack of diverse and repetitive text generation.
	
	In order to address this problem, we have increased the temperature $\tau$ of the $softmax$ $\exp(z_{i}/\tau )/\sum_{j}^{}(z_{j}/\tau)$ while decoding the output which is a simple way for trying to encourage more diversity in decoded outputs. We also set diverse penalty constraint while decoding the outputs from the decoder to constraint the decoder not to generate repetitive n-grams. 
	
	\textbf{Extensibility}
	
	NL2Code systems have the additional challenge of handling dynamic user instructions which change, evolve, and be of different data distribution from the initial data that the models were originally trained on. The earlier statistical semantic parsers were expensive to retrain and did not scale their capability to unseen data, hence, lack extensibility properties. In this thesis, we advocate an inexpensive alternative way to train models for semantic parsing by using a general-purpose popular programming language, Python. Additionally, the ubiquitous nature of Python has permitted us to swiftly solicit new annotations to periodically retrain deployed models to adapt to dynamic user input distributions. 
	
	\textbf{Evaluation Metric}
	
	For code synthesis, BLEU \citep{10.3115/1073083.1073135}, ROUGE \citep{lin-2004-rouge}, and perfect accuracy are the commonly used evaluation metrics. However, both BLEU and ROUGE metrics are not suitable as they neglect the important syntactic and semantic features of codes, and perfect accuracy is too strict to consider the different correct outputs with the same semantic logic.
Thus, we proposed to use the METEOR \citep{banarjee2005} metric for the Code2NL objective. METEOR considers syntactic and semantic features by looking for exact, stem, synonym, and paraphrase matches between words and phrases between translation and references.

	\section{Contribution}
	We have worked on this thesis topic for more than seven months since December 2020. The work involved designing the problem, literature survey, researching the state-of-the-art, designing and implementing the novel model architectures, performing experiments, analyzing the results, and finally drawing comparisons. In this section, we summarize the overall contribution as follows:
	\begin{itemize}
		\item We proposed a novel hybrid SeqSeq-RoBERTa architecture with RoBERTa as an encoder and decoder for the NL2Code objective.
		
		\item We proposed a bidirectional hybrid Seq2Seq-BART with BERT encoder and GPT decoder architecture that works for both NL2Code and Code2NL objectives. 
		
		\item We introduced a pretraining technique of the hybrid Seq2Seq architectures on the larger mined corpus  then used transfer learning to fine-tune the learned weights on the smaller annotated CoNaLa training set.
		
		\item We introduced a data augmentation technique using back-translation of Code2NL to generate new intents in natural language from code snippets and used them to augment the training set to make it 3$x$, 5$x$, 7$x$ in size of the original CoNaLa training set.

		\item We performed an empirical and comparative study of transformer-based architecture, vanilla Seq2Seq, GRU-based architecture, proposed hybrid Seq2Seq architectures with RoBERTa, and BART.
		
		\item We also performed ablation studies on the transformer and Seq2Seq-RoBERTa architecture regarding the ablation of the number of self-attention heads, and the number of layers.

		\item We also introduced a detailed qualitative and quantitative analysis for the CoNaLa challenge by calculating Sentence BLEU for each generated code snippet from natural language and classified the generated code snippets into subgroups of exact match, mostly correct, marginally correct, semantically equivalent, and wrong translations.
		
		\item We introduced a valid parsable count metric for the CoNaLa challenge by parsing the generated Python code snippets into AST parser that measures the number of the valid parsable generated code snippet.
		
		\item We developed a RoBERTa masked language model exclusively for Python source code for the code completion objective. 
		
	\end{itemize}
	
	\section{Structure}
	\hyperref[relatedwork]{Chapter 2} describes the contents of related work in reference to this work.
	Together with \hyperref[background]{Chapter 3}, these formulate the relevant theoretical background knowledge this thesis work is based on.
	\hyperref[approach]{Chapter 4} discusses the proposed NL2Code system architecture, empirical study design, dataset used, model configuration, implementation of the model design. 
	
	 In \hyperref[dataset]{Section 4.1}, \hyperref[curated-dataset]{Section 4.2} we describe dataset that we use and the methods for creating our own curated dataset. In \hyperref[system-design]{Section 4.3}, we describe the proposed NL2Code system architecture. In \hyperref[model-config]{Section 4.4}, we discuss the configuration of various deep learning architectures developed and implemented in this thesis, to translate the natural language intent into source code snippet. In \hyperref[implement]{Section 4.5}, we briefly discuss the implementation details of various deep learning architectures proposed to solve the NL2Code objective, and code completion tasks. 
	 
	 Correspondingly, in \hyperref[evaluation]{Chapter 5}, we discuss the results of various experiments conducted for the proposed NL2Code system, and for the empirical study evaluation. We pose research questions and also briefly answer the research questions through our evaluation and analysis of the results. 
	 We also document the deduction of key insights from the work described in this thesis. We have also included a comparison of different deep learning architectures developed in this thesis for the NL2Code objective, as well as a quantitative, qualitative view, and the corresponding comparison of the developed hybrid Seq2Seq architectures with the semantic neural parser, TranX. We conclude the thesis in \hyperref[conclusion]{Chapter 6} with a brief reflection on the work done and possible future work.
	
	
	\chapter{Related Work}\label{relatedwork}
		Machine learning has shown great promise towards improving automated software engineering across all stages such as in assistive code suggestion systems or even code auto-completion systems. The researchers were influenced to use statistical  N-gram language models to model source code by  \citep{10.5555/2337223.2337322} who showed that the programming source code shares mutual statistical properties with natural language. \citep{10.1145/2635868.2635875} pointed out traditional N-gram language models are unable to capture the ``localness'' property of source code. An improved cache-based N-gram model that addresses the localness property and unlimited vocabulary in source code was introduced by \citep{10.1145/3106237.3106290} for predicting source code tokens. 
		
		Source code generation from natural language has also been considered as a semantic parsing problem by various researchers in generating structured code snippets. Semantic parsing is the task of generating machine-executable meaning representations from natural language (NL) intent. \citep{10.1145/2983990.2984041} proposed a probabilistic language model based on decision trees and domain-specific grammar. They performed experiments on predicting AST nodes rather than directly performing on source code. Semantic parsing researchers have recently used deep learning models to automatically generate source code from natural language. One of the key challenges of code generation is to generate syntactically valid compilable code snippets. 
		
		To address the challenge of generating valid compilable code snippets from natural language, \citep{dong-lapata-2016-language} proposed a tree-structured Long Short Term Memory (LSTM) architecture named SEQ2TREE to directly generate Abstract Syntax Tree (ASTs) from source code, but SEQ2TREE often failed to guarantee the validness of generated trees to convert back into valid compilable source code. \citep{yin-neubig-2017-syntactic} addressed this issue by converting the generation of a code snippet into applying a sequence of actions which can either apply a grammar production rule or produce a terminal node as defined by the grammar of the programming language. The incorporation of Abstract syntax description language (ASDL) \citep{Wang1997TheZA} grammar into code generation using a tree-structured decoder was introduced by \citep{rabinovich-etal-2017-abstract}. \citep{yin-neubig-2018-tranx} proposed TranX, which replaces the context-free grammar in \citep{yin-neubig-2017-syntactic}  used to generate action sequences into ASDL grammars and further improved the code generation framework in \citep{yin-neubig-2017-syntactic} by expanding the framework into a wider range of languages including SQL, Python3, and regular expressions.
		
		Recently, deep learning-based language models have been applied to source code for the code completion task. \citep{liu2016neural} proposed a LSTM based model for code completion. A recurrent neural network (RNN) based model with a pointer mechanism aiming at copying tokens from the past was proposed by \citep{bhoopchand2016learning}. To address the Out-of-Vocabulary($OoV$) problems of RNN based models, \citep{Li2018CodeCW} proposed a pointer mixture network in predicting the type of the next AST node, for code completion and their approach outperformed \citep{10.1145/3022671.2984041} on both Python and JavaScript datasets. Recently, \citep{Karampatsis2020BigC} showed that instead of representing source code as ASTs, or action sequences, source code can be predicted by using a GRU \citep{chung2014empirical} language model with the use of byte-pair encoding (BPE) \citep{sennrich-etal-2016-neural} to overcome the open vocabulary problem by treating source code as unstructured text, which can be successfully used for code completion. 
		
		\citep{iyer2018mapping} used a two-step attention method over an LSTM based Seq2Seq architecture, one of which is a general attention mechanism and the other one is the attention over return types, variables, and methods from the environment. This improved the exact match accuracy and the BLEU score against the previous works based on RNNs. 
		However, RNNs share a common weakness: they are unable to efficiently capture the long-term dependencies in sequential data \citep{khandelwal-etal-2018-sharp}. 
		
		An efficient way of mitigating long-term dependency problems in neural network-based language models is to use the Transformer\citep{vaswani2017attention} based model. For code completion, \citep{liu2020selfattentional} adopted Transformer-XL \citep{dai-etal-2019-transformer} on the AST node completion task, and achieved state-of-the-art results. The state-of-the-art NLP models motivated us to explore more about the context-aware embeddings and the long-term contextual sequence-dependent deep learning models in sequence-to-sequence neural architecture framework to generate code from natural languages.

	\chapter{Background}\label{background}
	
	\section{Model architectures}
	\subsection{Seq2Seq}
	The sequence-to-sequence (Seq2Seq) model defines the conditional probability of discrete output sequence $Y = (y_{1}, \dots , y_{N})$ of length N, conditioned on the given sequence of discrete input $X = (x_{1}, \dots , x_{M})$ of length M using attention-based recurrent neural network (RNN) \citep{bahdanau2016neural}; \citep{luong2015effective}, Transformer \citep{vaswani2017attention}, or other neural sequence-generation architectures. A Seq2Seq model architecture is trained via Maximum Likelihood Estimation (MLE) to minimize the cross-entropy loss and learned parameters $\theta$ to estimate the conditional likelihood $P_{\theta}(Y \mid X)$,
	\begin{align*}
		L_{xe}(\theta) = - \log P_{\theta} (Y \mid X)
		= - \sum_{t=1}^{N} \log P_{\theta}(y_{t} \mid y_{1:t-1}, X)
	\end{align*}
	When the Seq2Seq architecture is applied to the NL-to-code translation problem, the input natural language and output commands are treated as sequences of tokens. At test time, the command sequences with the highest conditional probabilities were output as candidate translations.

	\subsection{Transformers}

\begin{figure}
	\centering
	\includegraphics[width=0.4\linewidth]{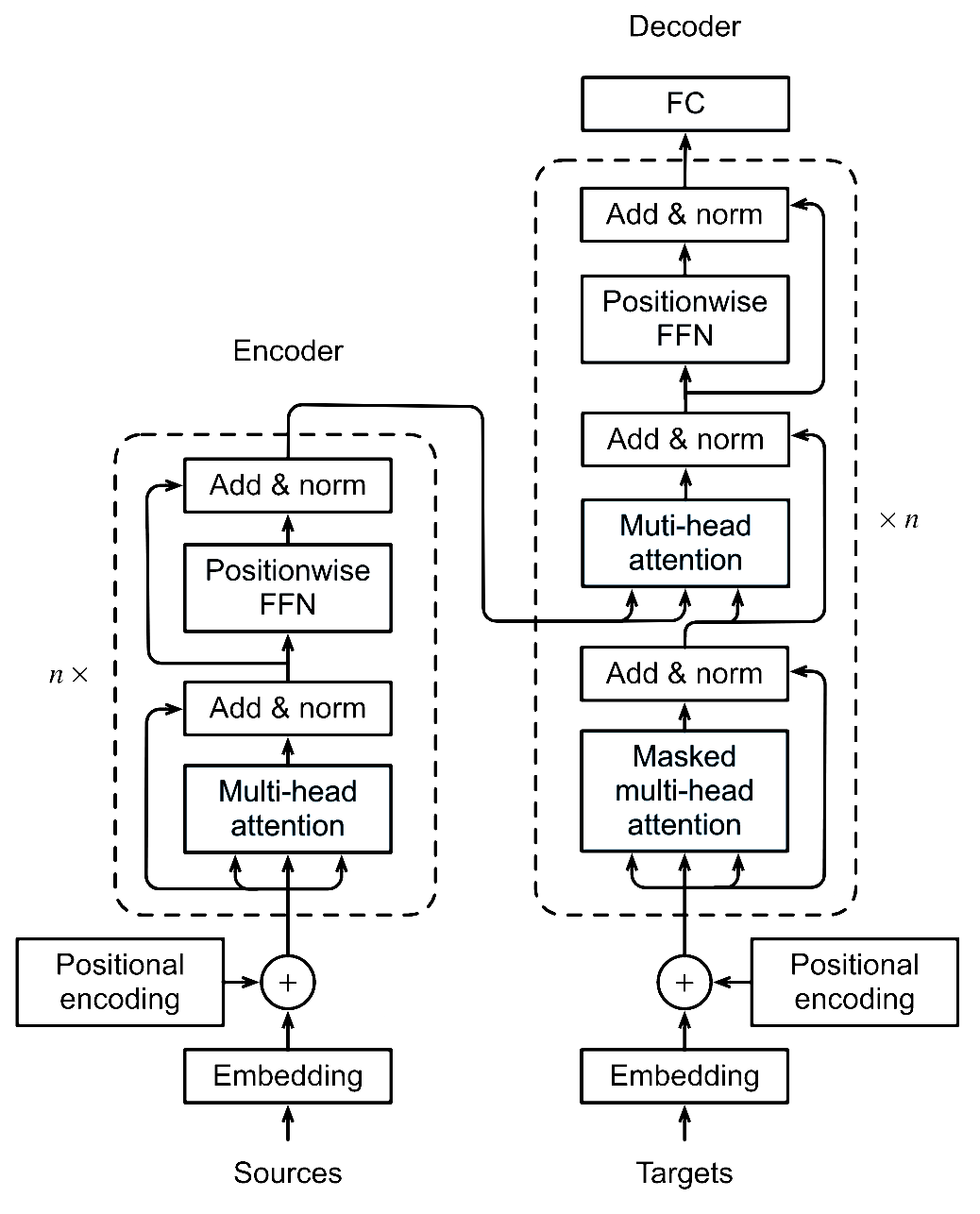}
	\caption[The Transformer model architecture.]{The Transformer model architecture. Figure drawn from \citep{vaswani2017attention}.}
	\label{fig:transf-1}
\end{figure}

A typical transformer \citep{vaswani2017attention} architecture for a sequence-to-sequence task has an encoder (a stack of transformer blocks) and a decoder stack which is composed of multi-head, self-attention layers, optionally containing residual connections and layer normalization \citep{ba2016layer}. The transformers are faster to train due to training parallelization as they do not need the input tokens to be processed in a specific order like the vanilla recurrent neural networks (RNNs) or their gated variants, including LSTM \citep{sutskever2014sequence} and GRU \citep{chung2014empirical}.

The Transformer follows this overall architecture using stacked self-attention and point-wise, fully connected layers for both the encoder and decoder, which are as follows:

Encoder: The encoder consists of a stack of 6 identical layers. Each layer is divided into two sub-layers. The first layer is a multi-head self-attention layer, followed by a simple, position-wise fully connected feed-forward network layer. The two sublayers are connected by a residual connection \citep{he2015deep}, followed by layer normalization \citep{ba2016layer}. This makes the output of each sub-layer as $LayerNorm(x + Sublayer(x))$, where $Sublayer(x)$ is the function implemented by the sub-layer itself. The sub-layers and the embedding layers produce outputs of dimension $d_{model}$ = 512.

Decoder: Similarly, the decoder consists of a stack of 6 identical layers. In addition to the two sub-layers in each encoder layer, the decoder has a multi-head attention layer which computes multi-head attention over the encoder output. The decoder layers also employ residual connections around each of the sub-layers, followed by layer normalization. 

\subsubsection{Scaled Product Attention}
The mapping of query vector with a set of key-value vector pairs to an output represents an attention function. The output of the attention function is computed as a weighted sum of the values, where the weight assigned to each value is computed by the dot product of the query vector with the corresponding key vector.
The input consists of queries and keys of dimension $d_{k}$, values of dimension $d_{v}$. The dot products of the query vector with all keys vector, divided each by $d_{k}$, and then the $softmax$ function is applied to obtain the weights on the value vectors.
In practice, the attention function is computed over a set of queries simultaneously packed together into a matrix $Q$. The keys and values are also packed together into matrices $K$ and $V$. The computation of the matrix of the output of the attention function is computed as:
\begin{align*}
	Attention(Q, K, V) = softmax(\frac{QK^{T}}{\sqrt{d_{k}}})V
\end{align*}

\begin{figure}
	\centering
	\includegraphics[width=0.6\linewidth]{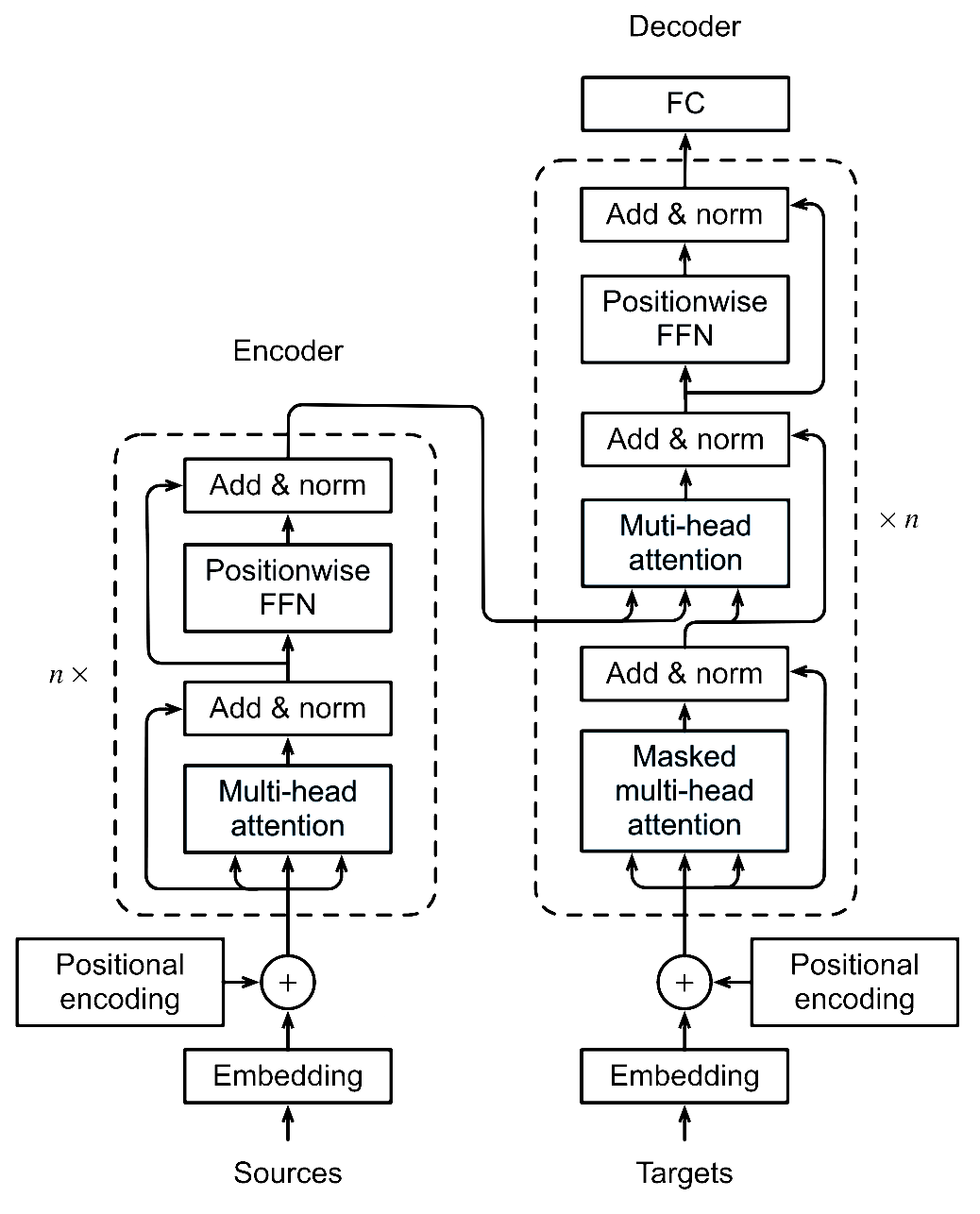}
	\caption[Scaled Dot-Product and Multi-Head Attention]{(left) Scaled Dot-Product Attention. (right) Multi-Head Attention consisting of several attention layers running in parallel. Figure drawn from \citep{vaswani2017attention}.}
	\label{fig:transf-2}
\end{figure}

\subsubsection{Multi-head attention}
Muti-head attention is a way to linearly project the $d_{model}$-dimensional  queries, keys and values $h$ times with different, learned linear projections to $d_{k}$, $d_{k}$ and $d_{v}$ dimensions, respectively. The attention function is computed in parallel over each of these projected queries, keys, and values, which yields $d_{v}$-dimensional output values. The output is concatenated and once again projected, resulting in the final values, as depicted in \autoref{fig:transf-2}.

Multi-head attention in the transformer architecture enables the transformer model to jointly attend to information from different representation subspaces at different positions. 
\begin{align*}
	\begin{split}
		MultiHead(Q, K, V ) = Concat(head_{1}, \dots, head_{h})W^{O} where \\
		head_{i} = Attention(QW_{i}^{Q}, KW^{K}_{i}, VW^{W}_{i} )
	\end{split}
\end{align*}
Where the projections are parameter matrices
$W_{i}^{Q}$  $\epsilon$  $\mathbb{R}^{d_{model} \times d_{k}} $, 
$W_{i}^{K}$ $\epsilon$ $\mathbb{R}^{d_{model} \times d_{k}}$
, $W_{i}^{V}$ $\epsilon$ $\mathbb{R}^{d_{model} \times d_{k}}$
and $W^{O}$ $\epsilon$ $\mathbb{R}^{hd_{v} \times d_{model}}$
In the original paper \citep{vaswani2017attention}, $h$ is set to 8 parallel attention layers, or self-attention heads, and $d_{k}$ = $d_{v}$ = $\frac{d_{model}}{h}$ = 64. The total computational cost for multi-head attention is similar to that of single-head attention with full dimensionality due to the reduced dimension of each attention head.

\subsection{How Transformer uses multi-head attention?}
The Transformer uses multi-head attention in three different ways:
\begin{itemize}
	\item In the ``encoder-decoder attention'' layers, the memory keys and the values come from the output of the encoder, and the queries come from the previous decoder layer. This allows every position in the decoder to attend all positions in the input sequence. 
	
	\item The self-attention layers in the encoder allow each position in the encoder to attend to all positions in the previous layer of the encoder. In a self-attention layer, all of the queries, keys, values, come from the output of the previous layer in the encoder.
	
	\item Each position in the decoder can attend to all positions in the decoder up to and including that position via self-attention layers. The leftward information flow in the decoder is prevented to preserve the auto-regressive property. Therefore, the authors masked out (setting to $-\infty$) all values in the input of the $softmax$ that correspond to illegal connections inside of the scaled dot-product attention as shown in \autoref{fig:transf-2}.
\end{itemize}
		
	\subsection{Bidirectional Encoder Representations from Transformers (BERT)}
	The BERT \citep{devlin-etal-2019-bert} model is an attention-based bidirectional language model. It is based on transformers architecture. When pretrained on a
	large language corpus, BERT has proven to be very effective for transfer learning to multiple natural language processing task. The
	BERT model operates on sequences of word tokens $w_{0}, . . . , w_{T}$ . These tokens are mapped to learned encodings and passed through
	passed through L ``encoder-style'' transformer blocks to produce final representations $h_{0} , . . . , h_{T}$ . 
	
	The computation of a single encoder-style transformer block consists of a multi-headed attention block followed by a small fully-connected network, both wrapped in residual adds. Let $H^l$ be a matrix with rows $h_{0} , . . . , h_{T}$ corresponding to the intermediate representations after the $l^{th}$ layer. The intermediate representation $H^l$ is used to compute three matrices – $Q$, $K$, and $V$ – corresponding to queries, keys, and values that drive the multi-headed attention block. Specifically, the dot-product similarity between queries and keys determines attentional distributions over value vectors. The resulting weight-averaged value vector forms the output of the attention block.
	
	\subsubsection{Text representation:} BERT operates over sequences of discrete tokens comprised of word pieces and a small set of special tokens:
	$SEP$, $CLS$, and $MASK$. For a given token, the input representation is a sum of a token-specific learned embedding, position embeddings (i.e., token’s index in the sequence), and segment embeddings (i.e., index of the token’s sentence if multiple sentences exist)
	
	\begin{figure}
		\centering
		\includegraphics[width=0.8\linewidth]{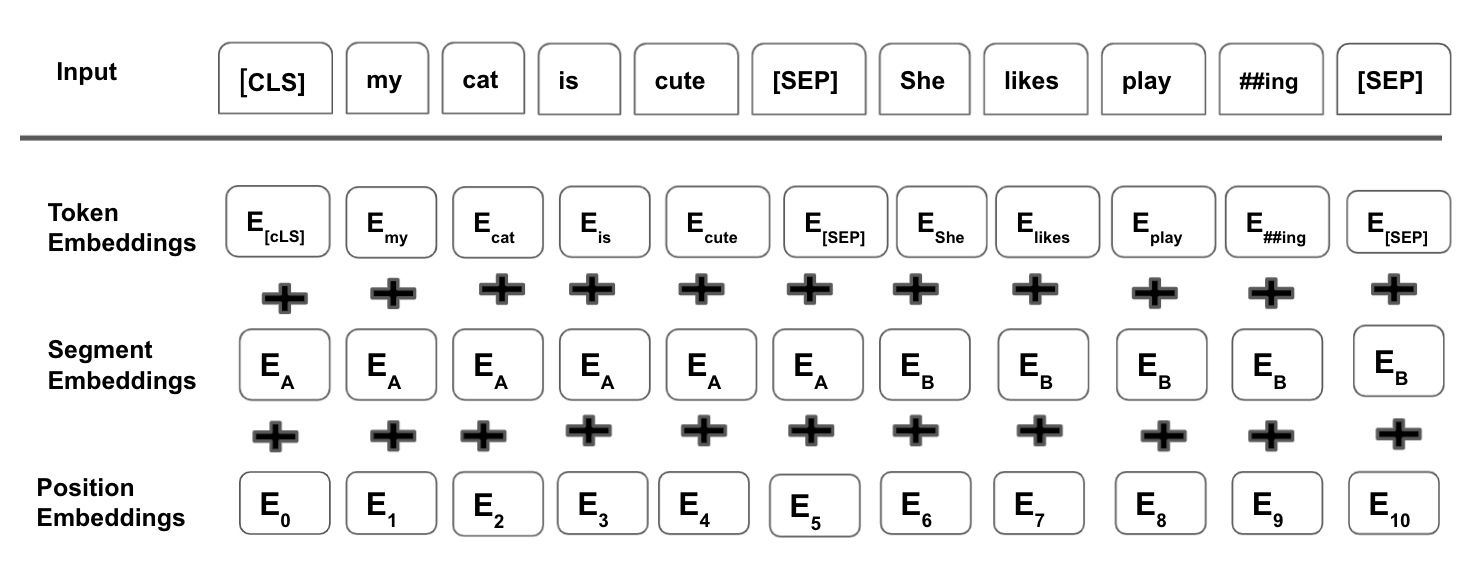}
		\caption[Encoding representation in BERT]{Encoding representation in BERT. Figure drawn from \citep{devlin-etal-2019-bert}.}
		\label{fig:bert-fig1}
	\end{figure}

	\subsubsection{Training Task and Objectives}
	The BERT model is trained end-to-end on a large language-corpus i.e., English Wikipedia and Book corpus under two tasks: masked language modeling and next sentence prediction.
	
	\begin{itemize}
		\item \textbf{Masked language modeling}: The masked language modeling task randomly divides input tokens into disjoint masked $X_{M}$ and observed $X_{O}$ tokens (approximately 15\% of tokens being masked). Masked tokens are replaced with a special $MASK$ token 80\% of the time, a random word 10\% of the time, and an unaltered 10\% of the rest. 
		The BERT model is then trained to reconstruct these masked tokens given the observed set. The final hidden vectors corresponding to the mask tokens are fed into an output $softmax$ layer over the vocabulary and the model is trained under a cross-entropy loss.
		
		\item \textbf{Next sentence prediction}: The model takes tuple of sentence segments $A$ and $B$ following the format {$CLS$, $w_{A1}$ , $\dots$ , $w_{AT}$ , $SEP$, $w_{B1}$ , . . . , $w_{BT}$ , $SEP$} and is trained to
		predict whether or not $B$ follows $A$ in the source text. $CLS$ token representation is fed into an output layer which is trained to minimise a binary cross-entropy loss on this label.
	\end{itemize}
	
	\subsubsection{Advantages/Limitations of BERT over Seq2Seq model:}
	BERT attends the entire input sequence as a whole sequence, unlike LSTMs where the sentence is processed sequentially. There is no Backpropagation through time in BERT alike LSTMs, where they backpropagate the error back in time through words, one word at a time.
	
		\subsection{RoBERTa: A Robustly Optimized BERT Pretraining Approach}
	The authors of RoBERTa \citep{liu2019roberta} found out that BERT was significantly undertrained and then they proposed RoBERTa which can match or exceed the performance of all of the post-BERT methods. The BERT architecture was modified to develop RoBERTa as follows: (1) training the BERT model longer over more data, with larger batches; (2) removing the next sentence prediction objective from BERT; (3) training on longer sequences; and (4) dynamically changing the masking pattern applied to the training data. Specifically, RoBERTa is trained with dynamic masking (\autoref{dmask}), FULL-SENTENCES without NSP loss(\autoref{nonsp}), and large mini-batches.
	
	RoBERTa uses a batch size of 2k, 8k sequences depending on the base or the large model, whereas BERT uses a batch size of 256 sequences. It is easier to parallelize large batch sizes via distributed data-parallel training.
	
	The removal of BERT's next sentence prediction pretraining objective; training the BERT with much larger batches and learning rates improved RoBERTa on the masked language modeling objective compared with BERT. RoBERTa was pretrained on 160GB of text data consisting of five English-language corpora of varying sizes and domains. In the following subsections, we discuss the various pretraining objectives proposed in the RoBERTa.
	
	\subsubsection{Pretraining Objectives}
	The following pretraining objectives were proposed and explored by the authors for RoBERTa which are as follows:
	\subsubsection{Dynamic Masking}\label{dmask}
	The original BERT implementation results in generating a single static mask by performing token masking during data preprocessing. In RoBERTa, dynamic masking had been performed where the masking pattern was generated every time for every input sequence to avoid using the same mask for each training instance in every epoch. 
	
	\subsubsection{Model Input Format and Training Format} \label{nonsp}
	\begin{itemize}
		\item \textbf{SEGMENT-PAIR with NSP}: This is the original
		input format used in BERT, with the Next Sentence Prediction (NSP) loss. Each input has a pair of segments and the total length of the sequence is less than 512 tokens.
		
		\item \textbf{SENTENCE-PAIR with NSP}: Pair of sentences sampled from the contiguous portion of a single document or separate documents for each input. As these inputs are significantly shorter than 512 tokens, the batch size was increased so that the total number of tokens remains similar to SEGMENT-PAIR with NSP. The NSP loss has been retained in this approach.
		
		\item \textbf{FULL-SENTENCES w/o NSP}: The input consists of full sentences sampled contiguously from one or more documents, such that the total length of the input sequence is at most 512 tokens. When the end of one document is reached while processing the sentences, the next sampling sentences are taken from the
		next document with an extra separator token
		between documents. The NSP loss has been removed.
		
		\item \textbf{DOC-SENTENCES w/o NSP}: Inputs are sampled near the end of a document and constructed in a way such that they may not cross the document boundaries. Inputs are generally shorter than 512 tokens. Thus, the batch size has been increased so that the total number of tokens remains similar to FULL-SENTENCES. The NSP loss was removed from the pretraining objective.
	\end{itemize}

	\subsection{BART}
	\citep{lewis2019bart} introduced BART as a denoising autoencoder for pretraining sequence-to-sequence models that applies to various seq2seq tasks such as machine translation, summarization, etc.  BART has achieved new state-of-the-art results on a range of abstractive dialogue, question answering, and summarization tasks.
	
	The denoising autoencoder architecture of the BART maps a corrupted document to the original document it was derived from. BART is implemented in the structure of sequence-to-sequence model with a bidirectional encoder which encodes the corrupted text and a left-to-right autoregressive decoder to reconstruct the original document from the corrupted document. BART is trained by corrupting documents and then optimizing the reconstruction cross-entropy loss between the original document and the decoder’s output.
	
	BART uses the standard sequence-to-sequence Transformer architecture from \citep{vaswani2017attention}, except, following GPT, BART uses GeLUs \citep{hendrycks2020gaussian} activation function and initializes parameters from N (0, 0.02). The BART-base model has 6 layers in the encoder and the decoder, and the BART-large model has 12 layers in each encoder and decoder. BART can be considered as generalizing BERT due to the presence of a bidirectional encoder and generalizing GPT due to the presence of a left-to-right decoder in its architecture. Thus, BART has been particularly effective when fine-tuned for text generation and it also works well for comprehension tasks.
	
	BART allows noise variability in its input by allowing arbitrary transformations that can be applied to the original text, including the change in the length of the input. After evaluating several noising approaches, the authors of BART found out the best performance by using a novel in-filling scheme,
	where a single mask token is used to replace any arbitrary length spans of text (including zero-length), and randomly shuffling the order of the original sentences. This approach in BART generalizes the next sentence prediction, and the original token masking objectives in BERT by forcing the model to make a longer range of input transformations and reason more about overall sentence length. BART is also able to reduce the mismatch between pre-training and generation tasks because the decoder is always trained on an uncorrupted context.
	
	The architecture of BART is similar to BERT \citep{devlin-etal-2019-bert}, with the following differences: (1) each layer of the left-to-right decoder in the BART additionally performs cross-attention operation over the final hidden layer of the encoder as in the transformer sequence-to-sequence architecture; and (2) BART does not use any additional feed-forward network before word prediction like BERT. In total, BART model has roughly 10\% more parameters than the equivalently sized BERT model.
	
	\subsubsection{Pretraining objectives of BART}
	In this section, we discuss the various pre-training tasks of BART.

\begin{figure}
	\centering
	\includegraphics[width=0.6\linewidth]{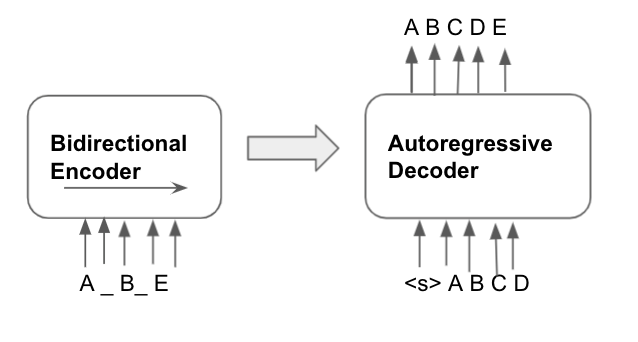}
	\caption[The BART architecture.]{BART: Inputs to the encoder do not need to be aligned with the decoder outputs, allowing arbitrary noise transformations. The corrupted document is encoded with a bidirectional encoder (left), then the likelihood of the original document is calculated with an auto-regressive decoder (right). Figure drawn from \citep{lewis2019bart}.}
	\label{fig:bart-1}
\end{figure}

\begin{figure}
	\centering
	\includegraphics[width=0.7\linewidth]{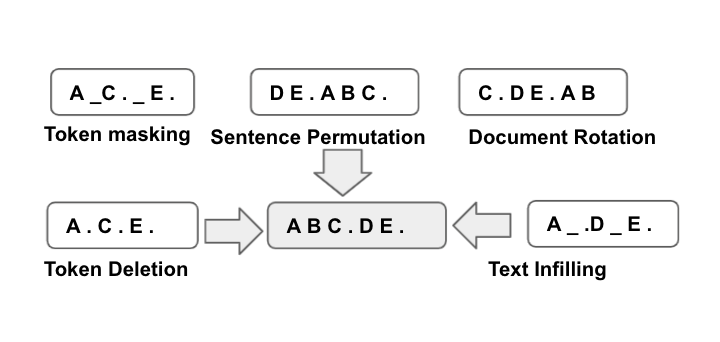}
	\caption[Pre-training objectives in BART]{Transformations as part of Pre-training objectives for noising the input in BART. Figure drawn from \citep{lewis2019bart}.}
	\label{fig:bart-2}
\end{figure}

	\subsubsection{Token Masking}
	This pretraining task follows BERT \citep{devlin-etal-2019-bert}, where random tokens are sampled and replaced with [MASK] tokens.
	
	\subsubsection{Token Deletion }
	In this pretraining task, random tokens are removed from the input. In contrast to token masking, the model infers for the positions where the inputs are missing.
	
	\subsubsection{Text Infilling} In this task, some text spans are sampled, with span lengths drawn from a Poisson distribution ($\lambda$ = 3). A single $[MASK]$ token is used to replace each span text. 0-length spans correspond to the insertion of $[MASK]$ tokens. Text infilling is inspired by SpanBERT \citep{joshi2020spanbert}, but SpanBERT samples span lengths from a clamped geometric distribution, and a sequence of $[MASK]$ tokens of the same length are used to replace each span. Text infilling task teaches the model to predict the number of tokens that are missing from a given span.
	
	\subsubsection{Sentence Permutation} In this task, sentences in the documents are shuffled in random order.
	
	\subsubsection{Document Rotation} In this task, a token is chosen uniformly at random, and the document is rotated in such a way that it begins with the selected token. This task trains the model to identify the beginning of the document.

	\section{Tokenization}
	\subsection{Byte Pair Encoding}
	Byte pair encoding (BPE) \citep{10.5555/177910.177914} is a data compression algorithm that iteratively finds the most frequent pair of bytes in the vocabulary appearing in a given sequence, then replaces it with a new unused entry. The same algorithm was adapted for word segmentation by \citep{sennrich2016neural}. 
	
	For word segmentation, the algorithm relies on a pre-tokenizer that splits the training data into words. The algorithm then creates a set of unique words and their frequency of occurrence in the training data after the pre-tokenization step. Then, a base vocabulary is created which consists of all symbols that occur in the set of unique words and the tokenizer learns the merge rules to form a new symbol from given pair of symbols in the base vocabulary. The tokenizer gets trained until the vocabulary size reaches the defined vocabulary size which is a hyperparameter assigned for the training of the tokenizer.
	
	\subsection{WordPiece}
	WordPiece is a subword tokenization algorithm introduced by \citep{6289079}. In subword tokenization algorithms, rare words are decomposed into meaningful subwords instead of frequently used words split into subwords.
	
	This algorithm is quite similar to BPE. WordPiece tokenizer is used in BERT, DistilBERT architectures. This algorithm first includes all the characters present in the training data in its vocabulary then learns the given number of merge rules progressively. WordPiece picks up the symbols which maximize the likelihood of the training data if those symbols are added to the vocabulary, whereas BPE picks up the most frequent symbol pairs.
	
	The training of WordPiece tokenizer aims to find the symbol pairs that maximize the likelihood of the training data and for which the probability of the merged pairs divided by the probabilities of its first symbol, followed by its second symbol is the highest among all symbol pairs. E.g., symbol pairs such as ``a'', followed by ``b'' will be merged if the probability of ``ab'' divided by ``a'' and ``b'' is the highest among all the symbol pairs.
	
	\subsection{Unigram}
	\citep{kudo2018subword}, introduced a subword tokenizer, Unigram, which considers multiple subword candidates. Unigram is not based on merge rules such as BPE, or WordPiece. This means that the algorithm has several ways of tokenizing new text after training. At first, Unigram initializes its base vocabulary to a large number of symbols with pre-tokenized words and the most common substrings then it progressively trims down each symbol to reduce the size of the vocabulary. However, the Unigram algorithm always keeps the base characters so that any word can be tokenized.
	
	Given the unigram language model and the vocabulary, the Unigram algorithm computes the log-likelihood loss over the training data at each training step. Then, for each symbol in the vocabulary, the algorithm computes the increase in the training loss if the selected symbol is removed from the vocabulary. This allows the algorithm to remove $p$ percent ($p$ usually being 10\% or 20\%) of symbols for which the increase in the training loss is the lowest. This process is repeated unless the vocabulary has reached the desired size.
	
	Assuming that the training data consists of the words $x_{1},\dots, x_{N}$ and that the set of all possible tokenizations for a word $x_{i}$ is defined as S($x_{i}$), then the overall loss is defined as $-\sum\limits_{i=1}^{N}\log\sum_{x \epsilon S(x_{i})}^{}p(x)$.
	
	\subsection{Byte-level BPE}
	GPT-2 \citep{Radford2019LanguageMA} uses bytes as a base vocabulary instead of including all possible Unicode base characters. This forces the base vocabulary to be of size 256 while ensuring that every base character is included in the vocabulary. GPT2’s tokenizer does not need the $<unk>$ token to tokenize every text. The vocabulary size of the GPT-2 is 50,257, which corresponds to the 256 bytes base tokens, a special end-of-text token. The tokenizer learns the symbols with 50,000 merges.
	
	\subsection{RoBERTa Tokenizer}
	HuggingFace's ``RoBERTaTokenizerFast'' is implemented from the GPT-2 tokenizer, using byte-level Byte-Pair-Encoding. This tokenizer is trained to treat spaces like parts of the tokens (a bit like SentencePiece) so that a word will be encoded differently independent of the position of the word, i.e., whether the word is at the beginning of the sentence (without space) or not.
	
	\subsection{BART Tokenizer}
The HuggingFace BART tokenizer uses byte-level Byte-Pair-Encoding. It is identical to the ``RobertaTokenizerFast'' tokenizer as discussed before.
	
	\subsection{CoNaLa code tokenizer}
	\citep{ling-etal-2016-latent} introduced a tokenizer that splits on all punctuation characters, except for the ``\_'' character. It also facilitates the task by splitting CamelCase words (e.g., class TirionFordring $\rightarrow$ class Tirion Fordring) so that all class names would be generated correctly. The implementation can be found here \citep{Conalacode}.
	
		\section{How TranX works?}
	TranX \citep{yin-neubig-2018-tranx} is a neural semantic parser having a transition system, and a neural encoder-decoder network to compute action probabilities. The transition system is extendable
	to new programming languages (PL) while the neural network is  PL agnostic, i.e., independent
	of the specific PL.
	\begin{figure}[h!]
		\centering
		\includegraphics[width=0.7\linewidth]{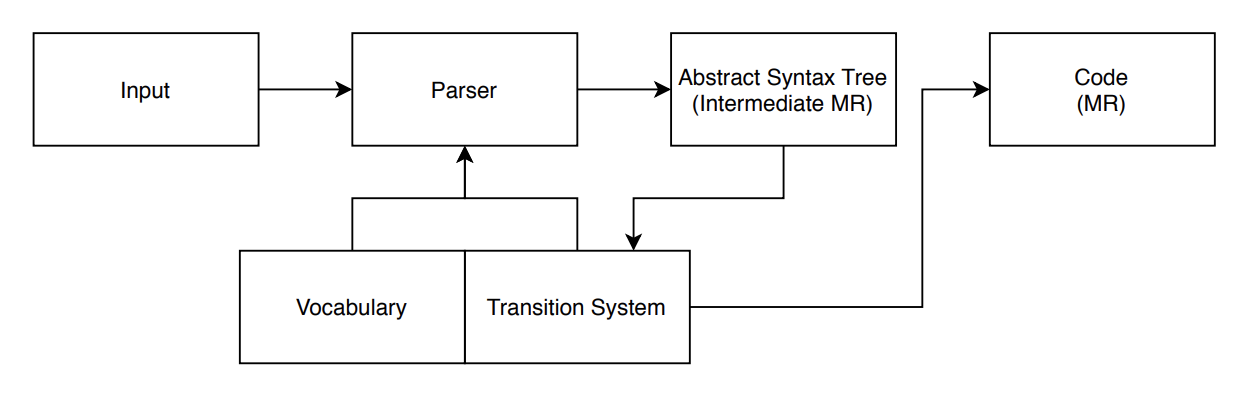}
		\caption[TranX overview.]{The overview of the workflow of TranX. Figure drawn from \citep{yin-neubig-2018-tranx}}
		\label{fig:tranx}
	\end{figure}
	
	The transition system of TranX maps the given input NL utterance $x$ into an AST $z$ using a series of three tree-construction actions. These actions are $APPLYCONSTR[c]$ ($c$ is a constructor; this action applies a constructor $c$ to the opening composite frontier field of the same type as $c$, and populates the opening node using the fields in $c$), $REDUCE$ (marks the completion of the generation of child values for a field with optional (?) or multiple (*) cardinalities.), and $GENTOKEN[v]$ ($v$ is a token; this action populates an empty primitive frontier field with a token $v$) as defined in \citep{yin-neubig-2018-tranx}. TranX utilizes ASTs as the intermediate meaning representations (MRs) to extract over the domain-specific structure of MRs. This parsing process follows the user-defined, domain-specific grammar specified under the ASDL formalism. The decomposition for AST into a sequence of tree-constructing actions begins from an initial derivation AST with a single root node and proceeds next with a top-down, left-to-right order traversal of the AST. The parsing process involves the conversion of the intermediate generated AST $z$ into a domain-specific meaning representation $y$. TranX also uses a probabilistic model $p(z \mid x)$, parameterized by a neural encoder-decoder network with augmented recurrent connections to score each hypothesis AST and to reflect the topology of ASTs such that $p(z \mid x) = \sum_{t} p(a_{t}\mid a_{< t}, x)$. 
	
	$p(z \mid x)$ is the probability of $z$ being decomposed into the probabilities of the sequence of actions used to generate $z$. \autoref{fig:tranx} shows an overview of how TranX works. The transition system, together with the vocabulary, presents possible continuations for an unfinished output sequence to the parser at each step of the parsing process. The parser generates an intermediate AST representation from the natural language input given these continuations. The AST is still PL agnostic and has to be transformed by the transition system into the target Meaning Representation (MR), i.e., Python source code in our case.

	\section{Evaluation Metrics}
	The commonly used metrics for the evaluation of the deep learning system for natural language to code translation are as follows: 
	\subsection{BLEU}
	
	BLEU (bilingual evaluation understudy) \citep{10.3115/1073083.1073135} is an algorithm for evaluating the quality of text that has been machine-translated from one natural language to another. In this case, it is the translation of natural language to programming language source code. The rationale behind BLEU is that translation closeness is defined by counting matches of n-grams in candidate and reference translation. 
	This metric does not take into account the intelligibility or grammatical correctness of a translated text.
	Equation 3.5 defines the formula for BLEU metric.
	\begin{align}
		BP &=\begin{cases} 1 & if\ c > r  \\
			\exp(1-\frac{r}{c}) &  if\ c \leq r \\
		\end{cases} \\
		BLEU &= BP \cdot \exp(\sum_{n=1}^{N}\frac{1}{N}\log p_{n}) \\
		p_{n} &= \frac{\sum_{c\ \epsilon\ \{candidates\}}\sum_{n-gram\ \epsilon\ c} count_{clip}(n-gram)}{\sum_{c'\ \epsilon\ \{candidates\}}\sum_{n-gram\ \epsilon\ c'} count(n-gram')}
	\end{align}
	where $p_{n}$ measures the modified n-gram precision between a document with
	candidate translations and a set of human authored reference documents, and
	the brevity penalty (BP) down-scales the score for outputs shorter than the
	reference. Candidates are the set of sentences to be evaluated. $count(n-gram')$ counts the number of times the n-gram appears in the candidate sentence, and $count_{clip}(n-gram)$
	is the same albeit clipped such that it does not exceed the number of times it
	appears in one of the reference sentences (which may be zero). 
	\subsubsection{Corpus-BLEU}
	It calculates a single corpus-level BLEU score (aka. system-level BLEU) for all the hypotheses and their respective references. The original BLEU metric \citep{10.3115/1073083.1073135} accounts for the micro-average precision (i.e., summing the numerators and denominators for each hypothesis-reference(s) pairs before the division).
	
	\subsubsection{Sentence-BLEU} 
It computes the BLEU metric on a single sentence pair. It calculates the averaging of the macro-average precision. However, the meaning of the sentence-level BLEU is more or less a special case of the corpus-level BLEU, and it can easily get zero value without smoothing function.
	
	\subsection{Rouge Scores}
	
	ROUGE (Recall-Oriented Understudy for Gisting Evaluation) metric \citep{lin-2004-rouge} measures the n-gram overlap between generated translation and its reference translation. It is a widely used evaluation metric for summarization. As the ROUGE score only measures token hard-match, in some cases, the ROUGE score penalizes two sentences that convey the same semantic information, but this metric highly rewards sentences with completely different semantics yet in similar surface forms.
	
	\textbf{ROUGE-N}: measures unigram, bigram, trigram, and higher-order n-gram overlap.
	
	\textbf{ROUGE-L}: measures the longest matching sequence of words using Longest Common Subsequence (LCS) algorithm. The advantage of LCS is that it reflects the sentence-level word order as it does not require consecutive matches but in-sequence matches. Since it automatically considers the longest in-sequence common n-grams, there is no need for predefined n-gram length.
	
	\subsection{Valid Compilable Code Snippets}
	A Python parser based on the Python AST module has been constructed, where the code generated from the deep learning model is parsed. This metric gives the count of the number of valid Python compilable code snippets generated from the deep learning models developed in this work.
	\subsection{Top-K accuracy}	
	Top-K Accuracy takes the K model predictions with higher probability. If one of them is a true label, it classifies the prediction as correct. Top-1 Accuracy is a special case, in which only the highest probability prediction is taken into account.

\section{One cycle training policy}
The one cycle policy follows the Cyclical Learning Rate (CLR) to obtain faster training time with regularization effect but with a slight modification, thus producing very fast results when training complex models. The original one cycle policy has three steps:
\begin{itemize}
	\item Initially, the learning rate is progressively increased from $lr_{max}/div_{factor}$ to $lr_{max}$ and at the same time, the momentum is decreased from $mom_{max}$ to $mom_{min}$.
	\item Then, the learning rate is decreased from $lr_{max}$ to $lr_{max}/div_{factor}$ and at the same time, the momentum is increased progressively from $mom_{min}$ to $mom_{max}$.
	\item  At last, the learning rate is decreased further from $lr_{max}/div_{factor}$ to $lr_{max}/(div_{factor} \times 100)$ and the momentum is kept steady at $mom_{max}$.
\end{itemize}

\section{Paired Bootstrap Significance Test}
Statistical significance testing is a standard statistical tool designed to ensure that experimental results are not coincidental. 
Parametric tests are preferred when the distribution of the test statistic is known, otherwise non-parametric tests are performed for the unknown test statistic distribution.

The paired bootstrap significance test is a non-parametric sampling-based significance testing. A sampling-based test is a computationally intensive procedure. The paired bootstrap test is quite similar to approximate randomization of the permutation test \citep{WILCOX2003237}, but with the sampling with replacements (i.e., an example from the original test data can occur more than once in a sample undertaken), unlike the permutation test.

The paired bootstrap test considers the absolute values of the evaluation measure, not restricted to higher-order properties (e.g., ranks) of the observed values. Hence, their statistical power is higher than the non-sampled tests.

\chapter{System Design}\label{approach}
In this section, we discuss the dataset used in the study, illustrate the data augmentation process, design of the empirical study, i.e, the proposed system design. In the remainder of this section, we discuss the configuration of the various model architectures, implementation details of the proposed system architecture, and pose research questions about the empirical study at the end.

\section{Dataset}\label{dataset}
CoNaLa (Code/Natural Language) is a dataset introduced by \citep{yin2018mining}. This dataset has two parts, one manually labeled small training dataset, and the other one is a big automatically mined large corpus.

Each part consists of code and natural language intent pairs scrapped from StackOverflow\footnote{\href{https://stackoverflow.com/}{https://stackoverflow.com/}} questions and answers forum. Each sample of the labeled dataset includes an intent, one line of code snippet, and a rewritten intent (\autoref{tab:dataset-sample}). The intent is the title of the  StackOverflow question. The line of code snippet represents the code blocks of the accepted answer to the specific question. The intent is then rewritten rephrasing to make it more specific and describing exactly what happens in the line of code and including variables and constants.

The mined corpus has been created by training a classifier on the above labeled dataset and applying it to all StackOverflow questions. For each question, the classifier ranks sets of answer source code snippets based on the likelihood of it being the correct solution to the question.

The resultant mined samples include the question title as intent, the selection of code lines, and a probability. In particular, mined samples do not include a rewritten intent.

\begin{table}[]
	\centering
	\begin{tabular}{ll}
		\hline
		\textbf{Intent} & How can I send a signal from a python program? \\ \hline
		\textbf{Code Snippet} & 	os.kill(os.getpid(), signal.SIGUSR1) \\ \hline
		\textbf{Rewritten Intent} & Send a signal `signal.SIGUSR1' to the current process. \\ \hline
	\end{tabular}
	\captionof{table}{CoNaLa Dataset Sample. \label{tab:dataset-sample}}
\end{table}

\begin{table}[]
	\centering
	\begin{tabular}{|c|c|}
		\hline
		\textbf{Dataset} & \textbf{Number of samples} \\ \hline
		Training    & 1903    \\ \hline
		Validation & 476     \\ \hline
		Test     & 500    \\ \hline
		Mined100k training set       & 96179    \\ \hline
		Mined100k validation set & 10687  \\ \hline
		Mined30k training set     & 31741  \\ \hline
		Mined30k validation set     & 7935 \\ \hline
	\end{tabular}
	\captionof{table}{CoNaLa Dataset. \label{tab:dataset}}
\end{table}

\begin{figure}
	\centering
	\includegraphics[width=\linewidth]{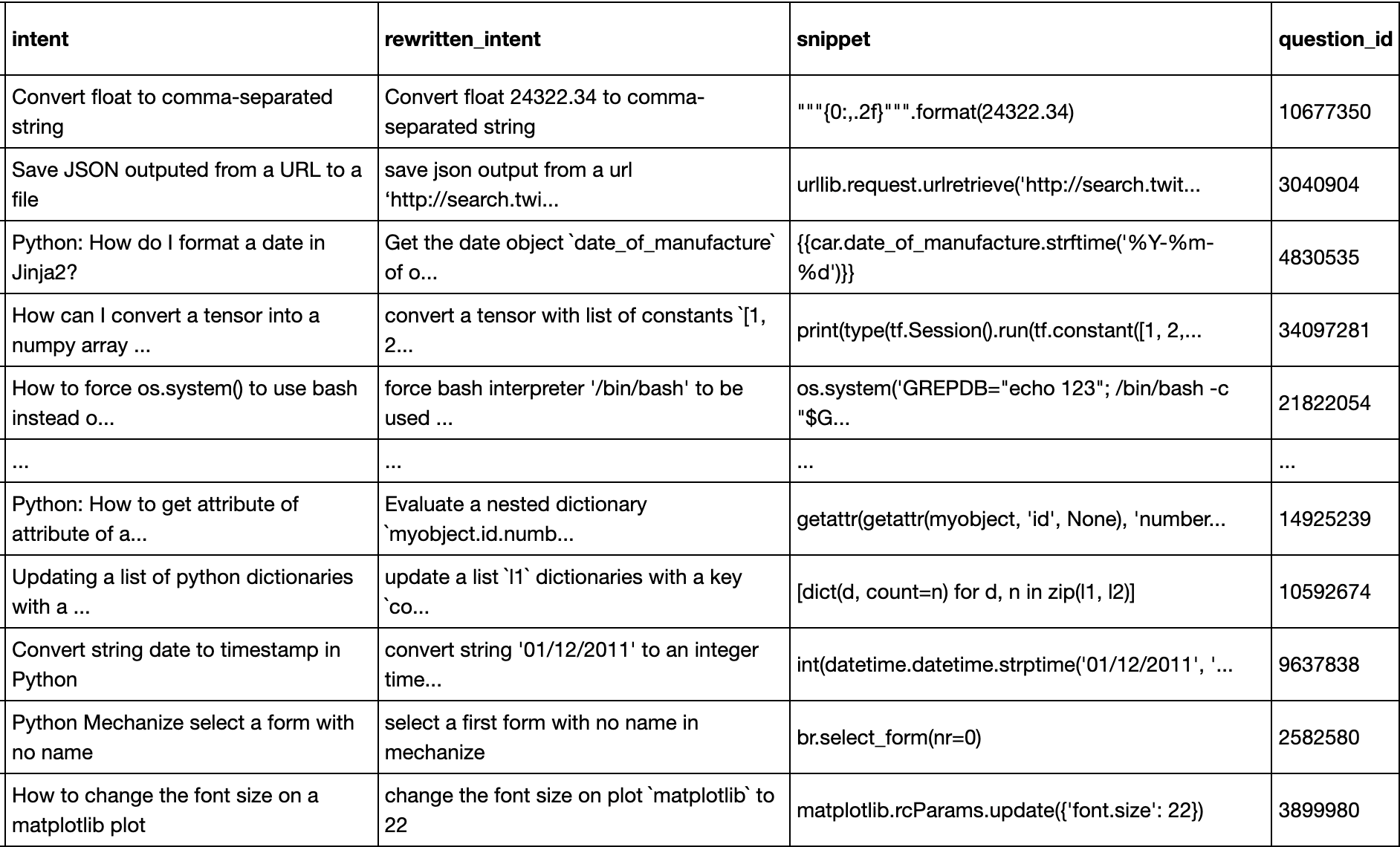}
	\caption{CoNaLa dataset snapshot.}
	\label{fig:dataset-snippet}
\end{figure}

\subsection{CoNaLa Dataset Statistics}
\autoref{tab:len-stats} shows the length statistics of the ``intent'' and code ``snippet'' field in the training set where we measure the mean, mode, and max length of the text for our Code2NL and NL2Code tasks.
\begin{table}[]
	\centering
	\begin{tabular}{|c|c|}
		\hline
		\textbf{Average length of nl intent}    & 46.53  \\ \hline
		\textbf{Max length of nl intent}        & 122.00    \\ \hline
		\textbf{Median length of nl intent}     & 45.00    \\ \hline
		\textbf{Mode length of nl intent}       & 46.00    \\ \hline
		\textbf{Average length of code snippet} & 39.77  \\ \hline
		\textbf{Max length of code snippet}     & 232.00 \\ \hline
		\textbf{Median length of code snippet}  & 38.00  \\ \hline
		\textbf{Mode length of code snippet}    & 33.00  \\ \hline
	\end{tabular}
	\captionof{table}[Length statistics of CoNaLa data attributes.]{Length statistics of the intent and snippet fields for the CoNaLa train dataset. \label{tab:len-stats}.}
\end{table}
\section{Augmented Dataset Creation}\label{curated-dataset}
To make the training data larger, we used the idea to generate back translation by reversing the training objective, i.e., Code2NL, generating natural language intent from the code snippets. We first trained a model for the Code2NL objective and using this model, we generated the predicted natural language intent from the code snippets for both the training and the validation set. This made us augment both the training and the validation set by 1$x$, 2$x$, or even k$x$ depending on the top-k predictions retrieved from the Code2NL model. Then, we trained the NL2Code model on the augmented training set to see how it performs. 

We picked up the top-2 Code2NL predictions each from the models trained on the ``intent'', and on the ``rewritten\_intent'' data fields respectively. Therefore, we had 4$x$ more training data generated from the Code2NL system, then, appended newly generated 4$x$ data to the original training dataset to make the original dataset grow by 5$x$ data samples.

Augmented curated datasets using back-translation: 

Top-1 from intent and top-1 from rewritten intent resulted into 3$x$ dataset. \\
Top-2 from intent and top-2 from rewritten intent resulted into 5$x$ dataset. \\
Top-1 from rewritten intent resulted into 2$x$ dataset. \\
Top-2 from rewritten intent resulted into 3$x$ dataset. \\

We used the following new datasets for training and for validation purposes for the Seq2Seq-BART model, then we fine-tuned the pretrained Nl2Code mined corpus model on this dataset to achieve the highest test metric scores.

\begin{table}[]
	\centering
	\begin{tabular}{|c|c|}
		\hline
		\textbf{Dataset} & \textbf{Number of samples} \\ \hline
		3$x$ Training set    & 5520  \\ \hline
		3$x$ Validation set        & 1380     \\ \hline
		5$x$ Training set     & 9200    \\ \hline
		5$x$ Validation set       & 2300   \\ \hline
		7$x$ Training set & 12880  \\ \hline
		7$x$ Validation set     & 3220 \\ \hline
	\end{tabular}
	\captionof{table}{Augmented Dataset. \label{tab:curateddataset}}
\end{table}

\section{Proposed System Architecture}\label{system-design}
\begin{figure}[!h]
	\centering
	\includegraphics[width=0.9\linewidth]{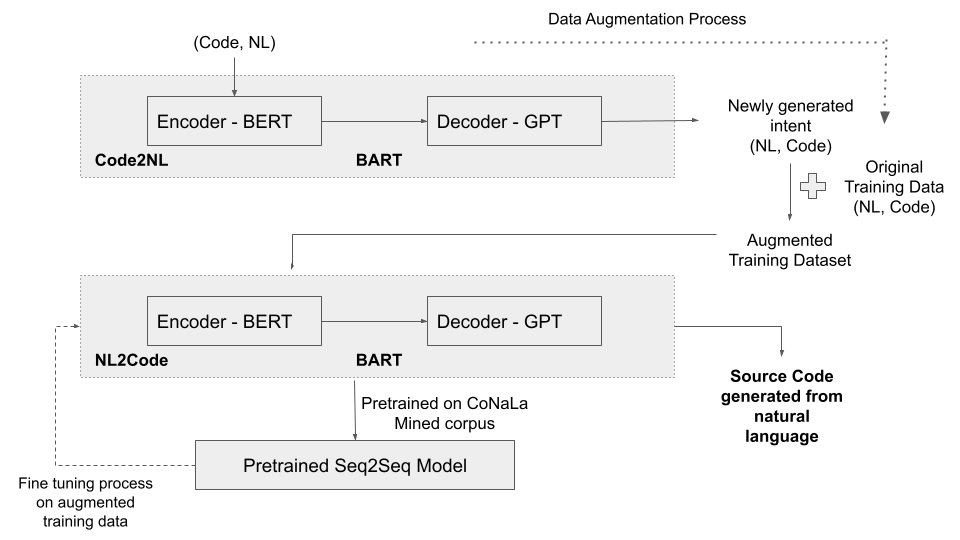}
	\caption[Proposed System Architecture.]{Proposed System Architecture. (Top) Code2NL system, (Down) NL2Code system}
	\label{fig:system-design}
\end{figure}

\autoref{fig:system-design} shows the main proposed system architecture for translating natural language to source code that performed to have the best results amongst all other architectures tried in this work. The system has two primary parts. The first part has Code2NL architecture which is based on Seq2Seq-BART architecture which has a BERT encoder and a GPT decoder. This Code2NL part is responsible for generating new natural language intent given a source code snippet. This part of the system is responsible for the data augmentation technique which augments the original training dataset with the newly created natural language intent data points, using the back-translation approach from the Code2NL. 

The second part is responsible for the pretraining objective as well as the fine-tuning process of the pretrained model on the augmented training dataset. The pretraining objective is served by pretraining the NL2Code system on the CoNaLa mined corpus.

\section{Model configuration}\label{model-config}
In this section, we discuss the configuration of each model's architecture by exclusively setting the values of the hyperparameters used to conduct the experiments in this work.
\subsection{Seq2Seq}
 \autoref{tab:seq-config} shows the values of the hyperparameters of the model used in our experiments for training of the vanilla Seq2Seq model.
\begin{table}[!htb]
	\centering
	\begin{tabular}{|c|c|}
		\hline
		\textbf{Model hyperparameters}                      & \textbf{Values}         \\ \hline
The input dimension of the model, INPUT\_DIM        & len(nl\_src.vocab)      \\ \hline
The output dimension of the model, OUTPUT\_DIM      & len(code\_target.vocab) \\ \hline
The encoder embedding dimension, ENC\_EMB\_DIM      & 256                     \\ \hline
The decoder embedding dimension, DEC\_EMB\_DIM      & 256                     \\ \hline
The encoder hidden layer dimension, ENC\_HID\_DIM   & 512                     \\ \hline
The decoder hidden layer dimension, DEC\_HID\_DIM   & 512                     \\ \hline
	\end{tabular}
	\captionof{table}{Seq2Seq model configuration. \label{tab:seq-config}}
\end{table}

\subsection{Transformer}
 \autoref{tab:transf-config} shows the values of the hyperparameters of the model used in our experiments for training of the transformer model.
\begin{table}[!htb]
	\centering
	\begin{tabular}{|c|c|}
		\hline
		\textbf{Model hyperparameters}                                & \textbf{Values} \\ \hline
		The dimension of the model, ``D\_MODEL''                       & 256             \\ \hline
		The number of self-attention heads, ``N\_HEADS''              & 3               \\ \hline
		The hidden size of the encoder, and decoder, ``HIDDEN\_SIZE'' & 512             \\ \hline
		The maximum length of the input sequence, ``MAX\_LEN''        & 50              \\ \hline
	\end{tabular}
	\captionof{table}{Transformer model configuration.\label{tab:transf-config}}
\end{table}

\subsection{Hybrid Seq2Seq with RoBERTa}
We set the following hyperparameters in the HuggingFace's Seq2Seq EncoderDecoder module where we loaded the pretrained RoBERTa base model, i.e, our pretrained model trained on the CoNaLa mined corpus. The parameters of the RoBERTa model configuration were set as in the \autoref{tab:roberta-config}.
 
 \begin{table}[!htb]
 	\centering
 	\begin{tabular}{|c|c|}
 		\hline
 		\textbf{Model hyperparameters}         & \textbf{Values}   \\ \hline
 		"encoder\_max\_length"            & 512               \\ \hline
 		"decoder\_max\_length"            & 256               \\ \hline
 		"is\_encoder\_decoder"            & True              \\ \hline
 		"length\_penalty"                 & 2.0               \\ \hline
 		"max\_length"                     & 64                \\ \hline
 		"model\_type"                     & "encoder-decoder" \\ \hline
 		"$no\_repeat\_ngram\_size$"       & 3                 \\ \hline
 		"num\_beams"                      & 4                 \\ \hline
 		"tie\_encoder\_decoder"           & True              \\ \hline
 		"vocab\_size"                     & 50265             \\ \hline
 		"attention\_probs\_dropout\_prob" & 0.1               \\ \hline
 		"hidden\_act"                     & "gelu"            \\ \hline
 		"hidden\_dropout\_prob"           & 0.1               \\ \hline
 		"hidden\_size"                    & 768               \\ \hline
 		"layer\_norm\_eps"                & 1e-05             \\ \hline
 		"max\_position\_embeddings"       & 514               \\ \hline
 		"min\_length"                     & 0                 \\ \hline
 		"model\_type"                     & "roberta"         \\ \hline
 		"num\_attention\_heads"           & 12                \\ \hline
 		"num\_beam\_groups"               & 1                 \\ \hline
 		"num\_hidden\_layers"             & 6                 \\ \hline
 		"num\_return\_sequences"          & 3                 \\ \hline
 		"top\_k"                          & 50                \\ \hline
 		"top\_p"                          & 1.0               \\ \hline
 	\end{tabular}
 	\captionof{table}{Seq2Seq-RoBERTa model configuration. \label{tab:roberta-config}}
 \end{table}

\subsection{Hybrid Seq2Seq with BART}
We used HuggingFace's BartForConditionalGeneration model architecture as a part to integrate into our hybrid Seq2Seq-BART architecture for NL2Code task. We set the following hyperparameters in the FastAI module where we loaded the pretrained BART base model, i.e, our pretrained model on CoNaLa mined corpus. The parameters with their values can be found in \autoref{tab:bart-config}.

\begin{table}[!htb]
	\centering
	\begin{tabular}{|c|c|}
		\hline
		\textbf{Model hyperparameters}          & \textbf{Values} \\ \hline
		"max\_length"                      & 128             \\ \hline
		"min\_length"                      & 12              \\ \hline
		"do\_sample"                       & True            \\ \hline
		"early\_stopping"                  & True            \\ \hline
		"length\_penalty"                  & 1.0             \\ \hline
		"temperature"                      & 1.0             \\ \hline
		"$no\_repeat\_ngram\_size$"        & 3               \\ \hline
		"num\_beams"                       & 4               \\ \hline
		"encoder\_no\_repeat\_ngram\_size" & 0               \\ \hline
		"repetition\_penalty"              & 1.0             \\ \hline
		"attention\_probs\_dropout\_prob"  & 0.1             \\ \hline
		"bos\_token\_id"                   & 0               \\ \hline
		"pad\_token\_id"                   & 1               \\ \hline
		"eos\_token\_id"                   & 2               \\ \hline
		"use\_cache"                       & True            \\ \hline
		"decoder\_start\_token\_id"        & 2               \\ \hline
		"output\_hidden\_states"           & False           \\ \hline
		"diversity\_penalty"               & 0.0             \\ \hline
		"output\_attentions"               & False           \\ \hline
		"num\_beam\_groups"                & 1               \\ \hline
		"output\_scores"                   & False           \\ \hline
		"num\_return\_sequences"           & 3               \\ \hline
		"top\_k"                           & 50              \\ \hline
		"top\_p"                           & 1.0             \\ \hline
	\end{tabular}
	\captionof{table}{Seq2Seq-BART model configuration. \label{tab:bart-config}}
\end{table}

\section{Implementation}\label{implement}
Initially, we implemented the Seq2Seq and Transformer architectures in Pytorch using the TorchText\footnote{\href{https://pytorch.org/text/stable/index.html}{https://pytorch.org/text/stable/index.html}} framework. We also implemented both greedy search and beam search decoding algorithms for decoding the code translations from the NL  
 and implemented top-k accuracy metric for evaluation of the vanilla Seq2Seq and Transformer architectures. 

Thereafter, we also implemented an early stopping mechanism for our vanilla Seq2Seq and Transformer architectures to control overfitting. We implemented and trained the SentencePiece \citep{kudo2018sentencepiece} tokenizer for the CoNaLa dataset for NL2Code task using TorchText Functional libraries\footnote{\href{https://text-docs.readthedocs.io/en/latest/data_functional.html}{https://text-docs.readthedocs.io/en/latest/data\_functional.html}} to implement subword tokenization algorithms such as Unigram, WordPiece, and BPE. 

For the hybrid-Seq2Seq architectures, we used HuggingFace Seq2Seq Trainer\footnote{\href{https://huggingface.co/transformers/main_classes/trainer.html\#seq2seqtrainer}{https://huggingface.co/transformers/main\_classes/trainer.html\#seq2seqtrainer}} library to set up the Seq2Seq framework for training and evaluation. We used HuggingFace's DistilRoBERTa\footnote{\href{https://huggingface.co/distilroberta-base}{https://huggingface.co/distilroberta-base}} base model configuration, tokenizer to implement our hybrid Seq2Seq-RoBERTa architecture for the NL2Code task. 

We used BLURR\footnote{\href{https://github.com/ohmeow/blurr}{https://github.com/ohmeow/blurr}} library that integrates Hugging Face's transformers framework with fastai version 2 API to implement the Seq2Seq-BART architecture and also to find the optimal learning rate for training the architecture. We used BLURR's datablock and dataloaders\footnote{\href{https://ohmeow.github.io/blurr/modeling-seq2seq-summarization/}{https://ohmeow.github.io/blurr/modeling-seq2seq-summarization/}} libraries to preprocess the data on-the-fly batch-time tokenization to fit into the hybrid Seq2Seq-BART architecture. We implemented HuggingFace's BartForConditionalGeneration\footnote{\href{https://huggingface.co/transformers/model_doc/bart.html\#transformers.BartForConditionalGeneration}{https://huggingface.co/transformers/model\_doc/bart.html\#transformers.BartForConditionalGeneration}} model for the conditional generation of decoding output given the conditional dependency of the input words into the proposed Seq2Seq-BART architecture for both the NL2Code and the Code2NL task. 

We used HuggingFace metrics library\footnote{\href{https://huggingface.co/metrics}{https://huggingface.co/metrics}} to implement metrics such as SacreBLEU, ROUGE metrics such as ROUGE-1, ROUGE-2, ROUGE-L, and METEOR metrics for the quantitative evaluation for the vanilla Seq2Seq, Transformer, hybrid Seq2Seq-RoBERTa, and hybrid Seq2Seq-BART architectures. SacreBLEU metric is the re-implementation of BLEU metric by wrapping the original BLEU \citep{10.3115/1073083.1073135} implementation. SacreBLEU scales the BLEU score between 0 and 100, whereas, in the original script, the score is between 0 and 1. We have considered the BLEU-4(ngram-overlap:4) metric score for our experiments. 

We used one cycle training policy to train the Seq2Seq-BART architecture on the training set. To have a faster convergence of the training objective, we scheduled the learning rate as described in \citep{smith2018superconvergence}  by using the one cycle policy (fit one cycle) in FastAI\footnote{\href{https://fastai1.fast.ai/train.html\#fit_one_cycle}{https://fastai1.fast.ai/train.html\#fit\_one\_cycle}}. Considering the transfer learning technique used, we trained the first ``one cycle'' on the top of the existing pretrained model to later unfreeze all the model layers and to do a more extended training (< 100 epochs) with early stopping mechanism depending on the training requirements to improve the final results. 

To apply 1 cycle training policy, we first found the most appropriate learning rate to use automatically by using the function lr\_find\footnote{\href{https://fastai1.fast.ai/callbacks.lr_finder.html\#lr_find}{https://fastai1.fast.ai/callbacks.lr\_finder.html\#lr\_find}} provided by FastAI. 
Thereafter, we applied the 1cycle policy using FastAI fit\_one\_cycle() method with the chosen learning rate as the maximum learning rate. 

Finally, we conducted the training of all the model architectures developed in our work on a single NVIDIA Tesla T4 16 GB GPU available on Google Cloud Platform\footnote{\href{https://console.cloud.google.com/}{https://console.cloud.google.com/}} subscription by SAP.

\chapter{Experimental Results and Discussion}\label{evaluation}	
In this chapter, we first design the experiments and give an overview of generated code snippets from the proposed architectures and evaluate them against the corresponding state-of-the-art neural semantic parser, TranX, the vanilla Seq2Seq, and the Transformer architectures respectively. 

We also answer the research questions posed in this chapter. The first two research questions have subquestions that outline the results and comparisons of various experiments conducted in this work. For each research question, we describe the motivation behind the question, the approach undertaken to answer the research question, and the findings derived from the analysis.

\section{Research Questions}
The following research questions guide the evaluation:

\textbf{RQ1. Results \& Analysis: How well does the developed architectures perform on the NL2Code objective in comparison to the state-of-the-art?}

\textbf{RQ2. What did we find from the ablation studies of the developed architectures?}

\textbf{RQ3. Do pretraining knowledge and transfer learning help to improve the results? Are the improvements to the fine-tuned models statistically significant?}

\textbf{RQ4. Does the SentencePiece tokenizer improve the performance of the Transformer model?}

\textbf{RQ5. Does byte-pair encoding of code tokens capture more information than the AST and the AMT?}

\textbf{RQ6. Does the proposed hybrid Seq2Seq-BART architecture work bidirectionally to reverse the hypothesis?}

\textbf{RQ7. Do data augmentation and pretraining techniques improve the results of the proposed hybrid Seq2Seq-BART architecture?}

\textbf{RQ8. How well does the code completion task work with the use of neural language model?}

\subsection{RQ1. Results \& Analysis: How well does the developed architectures perform on the NL2Code objective in comparison to the state-of-the-art?}
\subsubsection{RQ1a. Does the proposed Seq2Seq-BART architecture perform better than the neural semantic parser, TranX?}
\textbf{Motivation.}
The idea is to compare the developed hybrid architectures, the vanilla Seq2Seq framework, and the transformer architecture with the state-of-the-art neural semantic parser, TranX.

\textbf{Approach.}
We evaluated all the architectures on the test BLEU score and the valid compilable code snippet metrics for the same test dataset. We further went to parse the generated code snippet from the model by using Python ``ast'' module to see whether the generated code snippet is valid parsable code or not. We further classified the translations/predictions into 5 groups based on the different range of Sentence-BLEU score, namely, totally wrong(<= 0.2), semantically equivalent (>= 0.2 and <= 0.4), marginally correct (>= 0.4 and <= 0.6), mostly correct (>= 0.6 and <= 0.9), and exact match(>= 0.9) translation categories.

\textbf{Results.}
\autoref{tab:metrics-overall} shows the overall comparison of all the model architectures developed in this work on the BLEU metric score.

\textbf{Finding 1.} From \autoref{tab:metrics-overall}, we found that the proposed Seq2Seq-BART when fine-tuned on the augmented dataset of 3$x$ size of CoNaLa dataset, which was pretrained on the mined100k CoNaLa corpus, surpassed the test BLEU metric score of the state-of-the-art, TranX, a neural semantic parser, which is heavily featured engineered by 10.82\%. This model achieved the state-of-the-art test BLEU score of 27.8235 for the CoNaLa dataset, among all the models developed in this work.

\textbf{Finding 2.} We also found that the Seq2Seq-BART model when fine-tuned on CoNaLa dataset after pretraining on the CoNaLa minded100k corpus, exceeded the performance of TranX on BLEU metric by 5.7\%.

\textbf{Finding 3.} Interestingly, the proposed Seq2Seq-BART when trained on our augmented dataset $3x$ size of CoNaLa without pretraining, also exceeded the test BLEU score of TranX by 2.65\%. This signifies that with more available training data, the proposed architecture had the ability to perform better even without fine-tuning and pretraining techniques.

\textbf{Finding 4.} TranX produced 206 valid compilable code snippets via AST parsing from the test set, whereas the proposed Seq2Seq-BART model resulted in 412 valid compilable code snippets from the test set.

\begin{table}[!htbp]
	\centering
	\large
		\resizebox{\linewidth}{!}{
	\begin{tabular}{|c|c|c|}
		\hline
		\large
		\textbf{Models trained on Dataset/Augmented Datasets}             & \textbf{Test BLEU} \\ \hline
		Seq2Seq-BART on 3$x$ size of CoNaLa dataset                                      & 25.7710            \\ \hline
		Seq2Seq-BART on 5$x$ size of CoNaLa dataset                                     & 25.1601            \\ \hline
		Seq2Seq-BART on CoNaLa dataset                       & 24.2990             \\ \hline
		Fine-tuned Seq2Seq-BART on CoNaLa,, pretrained on mined100k corpus & 26.5379           \\ \hline
		Fine-tuned Seq2Seq-BART on 3$x$ CoNaLa, pretrained on mined100k corpus & \textbf{27.8235}            \\ \hline
		Fine-tuned Seq2Seq-BART on 5$x$ CoNaLa, pretrained on mined100k corpus & 25.3153            \\ \hline
		Vanilla Seq2Seq on CoNaLa & 13.3270 \\ \hline
		Transformer-CoNaLa code tokenizer on CoNaLa & 15.3834 \\ \hline
		Transformer-BPE on CoNaLa & 19.3402 \\ \hline
		Transformer-Unigram on CoNaLa &20.9678 \\ \hline
		Transformer-WordPiece on CoNaLa & 17.3237 \\ \hline
		Fine-tuned Seq2Seq-RoBERTa on CoNaLa, pretrained on mined30k corpus & 18.8853 \\ \hline
		TranX on CoNaLa & 25.1050 \\ \hline
	\end{tabular}}
	\captionof{table}[Overall Comparison of all models on Test set.]{Overall Comparison of all models on Datasets.\label{tab:metrics-overall}}
	
\end{table}

\subsubsection{RQ1b. How well does the vanilla Seq2Seq architecture perform? }

\textbf{Motivation.}
A Seq2Seq model framework model usually consists of an encoder to map an input sequence to hidden representations, and a decoder to decode hidden representations to generate a target sequence. We want to see whether a standard Seq2Seq model architecture of 1 layer with Bahdanu \citep{bahdanau2016neural} attention can be used to translate natural language intent into a valid compilable structured code snippet without any feature engineered semantic parser to parse code but rather considering code snippets as programming language tokens.

\textbf{Approach.}
The idea is to map the input natural language intent in the encoder with the respective source code snippet in the decoder allowing the model to learn to translate the intent into respective source code snippet, while training on the dataset. Thus, we fed the input of the Seq2Seq encoder with the tokenized code tokens using the Python code tokenizer for the CoNaLa challenge without employing the hard requirements of semantic parsing such as the abstract syntax tree (AST) representation or abstract meaning representations (AMT) of the source code. We performed hyperparameter tuning for the model to find the optimal values of the hyperparameters to achieve the optimal results. 

\textbf{Hyperparameter tuning}: We varied the batch size, the dropout probability, and the learning rate for the vanilla Seq2Seq architecture. We found out that the evaluation metrics decrease significantly as the batch size increases. We also found that with a dropout probability of more than 0.3, the test metrics score further decreases. We also implemented the early stopping mechanism to avoid overfitting and unnecessary training for higher epochs than the required.

\textbf{Results.}
We evaluated the system with BLEU, ROUGE, and token accuracy metrics respectively. We further classified the generated translations into 5 subgroups as discussed before in RQ1a. 

\begin{figure}
	\centering
	\includegraphics[width=.9\linewidth]{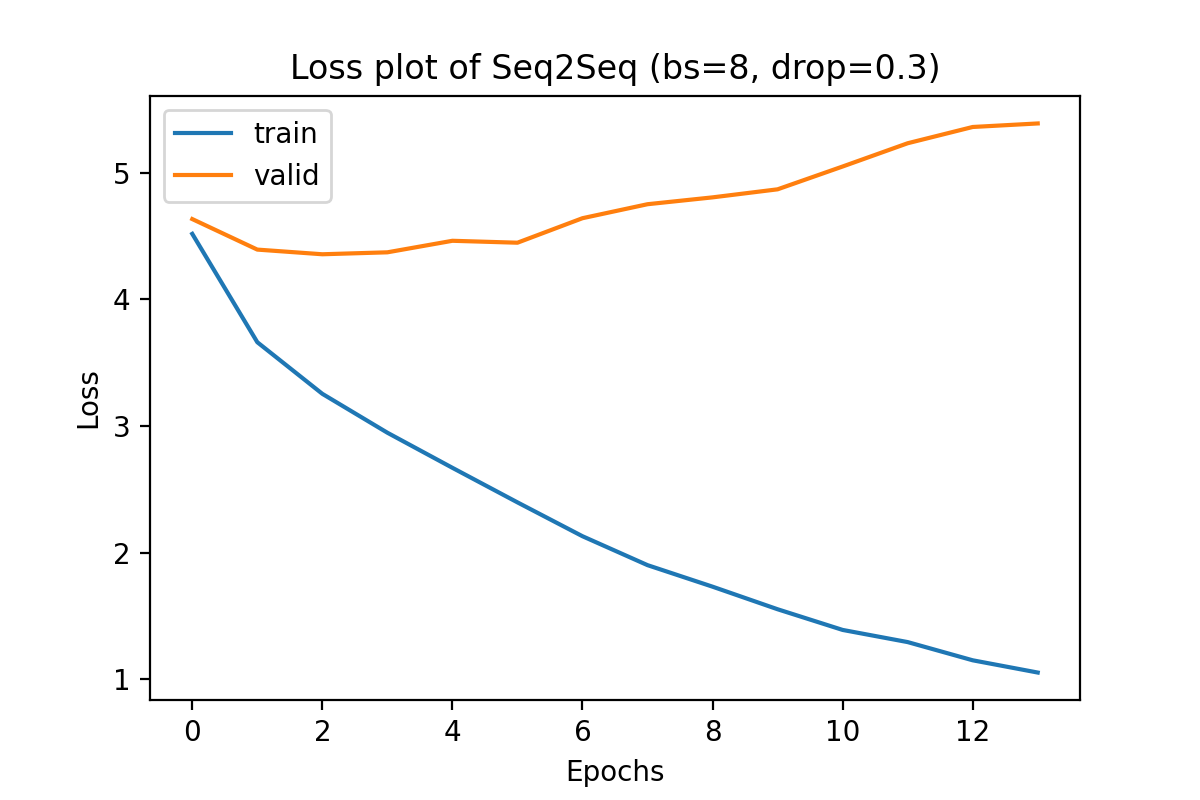}
	\caption{Loss plot for the vanilla Seq2Seq.}
	\label{fig:loss-seq2seq-b8-d0}
\end{figure}

\begin{figure}
	\centering
	\includegraphics[width=.9\linewidth]{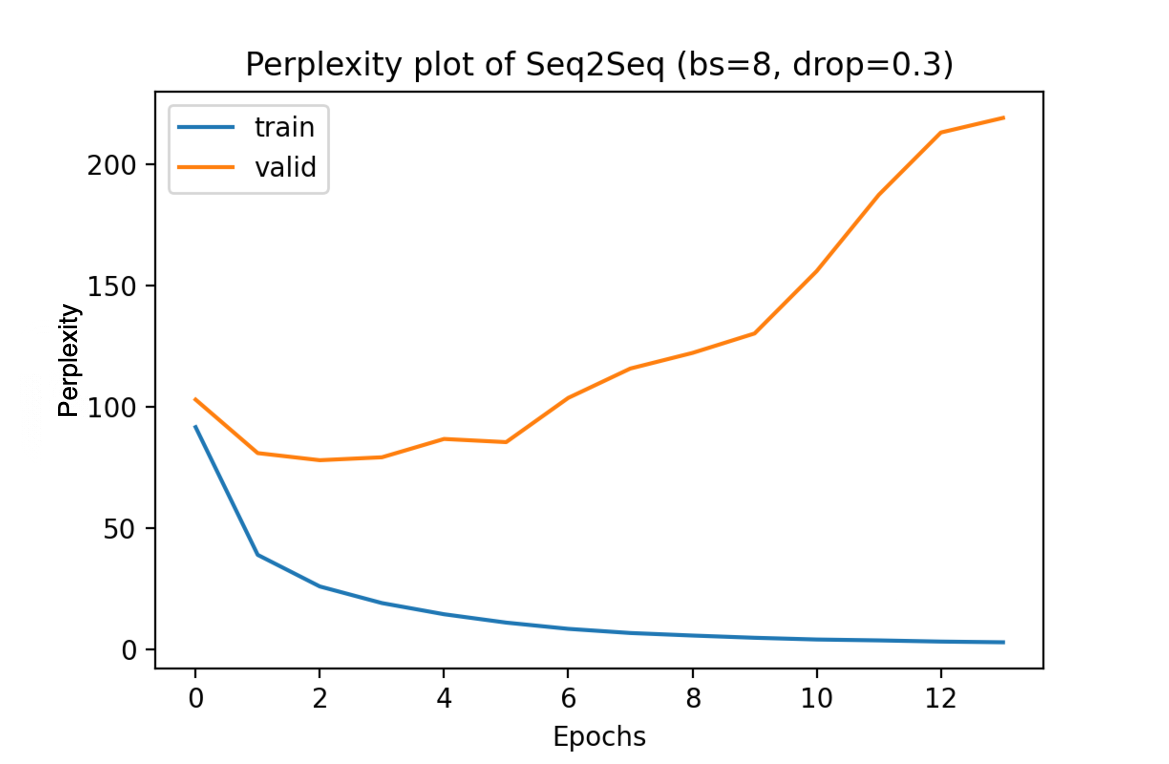}
	\caption{Perplexity plot for the vanilla Seq2Seq.}
	\label{fig:perplexity-seq2seq-b8-d0}
\end{figure}

\begin{figure}[]
	\centering
	\includegraphics[width=.9\linewidth]{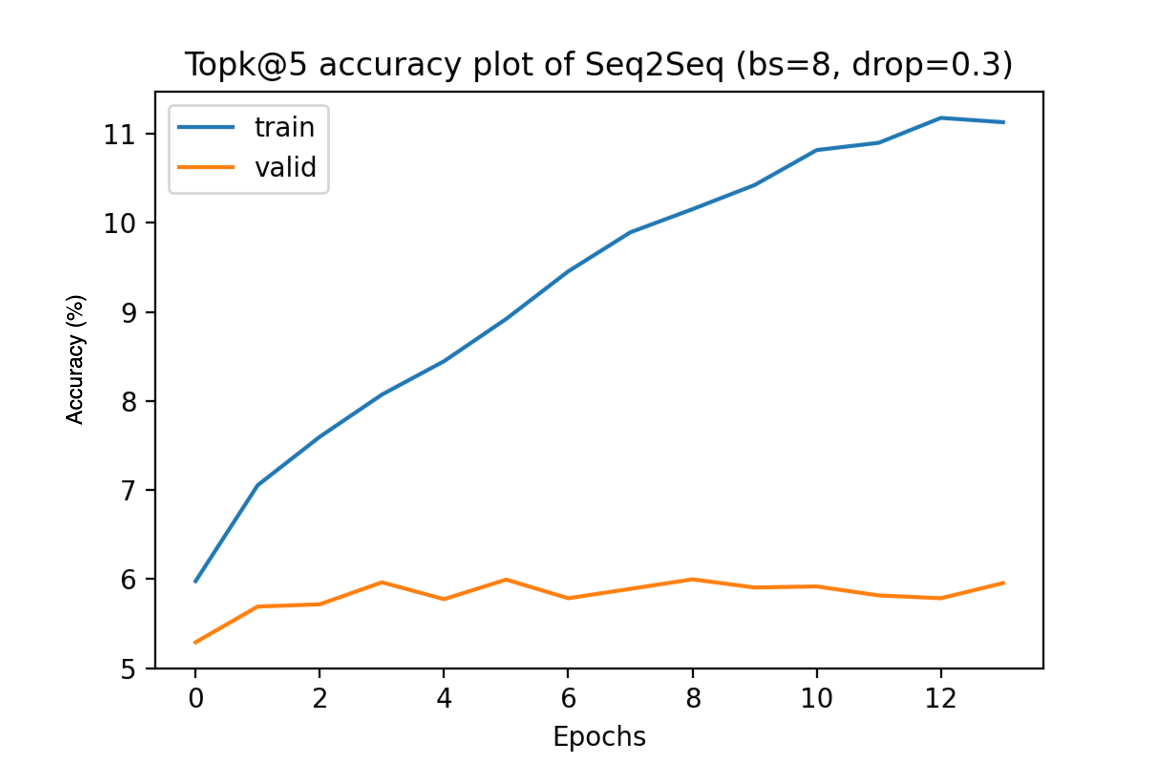}
	\caption{Top-k accuracy for the vanilla Seq2Seq.}
	\label{fig:acc-seq2seq-b8-d0}
\end{figure}

\autoref{fig:loss-seq2seq-b8-d0} shows the distribution plot of the training and the validation loss. \autoref{fig:perplexity-seq2seq-b8-d0} shows the training and the validation perplexity measure of the model architecture. \autoref{fig:acc-seq2seq-b8-d0} shows the top-5 accuracy for both training and validation.

\autoref{tab:metric-seq2seq} shows the metric scores of the optimal Seq2Seq model architecture after hyperparameter tuning for both validation and test datasets.  \autoref{tab:bleu-seq2seq} shows a finer Sentence-BLEU evaluation of predicted code snippets and the histogram of code translations sorted in interval range of Sentence-BLEU score. \autoref{tab:pred-seq2seq}  displays some test examples code snippet predictions from the test set with the ground truth. 

\textbf{Finding 1.} We observed that the validation loss diverges but generally gets in the range between 4 and 5 despite training for higher epochs without the early stopping mechanism. From the observation, the model is overfitting as the validation loss does not decrease on par with the training loss across multiple epochs of training. This may be due to the availability of less training data of around 2000 training examples.

\textbf{Finding 2.} From the \autoref{fig:acc-seq2seq-b8-d0}, we found that the top-5 training accuracy kept increasing, and then it got almost flat and for the validation top-5 accuracy, it seemed to have plateaued as the training progressed after 12 epochs. 

\textbf{Finding 3.} From the \autoref{tab:metric-seq2seq}, we found out that the vanilla Seq2Seq architecture reported a test BLEU-4 score of 13.327 and validation BLEU-4 score of 19.81 respectively.

\textbf{Finding 4.} From the \autoref{tab:bleu-seq2seq}, we noticed that the model had generated 3 code translations whose Sentence-BLEU score was greater than 0.8. (Sentence-BLEU is 1 for perfect translation and 0 for completely wrong translation.) There are only 1 exact match, 15 mostly correct, 75 marginally correct, and 191 semantically equivalent code translations generated from the model.

\textbf{Finding 5.} We found out that after parsing the generated code snippet, the model resulted in producing only 73 valid parsable code snippets from the test set. This reflects that the Seq2Seq model had failed to infer the context of majority of different NL intents given in the test set.

\textbf{Finding 6.} From \autoref{tab:pred-seq2seq} and \autoref{tab:pred-seq2seq-transf-sideways}, we observed that the Seq2Seq model had worked for the intents related to the function calls of changing directories, date-time calls, file operations like reading a file, erasing the contents of the file, string representation from the list. The model had failed for the intents related to regex operations, HTTP web API calls, dataframe operations, complicated string and substring operations, and various python data structure operations such as removing specific characters at a specified position, merging lists, sorting dictionary, and math operations such as addition, concatenation of vectors, creation of matrices, etc.

\begin{table}[]
	\centering
	\begin{adjustbox}{width=\linewidth}
		\begin{tabular}{|p{0.4\linewidth}|p{0.3\linewidth}|p{0.3\linewidth}|}
			\hline
			\textbf{Intent}                                               & \textbf{Prediction}                                        & \textbf{Snippet}                                                   \\ \hline
			Enable debug mode on flask application `app' & app.run(debug = True)  & app.run(debug = True) \\ \hline
			Erase all the contents of a file `filename' & open(`filename' , `' ).close() & open (filename , `w' ).close( ) \\ \hline
			Change working directory to the directory `owd'. & os.chdir(`) & os.chdir(owd) \\ \hline
			Convert list `list\_of\_ints' into a comma separated string & ` ` `  ` ` ` . join ( str ( i ) for i in list\_of\_ints ) ) & ` ` ` , ` ` ` . join({[} str ( i ) for i in list\_of\_ints{]}) \\ \hline
		\end{tabular}
	\end{adjustbox}
\captionof{table}{Cherry picked test results from the vanilla Seq2Seq.\label{tab:pred-seq2seq}}
\end{table}

\begin{table}[]
	\centering
	\begin{tabular}{|c|c|c|}
		\hline
		\textbf{Non fine-tuned model(Sentence-BLEU)} & \textbf{\#prediction} \\ \hline
		\textgreater{}0.9                           & 1                     \\ \hline
		\textgreater{}=0.8 and \textless{}=0.9      & 2                     \\ \hline
		\textgreater{}=0.7 and \textless{}=0.8      & 1                     \\ \hline
		\textgreater{}=0.6 and \textless{}=0.7       & 12                    \\ \hline
		\textgreater{}=0.5 and \textless{}=0.6      & 27                    \\ \hline
		\textgreater{}=0.4 and \textless{}=0.5      & 48                    \\ \hline
		\textgreater{}=0.3 and \textless{}=0.4      & 87                    \\ \hline
		\textgreater{}=0.2 and \textless{}=0.3      & 104                   \\ \hline
		\textgreater{}=0.1 and \textless{}=0.2      & 98                    \\ \hline
		\textless 0.1                               & 97                    \\ \hline
		\textgreater{}0.5                           & 43                    \\ \hline
	\end{tabular}
	\captionof{table}[Sentence-BLEU test metric for Seq2Seq]{Histogram of test predictions of Seq2Seq in the interval range of Sentence-BLEU scores.\label{tab:bleu-seq2seq}}
\end{table}

\begin{table}[]
\centering
\Large
\begin{adjustbox}{width=\textwidth}
	\begin{tabular}{|c|c|c|c|c|c|c|}
		\hline
		\Large
		\textbf{Dataset}    & \textbf{BLEU}     & \textbf{Rouge1 F1} & \textbf{Rouge2 F1} & \textbf{Rouge4 F1} & \textbf{RougeL F1} & \textbf{Token Accuracy} \\ \hline
		Validation & 19.8190 & 0.5604       & 0.3052       & 0.1755       & 0.5516        & 18.1031     \\ 
		 Test & 13.3270 & 0.5216       & 0.2379       & 0.1094       & 0.5128      & 14.6416    
		    \\ \hline
	\end{tabular}
\end{adjustbox}
\captionof{table}[Metrics score for Seq2Seq]{Test and Validation metrics for Seq2Seq. \label{tab:metric-seq2seq}}
\end{table}

\subsubsection{RQ1c. How well does the Transformer architecture perform in comparison to vanilla Seq2Seq?}

\textbf{Motivation.}
The recent success of Transformer architecture in the machine translation(MT) system motivated us to implement this architecture for the NL2Code objective, and report its performance compared to the vanilla Seq2Seq model.

\textbf{Approach.}
The transformer architecture has both encoder and decoder layers which are made of self-attention layers with multiple attention heads. We performed both ablation studies about the the number of self-attention heads and the depth of the transformer needed to study the importance of the crucial components of the Transformer architecture as discussed in \autoref{RQ2} (RQ2a and RQ2b). We trained the transformer architecture on the training set and validate the validation set and used beam search decoding with varying beam sizes to derive the results. We further performed hyperparameter tuning and implemented an early-stopping mechanism to further fine-tune the model to achieve the optimal results on the test dataset.

\textbf{Hyperparameter Tuning}: We varied the batch size from 8 to 256 and found out that the model resulted best test metric scores for the batch size of 64. We saw that for smaller batch sizes of 8 and 16, the model reported worst test metric scores. After varying the dropout probability in both the encoder and decoder layers, we observed that the test metric scores drop significantly with the increase of dropout probability for a fixed number of three layers and eight self-attention heads. We also found that when the dropout probability was 0.25, the model reported the highest test metric scores. We also varied the learning rate in [1e-03, 1e-04, 2e-04] to monitor the validation loss and evaluation metrics.

\begin{figure}
	\centering
	\includegraphics[width=0.7\linewidth]{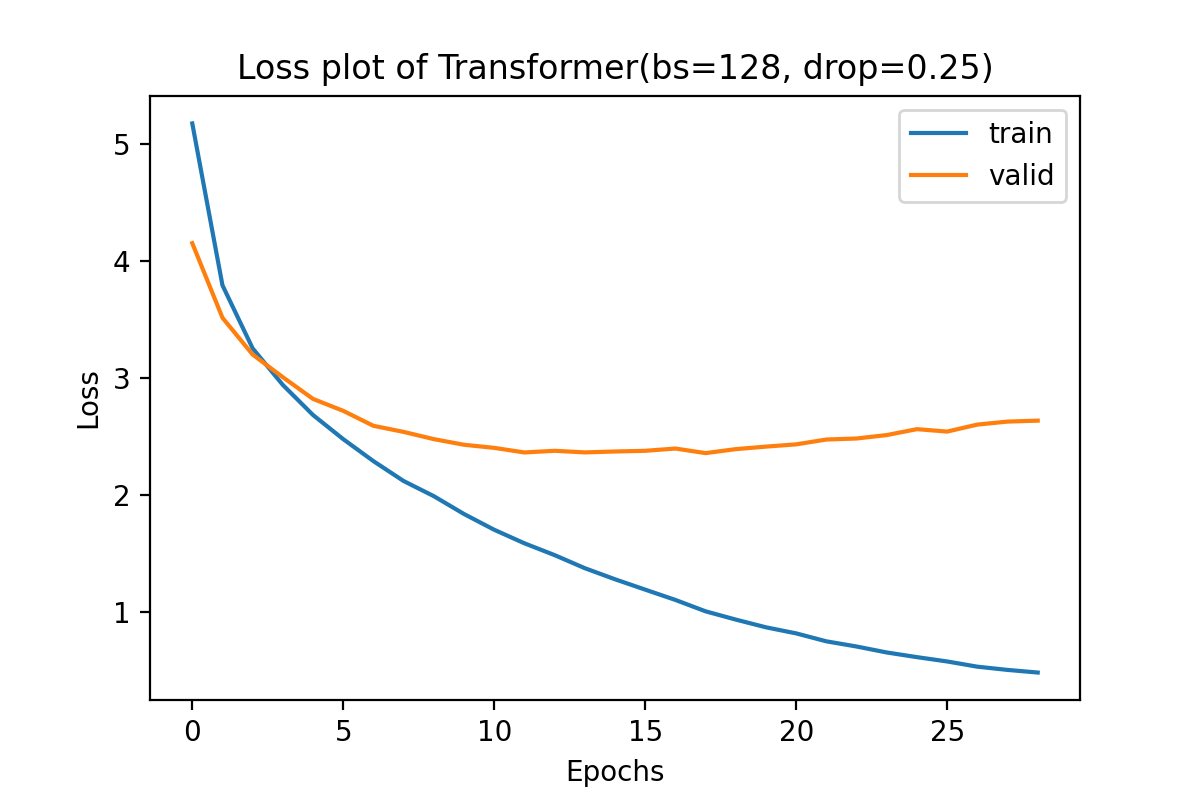}
	\caption{Loss plot for the Transformer.}
	\label{fig:loss-transf-b128-d0}
\end{figure}

\begin{figure}
	\centering
	\includegraphics[width=0.9\linewidth]{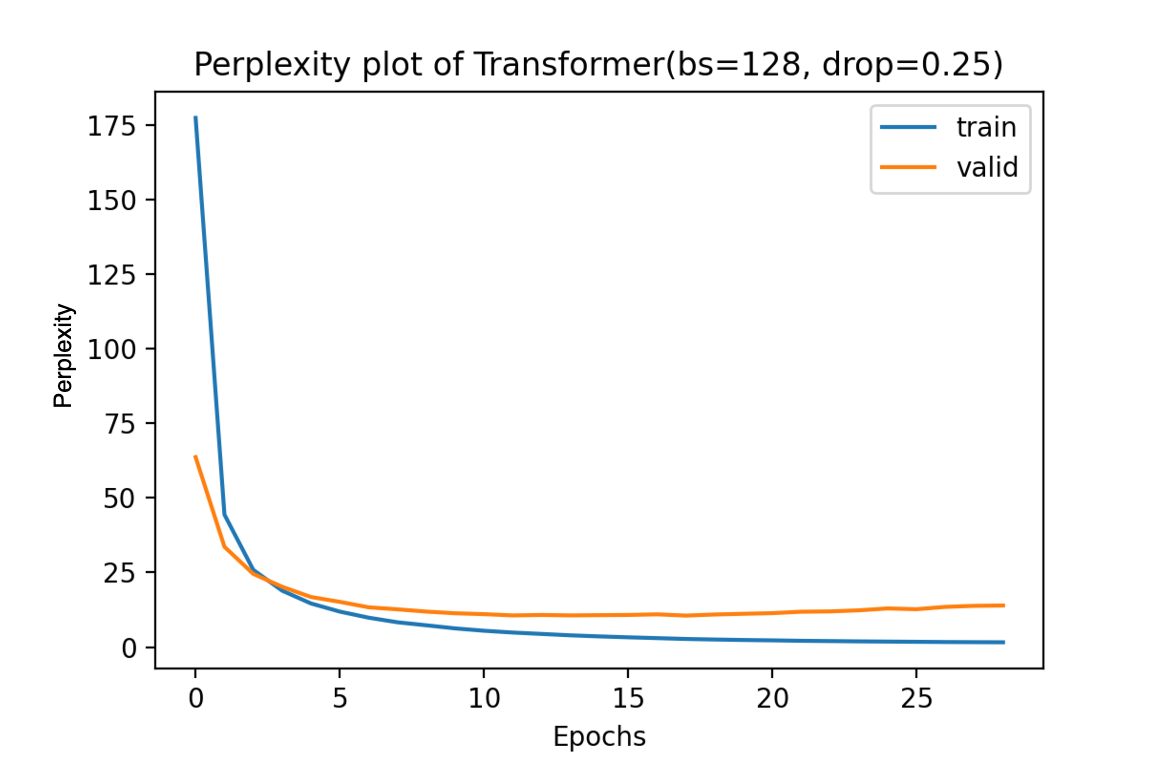}
	\caption{Perplexity plot for the Transformer.}
	\label{fig:perplex-transf-b128-d0}
\end{figure}

\begin{figure}
	\centering
	\includegraphics[width=0.9\linewidth]{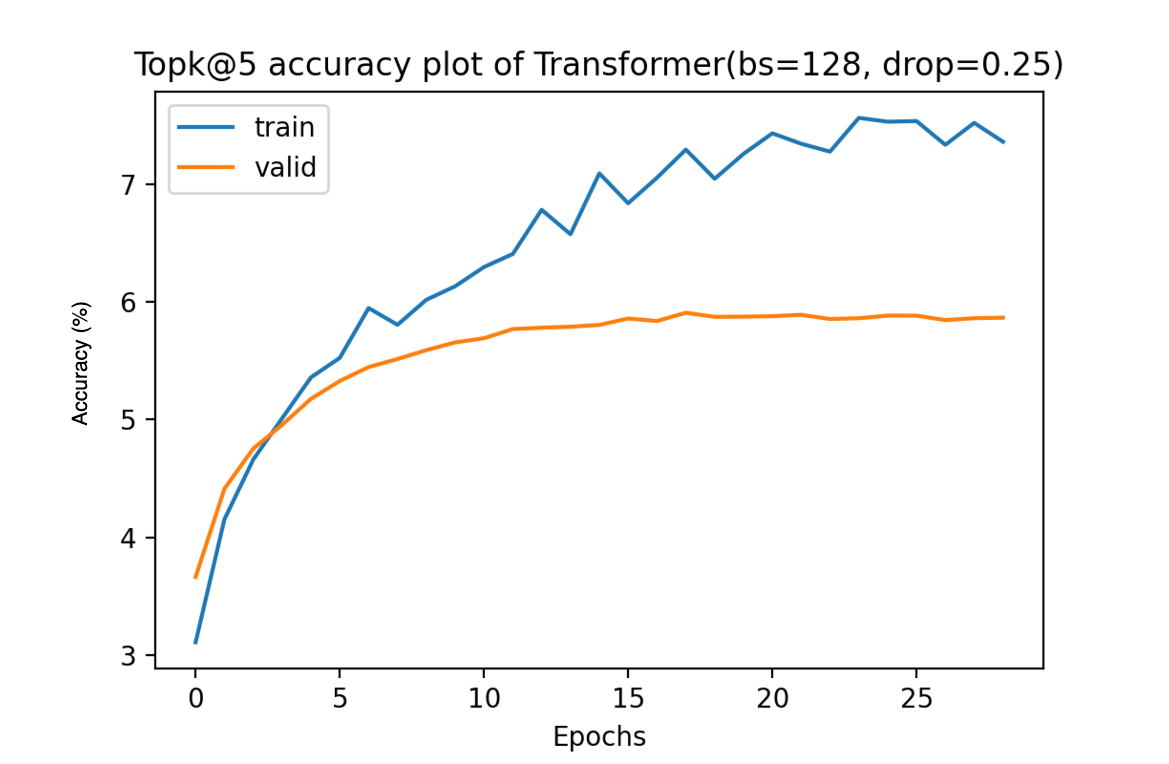}
	\caption{Top-k accuracy plot for the Transformer.}
	\label{fig:acc-transf-b128-d0}
\end{figure}

\textbf{Results.}
\autoref{fig:loss-transf-b128-d0} shows the plot of the training and validation loss across epochs. \autoref{fig:perplex-transf-b128-d0} shows the training and validation perplexity measure of the model architecture. \autoref{fig:acc-transf-b128-d0} shows the top-5 accuracy for both training and validation. 

\autoref{tab:metric-transf} shows the evaluation metric scores of the Transformer model after hyperparameter tuning for both validation and test datasets respectively. \autoref{tab:bleu-transf} shows a finer Sentence-BLEU evaluation of generated code snippets from the model and the histogram of code translations sorted in interval range of Sentence-BLEU score. \autoref{tab:metric-transf-seq2seq} shows the comparison of the metric scores of the Seq2Seq with the Transformer model architecture on the test dataset.  \autoref{tab:pred-seq2seq-transf-sideways} compares the translations generated by both the Transformer and the vanilla Seq2Seq on the test set.

\textbf{Finding 1.}  From the \autoref{fig:loss-transf-b128-d0}, we observed that the validation loss graph looked flat and generally the loss reported was between in the range of 2 and 3 while training for higher epochs without early stopping mechanism. We could say that the transformer model reported lower training and validation loss as compared to the vanilla Seq2Seq model. As we observed in the  \autoref{fig:perplex-transf-b128-d0}, the validation perplexity got decreased along with the training perplexity across epochs. Therefore, we could infer that the transformer model had been less prone to overfitting as compared to the vanilla Seq2Seq model.

\textbf{Finding 2.} From the \autoref{fig:acc-transf-b128-d0}, we found that the top-5 training accuracy kept increasing in a zigzag fashion, whereas the validation top-5 accuracy seemed to have plateaued as the training progressed after 15 epochs. 

\textbf{Finding 3.} From the \autoref{tab:bleu-transf}, we observed that the model generated 24 code translations whose Sentence-BLEU score is greater than 0.6 compared to 16 code translations from the Seq2Seq model. The transformer model generated only 2 exact matches, 22 mostly correct, 80 marginally correct, and 179 semantically equivalent code translations, much higher than the vanilla Seq2Seq model.

In comparison to the vanilla Seq2Seq, the transformer architecture had generated 138 valid compilable code snippets, whereas only 73 test predictions from the vanilla Se2Seq were valid compilable code snippets.

\textbf{Finding 4.} From \autoref{tab:metric-transf-seq2seq}, we observed that the Transformer with CoNaLa code tokenizer had exceeded the performance of the vanilla Seq2Seq on BLEU metric by 15.43\%, on Rouge1 F1-score by 4.45\%, on Rouge2 F1-Score by 10.45\%, and on RougeL F1-Score by 4.35\%. This explains that the Transformer model had resulted in generating better overall code translations with a higher quantitative metric score than the vanilla Seq2Seq model.

\textbf{Finding 5.} From \autoref{tab:pred-seq2seq-transf-sideways}, we inferred that the transformer model had worked well for the intents related to the os path and system calls, regex calls such as finding and matching requests, dictionary operations such as sorting dictionaries, removing none values from dictionaries, serialization operations like encoding and decoding of objects to ASCII/Unicode set, web API calls like downloading files from URL, datatype conversion operations, string, and list slicing operations. 

The transformer had failed to infer the variable, parameters, function names, and types from the NL intent and could not translate them into the correct code snippet in many test examples. The model failed for the intent like complicated Pandas dataframe and Numpy commands, Matplotlib command, complicated matrix and regex operations, string formatting operations, and database CRUD requests. 

\begin{table}[]
	\centering
	\Large
	\begin{adjustbox}{width=\textwidth}
		\begin{tabular}{|c|c|c|c|c|c|c|}
			\hline
			\Large
			\textbf{Dataset}    & \textbf{BLEU}     & \textbf{Rouge1 F1} & \textbf{Rouge2 F1} & \textbf{Rouge4 F1e} & \textbf{RougeL F1} & \textbf{Token Accuracy} \\ \hline
			Validation & 20.5571 & 0.5769       & 0.3273        & 0.2004       & 0.5682       & 18.0473       \\ 
			Test       & 15.3834 & 0.5449       & 0.2628       & 0.1311       & 0.5347      & 14.7041       \\ \hline
		\end{tabular}
	\end{adjustbox}
\captionof{table}[Metrics score for Transformer]{Validation and Test metrics for Transformer (best model). \label{tab:metric-transf}}
\end{table}

\begin{table}[]
	\centering
	\begin{tabular}{|c|c|}
		\hline
		\textbf{Non fine-tuned model(Sentence-BLEU)} & \textbf{\#prediction} \\ \hline
		\textgreater{}0.9                           & 2                     \\ \hline
		\textgreater{}=0.8 and \textless{}=0.9      & 1                     \\ \hline
		\textgreater{}=0.7 and \textless{}=0.8      & 7                     \\ \hline
		\textgreater{}=0.6 and \textless{}=0.7       & 14                    \\ \hline
		\textgreater{}=0.5 and \textless{}=0.6      & 29                    \\ \hline
		\textgreater{}=0.4 and \textless{}=0.5      & 51                    \\ \hline
		\textgreater{}=0.3 and \textless{}=0.4      & 79                    \\ \hline
		\textgreater{}=0.2 and \textless{}=0.3      & 100                   \\ \hline
		\textgreater{}=0.1 and \textless{}=0.2      & 91                    \\ \hline
		\textless 0.1                               & 103                   \\ \hline
		\textgreater{}0.5                           & 53                    \\ \hline
	\end{tabular}
\captionof{table}[Sentence-BLEU test metric score for Transformer]{Histogram of test predictions of Transformer in the interval range of Sentence-BLEU scores. \label{tab:bleu-transf}}
\end{table}

\begin{table}[]
	\centering
	\Large
			\resizebox{\linewidth}{!}{
		\begin{tabular}{|c|c|c|c|c|c|c|}
			\hline
			\Large
			\textbf{Model} & \textbf{BLEU}     & \textbf{Rouge1 F1} & \textbf{Rouge2 F1} & \textbf{Rouge4 F1} & \textbf{RougeL F1} & \textbf{Token Accuracy} \\ \hline
			Vanilla Seq2Seq  & 13.3270 & 0.5216       & 0.2379       & 0.1094       & 0.5128      & 14.6416        \\ 
			Transformer-code tokenizer       & \textbf{15.3834} & \textbf{0.5449}       & \textbf{0.2628}       & \textbf{0.1311}       & \textbf{0.5347}      & \textbf{14.7041}       \\ \hline
		\end{tabular}}
	\captionof{table}[Transformer v/s Seq2Seq]{Test set metrics comparison: Transformer v/s Seq2Seq. \label{tab:metric-transf-seq2seq}}
\end{table}

	\begin{sidewaystable}
			\captionof{table}{Comparison of Test set predictions of Seq2Seq v/s Transformer \label{tab:pred-seq2seq-transf-sideways}}
		\centering
		\begin{tabularx}{\textwidth}{|*{4}{>{\RaggedRight\arraybackslash}X|}}
			\hline
			\textbf{Intent}                                                                                     & \textbf{Seq2Seq pred}                                                                                                                                                                                                                                                                                                                                                                                                                                                                                                                                                                                                                                                                                                 & \textbf{Transformer Pred}                                                                                                                                                                     & \textbf{Snippet}                                                                                                                                     \\ \hline
			Download the file   from url `url' and save it under file `file\_name' & file\_name = requests.download (url)                                                                                                                                                                                                                                                                                                                                                                                                                                                                                                                                                                                                                                                                            & urllib.request.urlretrieve(` http : / / / randomsite . com / file . gz' , `file.gz') & urllib.request.urlretrieve(url , file\_name) \\ \hline
			Convert unicode   string `s' to ascii. & s.decode(`unicode\_escape')                                                                                                                                                                                                                                                                                                                                                                                                                                                                                                                                                                                                                                                                                  & s.encode(`ascii' , `ignore')                                                                                                                                                       & s.encode(`iso- 8859 - 15') \\ \hline
			Write dataframe `df', excluding index, to a csv file.                                          & df.df.index(df.values)                                                                                                                                                                                                                                                                                                                                                                                                                                                                                                                                                                                                                                                                                     & df.to\_csv (`a' , columns)                                                                                                                                                            & df.to\_csv(filename , index = False) \\ \hline
			Split string `words, words, words.' using a regex ` ( \textbackslash{}\textbackslash{}w+ ) ' & re.sub(` \textbackslash \textbackslash \textbackslash   \textbackslash \textbackslash\textbackslash{}') & re.split( ` \textbackslash \textbackslash   \textbackslash W + ` , ` \textbackslash \textbackslash \textbackslash \textbackslash \textbackslash 1 ` , a \textbackslash \textbackslash 1 ' ) & re.split(`( \textbackslash   \textbackslash W + ) ' , `Words, words, words.' )                                                    \\ \hline
			Open a file `output.txt' in append mode. & try = open ( `test.txt' , `rb')                                                                                                                                                                                                                                                                                                                                                                                                                                                                                                                                                                                                                                                                               & with open(`test.txt' , `r' ) as f : myfile.read()                                                                                                                               & file = open (`Output.txt' , `a')                                                                                                  \\ \hline
			Decode json string   `request.body' to python dict.                                                & json.decode(`\% 0 : \% m - \% d ` \% f . \% m - \% d \% H : \% M : \% S . \%                                                                                                                                                                                                                                                                                                                                                                                                                                                                                                                                                                                                                                   & json.load(json\_string))                                                                                                                                                               & json.loads(request.body)                                                                                                           \\ \hline
			Remove last comma character `, ' in string ` s'                                                 & re.sub(` {[} : A   - Z {]}' , ` ' , s)                                                                                                                                                                                                                                                                                                                                                                                                                                                                                                                                                                                                                                                                          & s.replace(`,` ,` , 1)                                                                                                                                                               & ` ` ` ` ` ` . join(s.rsplit(` , ` , 1 ))                                                                                           \\ \hline
			Check characters of string `string'  are true predication of function `predicate'. & ` ` ` . . join(string.isdigit( ) )                                                                                                                                                                                                                                                                                                                                                                                                                                                                                                                                                                                                                                                                             & print(re.findall ( ` {[} \textbackslash \textbackslash \textbackslash \textbackslash \textbackslash \textbackslash \textbackslash \textbackslash \textbackslash 1 {]} * ` , string ) )  & all( predicate(x ) for x in string )                                                                                                   \\ \hline
			Sort a list of   dictionary ` mylist ` by the key `title'. & mylist.sort(key = lambda.get ( ` .' ). \_\_dict\_\_) )                                                                                                                                                                                                                                                                                                                                                                                                                                                                                                                                                                                                                                                     & mylist.sort(key = lambda x : x {[} 1 {]} )                                                                                                                                              & mylist.sort(key = lambda x : x {[} ` title ` {]} )                                                                                    \\ \hline
			Get rid of none values in dictionary `kwargs'. & print((v for v in list(subset.items()))                                                                                                                                                                                                                                                                                                                                                                                                                                                                                                                                                                                                                                                                  & \{ k : v for k , v   in list ( d . items ( ) ) if v ! = k ! = ` \})                                                                                                                           & res = \{ k : v for k, v in list (kwargs.items()) if v is not None \}                                                               \\ \hline
			Match urls whose  domain doesn't start with ` t'  from string `document' using regex. & re. sub(`{[} \textasciicircum \textbackslash   \textbackslash \textbackslash {]} ` ` , ` ` , ` ` , ` )                                                                                                                                                                                                                                                                                                                                                                                                                                                                                                                                                                                                            & re.findall (` {[}   \textbackslash \textbackslash \textbackslash \textbackslash {]} ` , ` ` ` ` ` , text )                                                                                 & re.findall(`http : / / {[} \textasciicircum t {]} {[} \textasciicircum s ` {]} + \textbackslash \textbackslash . html ` , document) \\ \hline
			Sum columns of a list `array'. & numpy.array(array)                                                                                                                                                                                                                                                                                                                                                                                                                                                                                                                                                                                                                                                                                             & sum(sum(j) for i in structure)                                                                                                                                                        & {[} sum ( row {[} i {]}   for row in array ) for i in range ( len ( array {[} 0 {]} ) ) {]}                                                 \\ \hline
		\end{tabularx}
	\end{sidewaystable}
	

\subsubsection{RQ1d. How well the proposed hybrid Seq2Seq-RoBERTa architecture performs w.r.t. the Transformer?}

\textbf{Motivation.}
Given that BERT \citep{devlin-etal-2019-bert} has achieved great success in natural language understanding tasks, we were motivated to study how we can solve our task by incorporating BERT-based architectures such as RoBERTa in the Seq2Seq framework. Due to unavailability of high computational resources, we could not train a RoBERTa model from scratch on the dataset. Hence, we focused on leveraging a pretrained BERT-based model (instead of training a BERT-based model from scratch) for translating natural language to a code snippet by proposing to build a hybrid Seq2Seq architecture framework, where we could leverage the pretrained weights of the DistilRoBERTa\footnote{https://huggingface.co/distilroberta-base}, a lighter architecture of RoBERTa. The hybrid Seq2Seq-RoBERTa architecture consists of a DistilRoBERTa encoder and a DistilRoBERTa decoder. The DistilRoBERTa model consists of six layers in both the encoder and the decoder respectively.

\textbf{Approach.}
Given that there is limited work leveraging RoBERTa for Code2NL task, our first attempt was to try two strategies: (1) using RoBERTa to initialize Seq2Seq model,  pretrain RoBERTa on the large mined CoNaLa corpus, then fine-tune the pretrained model on the smaller training CoNaLa dataset; and (2) using RoBERTa directly without pretraining on the CoNaLa mined dataset to initialize the Seq2Seq architecture and train on the smaller dataset for NL2Code task while using the original pretrained weights of the RoBERTa, which was pretrained on the large unstructured  English corpus instead.

Compared to the standard Seq2Seq framework, in the RoBERTa-encoder attention and RoBERTa-decoder attention framework, the input sequence is first transformed into BERT representations processed by RoBERTa architecture. Then, by the RoBERTa-encoder attention module, each Seq2Seq encoder layer interacts with the representations obtained from RoBERTa and eventually outputs fused representations leveraging both RoBERTa and the Seq2Seq encoder. The decoder works similarly and fuses BERT representations and Seq2Seq encoder representations.

The Seq2Seq-RoBERTa has both DistilRoBERTa's pretrained encoder and decoder respectively. Initially, Seq2Seq-RoBERTa was trained on the CoNaLa mined30k corpus then the pretrained model was fine-tuned on the training dataset. The pretraining step involved training the model on the mined30k corpus for the first 10 epochs with a warmup\_step = 50 then it was further trained for another 40 epochs with a warmup\_step = 10 and with a virtual batch size of 32 by gradient accumulation strategy\footnote{https://towardsdatascience.com/what-is-gradient-accumulation-in-deep-learning-ec034122cfa} with a gradient accumulation step size of 4. The learning rate and the optimizer were used as in  \citep{liu2019roberta}. We used SacreBLEU \citep{post-2018-call} to compute the corpus BLEU scores of the generated code translations.

We trained the Seq2Seq-RoBERTa architecture on the training set, validated the validation set, used beam search decoding with varying beam sizes, and greedy decoding to derive the results. We also performed hyperparameter tuning to further fine-tune the architecture to derive the best results on the test dataset. We also performed both ablation studies about the number of self-attention heads and the depth of the Seq2Seq-RoBERTa to understand the crucial and influential components of the proposed architecture as discussed in \autoref{RQ2} (RQ2c and RQ2d).

\textbf{Hyperparameter Tuning}: We varied the batch size from 8 to 64 and found out that the model resulted in best test metric scores for batch size of 8. We used a gradient accumulation strategy to increase the batch size virtually due to limited memory resources. We varied the number of self-attention layers, the number of encoder and decoder layers in the Seq2Seq-RoBERTa architecture, and the beam size of [3, 5, 7, 10, 15] for beam search decoding of the output from the decoder. We eventually observed that for smaller batch sizes of 8 and 16, the model reported the worst test metric scores.

\begin{figure}[h!]
	\centering
	\includegraphics[width=0.7\linewidth]{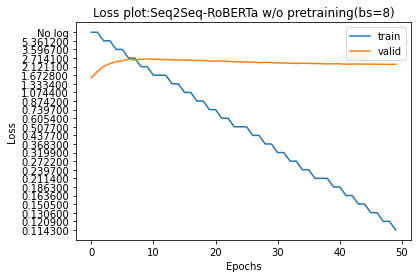}
	\caption{Loss plot for the Seq2Seq-RoBERTa without pretraining.}
	\label{fig:loss-robertaseq2seq-ep50-b8}
\end{figure}

\begin{figure}[h!]
	\centering
	\includegraphics[width=0.8\linewidth]{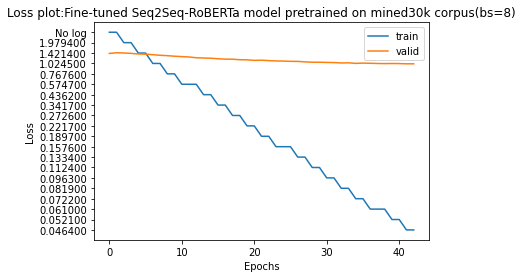}
	\caption[Loss plot for the fine-tuned Seq2Seq-RoBERTa.]{Loss plot for the fine-tuned Seq2Seq-RoBERTa model, pretrained on mined30k corpus.}
	\label{fig:loss-robertaseq2seq-finetuned-ep50-b8}
\end{figure}

\begin{figure}[h!]
	\centering
	\includegraphics[width=0.7\linewidth]{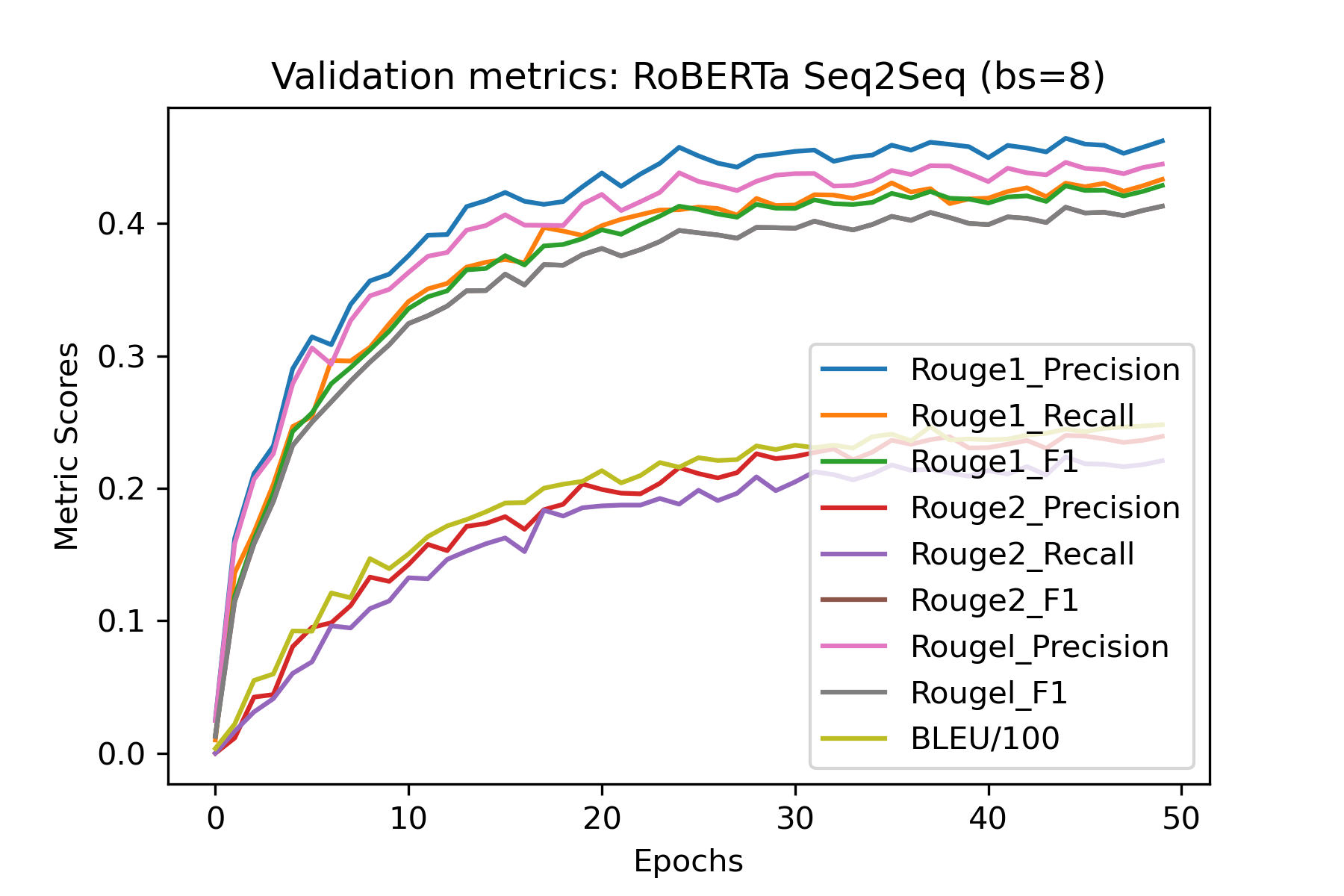}
	\caption[Validation metrics for the non fine-tuned Seq2Seq-RoBERTa.]{Validation metrics for the non fine-tuned Seq2Seq-RoBERTa without pretraining(bs=8).}
	\label{fig:valid-scores-roberta-ep50-b8}
\end{figure}

\begin{figure}[h!]
	\centering
	\includegraphics[width=\linewidth]{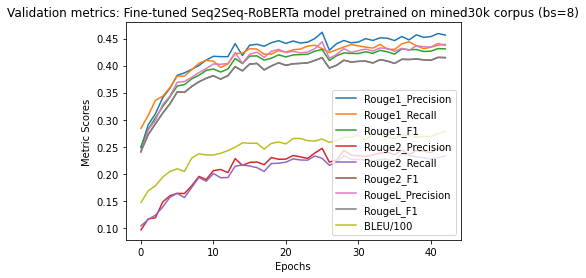}
	\caption[Validation metrics plot for the fine-tuned Seq2Seq-RoBERTa.]{Validation metric scores for the fine-tuned Seq2Seq-RoBERTa with pretrained on mined30k corpus.}
	\label{fig:valid-scores-roberta-seq2seq-finetuned30k-ep50-b8}
\end{figure}

\textbf{Results.}
\autoref{fig:loss-robertaseq2seq-ep50-b8} shows the plot of the training and the validation loss across epochs for the Seq2Seq-RoBERTa model without pretraining on the mined corpus.
\autoref{fig:loss-robertaseq2seq-finetuned-ep50-b8} shows the training and validation loss across epochs for the fine-tuned Seq2Seq-RoBERTa model with pretraining on mined30k corpus.
\autoref{fig:valid-scores-roberta-ep50-b8} shows the validation metric scores for the Seq2Seq-RoBERTa without pretraining. 
\autoref{fig:valid-scores-roberta-seq2seq-finetuned30k-ep50-b8} shows the validation metric scores for the fine-tuned Seq2Seq-RoBERTa with pretraining on mined30k corpus. 

\autoref{tab:bleu-roberta} displays the histogram of generated test code predictions of Seq2Seq-RoBERTa without pretraining in the interval range of Sentence-BLEU test set scores.
\autoref{tab:bleu-roberta-finetuned} shows the histogram of test code predictions of the fine-tuned Seq2Seq-RoBERTa, pretrained on mined30k corpus given the interval range of Sentence-BLEU metric test scores.
\autoref{tab:metric-transf-roberta} compares the test set metrics of Transformer with Seq2Seq-RoBERTa.
\autoref{tab:pred-roberta} shows the translations generated from the Seq2Seq-RoBERTa without pretraining.
\autoref{tab:pred-roberta-compare-sideways} lists down the test code output and compares the test code predictions qualitatively of Seq2Seq-RoBERTa without pretraining with the fine-tuned Seq2Seq-RoBERTa model with pretraining.

\textbf{Finding 1.} From \autoref{fig:loss-robertaseq2seq-ep50-b8}, we observed that both the training and the validation loss decreased rapidly till the 10th epoch; thereafter, both the training and the validation loss slowly converged. The validation loss dropped from 5.035 to 3.0822 on training for 10 epochs then it increased to 3.727, whereas the training loss kept decreasing to 0.1143 till the end of 50 epochs of training. This showed that the Seq2Seq-RoBERTa had overfitted on the dataset with much less bias compared to the training of the vanilla Seq2Seq architecture. 

From \autoref{fig:loss-robertaseq2seq-finetuned-ep50-b8}, we observed that for the fine-tuning process, the pretrained model initially started training with a very small loss compared to the non-pretrained model. Then, after 20 epochs of training, the validation loss kept oscillating between 0.63584 and 0.685394 and became flat, whereas the training loss kept converging. This showed that the pretrained model did not require to be trained over 20 epochs with much lower bias compared to the non-pretrained model.

\textbf{Finding 2.} From \autoref{fig:valid-scores-roberta-ep50-b8}, we observed that for the Seq2Seq-RoBERTa without pretraining, the validation scores of ROUGE metrics including the Precision, Recall, F1-Score steadily increased till 20 epochs of training and then the rate of increment almost flattened. We also observed that the validation BLEU score increased from 0.35 to 20.5313 after 20 epochs of training then the validation BLEU score slowly increased to 24.81 till the end of 50 epochs of training. 

\textbf{Finding 3.} From \autoref{fig:valid-scores-roberta-seq2seq-finetuned30k-ep50-b8}, we noticed that for the fine-tuned model with pretraining, the validation BLEU score increased from 14.690 to 25.611 after 20 epochs of training then the validation BLEU score slowly increased to 27.8190 till the end of 50 epochs of training. This showed that the fine-tuned model had resulted in a higher BLEU score than the non fine-tuned one without pretraining.

\textbf{Finding 4.} In \autoref{tab:bleu-roberta-finetuned}, we reported that there was an improvement in the generation of number of high-quality code translations for the fine-tuned model. This is due to the presence of the pretrained knowledge in the fine-tuned model. We reported 5 exact matches, 35 mostly correct, 89 marginally correct, and 148 semantically equivalent code translations generated for the fine-tuned model, much higher than the non fine-tuned model without pretraining.

\textbf{Finding 5.} Interestingly, the number of valid parsable generated code snippets increased to 425 from 277 for the fine-tuned model. This gain was significant not only for the generated syntactically correct code translation but also for the correctness of the translation with the ground truth.

\textbf{Finding 6.} In \autoref{tab:pred-roberta}, we listed down some code translations obtained from the non-pretrained SeqSeq-RoBERTa model for the test set. This model had been able to work on translating intents like, performing operations on the list, tuple such as finding the length of a list, converting a list into a tuple, concatenating arrays, sorting lists, clock operations, and file operations such as read and open. The non fine-tuned model didn't work well for complicated regex intents like replacing characters with ASCII letters in a list, os system calls, multiple operations with dictionaries, Pandas dataframe operations like merging dataframes, selecting particular column from a dataframe, etc. 

\textbf{Finding 7.} From \autoref{tab:pred-roberta-compare-sideways}, we saw that the code generated was of higher quality for the fine-tuned Seq2Seq-RoBERTa with pretraining than the Seq2Seq-RoBERTa without pretraining. This table also pointed out for which type of intents the fine-tuned model worked better than the non fine-tuned model. We found that the fine-tuned model worked better for complicated regex calls, dictionary operations like sorting list of dictionaries, system calls, web API calls such as, sending http header over HTTP, pandas data frame, Matplotlib intents like plotting figures, database CRUD requests like inserting values in a table, lambda expression intents, etc.

\textbf{Finding 8.} From \autoref{tab:metric-transf-roberta}, we saw that the fine-tuned Seq2Seq-RoBERTa with pretraining on mined30k corpus performed better than the non fine-tuned Seq2Seq-RoBERTa without pretraining by 10.9\%. The same fine-tuned Seq2Seq-RoBERTa model pretrained on mined30k corpus also surpassed the Transformer with CoNaLa code tokenizer by 22.767\% on the BLEU metric.


\begin{table}[!htbp]
	\centering
	\begin{tabular}{|c|c|c|}
		\hline
		\textbf{Non fine-tuned model(Sentence-BLEU)} & \textbf{\#prediction} \\ \hline
		\textgreater{}0.9                           & 3                     \\ \hline
		\textgreater{}=0.8 and \textless{}=0.9      & 1                     \\ \hline
		\textgreater{}=0.7 and \textless{}=0.8      & 4                     \\ \hline
		\textgreater{}=0.6 and \textless{}=0.7       & 11                    \\ \hline
		\textgreater{}=0.5 and \textless{}=0.6      & 26                    \\ \hline
		\textgreater{}=0.4 and \textless{}=0.5      & 42                    \\ \hline
		\textgreater{}=0.3 and \textless{}=0.4      & 67                    \\ \hline
		\textgreater{}=0.2 and \textless{}=0.3      & 103                   \\ \hline
		\textgreater{}=0.1 and \textless{}=0.2      & 98                    \\ \hline
		\textless 0.1                               & 122                   \\ \hline
		\textgreater{}0.5                           & 45                    \\ \hline
	\end{tabular}
	\captionof{table}[Sentence-BLEU metric for the non fine-tuned Seq2Seq-RoBERTa]{Histogram of test predictions of the non fine-tuned Seq2Seq-RoBERTa in the interval range of Sentence-BLEU scores.\label{tab:bleu-roberta}.}
\end{table}

\begin{table}[!htbp]
	\centering
	
	\begin{tabular}{|c|c|c|}
		\hline
		\textbf{Fine-tuned model after pretraining (Sentence-BLEU)} & \textbf{\#prediction} \\ \hline
		\textgreater{}0.9                           & 5                     \\ \hline
		\textgreater{}=0.8 and \textless{}=0.9      & 7                    \\ \hline
		\textgreater{}=0.7 and \textless{}=0.8      & 10                     \\ \hline
		\textgreater{}=0.6 and \textless{}=0.7       & 18                    \\ \hline
		\textgreater{}=0.5 and \textless{}=0.6      & 38                    \\ \hline
		\textgreater{}=0.4 and \textless{}=0.5      & 51                     \\ \hline
		\textgreater{}=0.3 and \textless{}=0.4      & 75                  \\ \hline
		\textgreater{}=0.2 and \textless{}=0.3      & 73                 \\ \hline
		\textgreater{}=0.1 and \textless{}=0.2      & 97                  \\ \hline
		\textless 0.1                               & 106                   \\ \hline
		\textgreater{}0.5                           & 78                    \\ \hline
	\end{tabular}
\captionof{table}[Sentence-BLEU metric for the fine-tuned Seq2Seq-RoBERTa]{Histogram of test predictions of the fine-tuned Seq2Seq-RoBERTa pretrained on mined30k corpus in the interval range of Sentence-BLEU scores.	\label{tab:bleu-roberta-finetuned}}
\end{table}

\begin{table}[!htbp]
	\centering
	\Large
	\begin{adjustbox}{width=\textwidth}
		\begin{tabular}{|c|c|c|c|c|c|c|}
			\hline
			\Large
			\textbf{Model}    & \textbf{BLEU}     & \textbf{Rouge1 F1} & \textbf{Rouge2 F1} & \textbf{RougeL F1}  \\ \hline
			Fine-tuned Seq2Seq with pretrained RoBERTa & \textbf{18.8853} & 0.33110 &  0.13540 & 0.31740  \\ 
			Seq2Seq-RoBERTa w/o pretraining & 17.0032 &  0.34790       & 0.13360  &  0.33360     \\ 
			Transformer with CoNala code tokenizer & 15.3834 & \textbf{0.54488}       & \textbf{0.26279}  & \textbf{0.53466}         \\ \hline
		\end{tabular}
	\end{adjustbox}
\captionof{table}[Test set metrics: Transformer v/s Seq2Seq-RoBERTa]{Test set metrics comparison: Transformer v/s Seq2Seq-RoBERTa\label{tab:metric-transf-roberta}.}
\end{table}

		\begin{sidewaystable}[!htbp]
			\captionof{table}{Test set predictions of RoBERTa without pretraining\label{tab:pred-roberta}.}
			\centering
		
		{\small\setlength{\tabcolsep}{2pt}
			\footnotesize
		\resizebox{\columnwidth}{!}{\begin{tabular}{|p{0.25\linewidth} | p{0.5\linewidth} | p{0.25\linewidth}|}
		\hline
		\textbf{Intent}                                                                               & \textbf{RoBERTa Prediction (top 3)}                                                                                                                                                                                                                                                                                                                                                                                                                                                                                                                                    & \textbf{Snippet}                                                                \\ \hline
		Enable debug mode on Flask application `app'.                                                & {[}`app.run(debug=True)',   `app.run(app.debug(debug=True)'                                                                                                                                                                                                                                                                                                                                                                                                                                                                           & app.run(debug=True)                                                             \\ \hline
		Make a delay of 1 second.                                                                    & {[}`time.sleep(1)',   `sleep(1)'                                                                                                                                                                                                                                                                                                                                                                                                                                                                                                                  & time.sleep(1)                                                                   \\ \hline
		Getting the length of array `s'.                                                             & {[}`len(s)', `length   = len(s)'                                                                                                                                                                                                                                                                                                                                                                                                                                                                                                                & len(s)                                                                          \\ \hline
		Erase all the   contents of a file `filename'.                                                 & {[}"open(`filename',   `w').close()", "open(`filename', `w').close() as f:\textbackslash{}n   f.close"                                                                                                                                                                                                                                                                                                                                                                                              & open(filename,   `w').close()                                                   \\ \hline
		Convert list `t'   to tuple                                                                   & {[}`tuple(t)',   `tuple(i) for i in s.split()'                                                                                                                                                                                                                                                                                                                                                                                                                                                                                 & tuple(l)                                                                        \\ \hline
		Numpy concatenate two arrays `a' and `b' along the first axis.                               & {[}`b =   np.concatenate((a, b), axis=0)', `b = np.concatenate((a, b){]}, axis=0)', `b =   np.concatenate((a, b))'{]} axis=0)'                                                                                                                                                                                                                                                                                                                                                                                                                                                  & print(concatenate((a,   b), axis=0))                                            \\ \hline
		Sort a list of tuples `b` by third item in the tuple .                                       & {[}`list.sort(key=lambda   x: b{[}1{]})', `list.sort(key=operator.itemgetter(1, 2))',   `print(sorted(key=lambda x: b.index(x{[}1{]}))'{]}                                                                                                                                                                                                                                                                                                                                                                                                                             & b.sort(key=lambda   x: x{[}1{]}{[}2{]})                                         \\ \hline
		Open a file `Output.txt' in append mode                                                     & {[}"file = open(`file.txt', `r')", "file = open(`data.txt', `r')", "file = open(`file.txt', `r') as myfile.txt')"{]}                                                                                                                                                                                                                                                                                                                                                                                                                                                 & file = open(`Output.txt', 'a')                                                \\ \hline
		Download a file   `url' over HTTP & {[}`r = requests.download(url)', `urllib.download(url)', `urllib.request.urlpost(url)'{]} & r = requests.get(url)              \\ \hline
		Combine list of dictionaries `dicts' with the same keys in each list to a single dictionary & {[}`dict((k, v) for   k, v in dicts for k, v in dict', `dict((k, v) for d in dicts for k, v in   dicts for', `dict((k, v) for k, v in dicts for k, v in list'{]}                                                                                                                                                                                                                                                                                                                                                                                                       & dict((k, {[}d{[}k{]} for d in dicts{]}) for k in dicts{[}0{]})                \\ \hline
		Create new matrix object by concatenating data from matrix A and matrix B.                   & {[}`numpy.concatenate(a,   B))', `numpy.concatenate(a, B{[}0{]})', `numpy.dot(a, B{[}0{]})'{]}                                                                                                                                                                                                                                                                                                                                                                                                                                                                         & np.concatenate((A,   B))                                                        \\ \hline
		Remove all special   characters, punctuation and spaces from a string `mystring' using regex.  & {[}"re.sub(`\textbackslash{}\textbackslash{}\textbackslash{}\textbackslash{}w+|{[}\textasciicircum{}\textbackslash{}\textbackslash{}\textbackslash{}\textbackslash{}s{]}',   '', mystring)",
		"re.sub(`\textbackslash{}\textbackslash{}\textbackslash{}\textbackslash{}w+|{[}\textasciicircum{}\textbackslash{}\textbackslash{}\textbackslash{}\textbackslash{}s{]}', '', s)", "re.sub(`\textbackslash{}\textbackslash{}\textbackslash{}\textbackslash{}w+|{[}\textasciicircum{}\textbackslash{}\textbackslash{}\textbackslash{}\textbackslash{}s{]}', '', text, re"{]} & re.sub(`{[}\textasciicircum{}A-Za-z0-9{]}+',   '', mystring)                    \\ \hline
		Get a list of integers by splitting a string `user' with comma.                              & {[}"{[}int(x) for x in input.split(`,'){]}","{[}int(x) for x in example\_string.split(`,'){]}","{[}int(x) for x in input.split(`,')){]}"{]}                                                                                                                                                                                                                                                                                                                                                                                                                      & {[}int(s) for s in user.split(`,'){]}                                         \\ \hline
		Parse tab-delimited CSV file `text.txt' into a list. & {[}"numpy.loadtxt(open(`test.csv', `rb'), delimiter=`,'", "numpy.loadtxt(open(`myfile.csv', `rb'), delimiter=`,'"                                                                                                                                                                                                                                                           & lol = list(csv.reader(open(`text.txt', `rb'), delimiter=`\textbackslash{}t')) \\ \hline
		Get rid of None values in dictionary `kwargs'.                                              & {[}`{[}k for x in L if  x is not None{]}', "{[}(k, x) for k, v in list(Y) if x!= '", `{[}k for k,v in list(d.items() if x!= 0{]}'{]} & res = \{k: v for k,   v in list(kwargs.items()) if v is not None\}              \\ \hline
		Send a signal `signal.SIGUSR1` to the current process. &{[}`sys.chviron.inviron.get\_CRUSDR(1,0,1')', `sys.chviron.inviron.get\_WRITE(1, 0, 1)', 
		`os.chviron.inviron.
		get\_CRUSDR(1,0,1'){]} & os.kill(os.getpid(),   signal.SIGUSR1)      \\ \hline
		Decode a hex string `4a4b4c' to UTF-8.                                                      & {[}"print(decode(`utf8'))","print(c.decode(`utf8')).encode(`utf8')"           & bytes.fromhex(`4a4b4c').decode(`utf-8'),  "print(decode(`utf8')).decode
		(`utf8')"{]}                                           \\ \hline
		Check if all   elements in list `myList' are identical                                        & {[}`len(set(mylist))   == 1', `len(set(myList)) == 1', `all(set(mylist)) == 1'{]}                                                                                                                                                                                                                                                                                                                                                                                                                                                                                      & all(x == myList{[}0{]} for x in myList)                                       \\ \hline
	\end{tabular}}}
\end{sidewaystable}

		\begin{sidewaystable}[!htbp]
		\captionof{table}[Test predictions of Seq2Seq-RoBERTa v/s fine-tuned Seq2Seq-RoBERTa]{Seq2Seq-RoBERTa v/s fine-tuned Seq2Seq-RoBERTa test predictions (beam size=15).\label{tab:pred-roberta-compare-sideways}}
		\centering
		{\small\setlength{\tabcolsep}{2pt}
		\footnotesize
		\resizebox{\columnwidth}{!}{\begin{tabular}{|p{0.2\linewidth} | p{0.3\linewidth} | p{0.3\linewidth} | p{0.2\linewidth}|}
			\hline
			\textbf{Intent}                                                                                                                                     & \textbf{RoBERTa Prediction}                                                                                                                                                                                                                                                                                                     & \textbf{Fine-tuned RoBERTa Prediction}                                                                                                                                                                                                                                                                & \textbf{Snippet}                                                                                                                                       \\ \hline
			Download a file `url' over HTTP. & requests.download(url)                                                                                                                                                                                          & `r = requests.get(url)'                                                                                                                                                                                                                                                                               & r = requests.get(url)                                                                                                                                  \\ \hline
			Retrieve each line from a file `File.txt' as a list.                                                                                                 & {[}with open(''.join(txt.txt', 'r') as ins:\textbackslash{}n "'                                                                                                                                                                                                                                                                 & {[}line for line in open(`file.txt') if 'r'){]}"                                                                                                                                                                                                                                                      & {[}line.split() for line in open(`File.txt'){]}                                                                                                        \\ \hline
			Swap keys with values in a dictionary `a'. & {[}(k, v) for k, v in a.items(){]}                                                                                                                                                                                                                                                                                              & `dict((k, v) for k, v in a.items()))                                                                                                                                                                                                                                                                  & res = dict((v, k) for k, v in a.items())                                                                                                               \\ \hline
			Add an entry `wuggah' at the beginning of list `myList'.                                                                                             & {[}"mylist.startswith(`$\sim$a'){[}0{]}", "mylist.startswith(`$\sim$a'){[}0{]} for i in mylist{]}", "mylist.findall(`\textbackslash{}\textbackslash{}w+', `yourlist')"{]}                                                                                                                                                         & {[}"mylist.insert(0, `a')", "newlist.insert(0, `a')","'Hello\textbackslash{}\textbackslash{}n'.join(list(mylist))"{]}                                                                                                                                                                                  & myList.insert(0, `wuggah')                                                                                                                             \\ \hline
			Concatenate array of strings `{[}`A', `B', `C', `D'{]}` into a string.                                                                               & `"""".join({[}\textbackslash{}'\textbackslash{}'.join(lambda x: x) for x in `B'                                                                                                                                                                                                                            & {[}'"""""".join({[}\textbackslash{}`a\textbackslash{}', \textbackslash{}`b\textbackslash{}', \textbackslash{}`c\textbackslash{}', \textbackslash{}`d\textbackslash{}'{]})'                                                                                                                            & """""".join({[}`A', `B', `C', `D'{]})                                                                                                                  \\ \hline
			Sort a list of dictionary `list' first by key `points' and then by `time'. & {[}`sorted(list.items()), key=lambda item: 
			datetime.index(`x', `sorted(list, key=lambda item: datetime.index(x.index())', `sorted(list.items()), key=lambda item: item{[}1{]}{[}1{]})'{]}                                                                                                                                        & {[}"list.sort(key=lambda item: item{[}`date'{]}, reverse=True)", "pd.list\_of\_sort(key=lambda item: item{[}`date'{]}, reverse=True)", `pd.merge(list(my\_dictionary.items()), key=lambda item: item{[}1{]}{[}1{]})'{]}                                                                                & list.sort(key=lambda item: (item{[}`points'{]}, item{[}`time'{]}))                                                                                     \\ \hline
			Sending http headers with python. & `r = requests.get(url, headers=headers)', `r = requests.get(url)', "r = requests.get(url, headers=json.org')"{]}                                                                                                                                                                                                                 & {[}"connection.send(`HTTP/1.0 200 OK')", `s.setsockopt(url, headers=headers)',"connection.send(`HTTP/1.0 200 established')"{]}                                                                                                                                                                        & client.send(`HTTP/1.0 200 OK\textbackslash{}r\textbackslash{}n')                                                                                       \\ \hline
			Create list`lst' containing 100 instances of object `Object'. & {[}`{[}dict(x) for x in lst{]}', `{[}dict(x) for x in Lst if x{[}0{]} for x in range', `{[}dict(x) for x in lst if x{[}0{]} for x in range'{]} & {[}`{[}(i * 2) for i in range(10){]}', `{[}(i, j) for i in range(10){]}', `{[}(i \textless 2) for i in range(10){]}'{]}                                                                                                                                                                               & lst = {[}Object() for i in range(100){]}                                                                                                               \\ \hline
			Sort a list `l' by length of value in tuple. & {[}`print(sorted(l, key=lambda x: len(len(x))', `print(l, key=lambda x: len(l))', `print(l, key=lambda x: len(len(x))'{]}                                                                                                                                                                                                       & {[}`print(sorted(l, key=(lambda x: x{[}1{]}), reverse=True)', `print(sorted(l, key=(lambda x: x{[}1{]}), reverse=True)))', `print(sorted(l, key=(lambda x: x{[}2{]}), reverse=True)'{]}                                                                                                               & l.sort(key=lambda t: len(t{[}1{]}), reverse=True)                                                                                                      \\ \hline
			Get rid of None values in dictionary `kwargs'. & {[}`{[}k for x in L if x is not None{]}',"{[}(k, x) for k, v in list(Y) if x!= '", `{[}k for k, v in list(d.items() if x!= 0{]}'{]}                                                                                                                                                                                             & {[}`dict((k, v) for k, v in list(lambda x: x.items()))', `\{k: v for k, v in list(hand.items()) if v\}', `dict((k, v) for k, v in list(range(5, 6))'{]}                                                                                                                                               & res = \{k: v for k, v in list(kwargs.items()) if v is not None\} \\ \hline
			Sorted a list `l' of dicts by dict value `title' & {[}"sorted(l, key=lambda x: d{[}'name'{]}, reverse=True", "sorted(l, key=lambda x: int(x{[}'title'{]}, reverse=", "sorted(l, key=lambda x: d{[}'title'{]}, reverse=True"{]}                                                                                                                                                       & {[}``sorted(l, key=lambda x: x{[}'title'{]}, reverse=True)", ``sorted(l, key=lambda x: x{[}'weight'{]}, reverse=True)", ``sorted(l, key=lambda x: x{[}`rank'{]}, reverse=True)"{]}                                                                                                                         & l.sort(key=lambda x: x{[}`title'{]})                                                                                                                   \\ \hline
			Concatenate items of list `l' with a space. & "l = {[}{[}`'.join(l) for l in l{]}","l = {[}{[}`'.join(l) for l in l) for l in l{]}","l = {[}{[}`'.join(l) for l in l{]} for l in l{]}"{]}                                                                                                                                                                                       & "` '.join(map(lambda x: x * y, l))","` '.join(map(lambda x: x + y, l))","{[}` '.join(map(str, x)) for x in l{]}"{]}                                                                                                                                                                                   & print(` '.join(map(str, l)))                                                                                                                           \\ \hline
\end{tabular}}}
\end{sidewaystable}

 \subsubsection{RQ1e. Does the proposed hybrid Seq2Seq-BART(BERT encoder, GPT decoder) perform well?} 
\textbf{Motivation.}
BART has achieved the state-of-the-art results in a range of text generation tasks such as abstractive dialogue generation, question answering, and summarization tasks. This motivated us to use this architecture especially designed for conditional text generation to solve for our NL2Code objective.

\textbf{Approach.}
Given that there is limited work leveraging BART for Code2NL task, our first attempt was to try three strategies: (1) using BART to initialize Seq2Seq model, pretrain BART on the large mined100k corpus and then fine-tune the pretrained model on the smaller training (NL, code) CoNaLa dataset; (2) using BART directly without pretraining on the mined corpus to initialize the Seq2Seq architecture and training on the smaller dataset for NL2Code task while using the original pretrained weights of BART which was pretrained on the large unstructured  English corpus instead; and (3) pretraining the Seq2Seq-BART model on the CoNaLa mined100k corpus then fine-tuning on our augmented curated training dataset that we had created in this thesis. The last approach allowed the model to first learn on larger amounts of potentially noisy data, while finally being tailored to the actual NL and code we wanted to model at the test time.


The Seq2Seq-BART architecture has a BERT-encoder and a GPT decoder respectively. Initially, Seq2Seq-BART was trained on the CoNaLa mined100k corpus and the pretrained model was fine-tuned on the training dataset.  

We trained the Seq2Seq-BART architecture on the training set and validated the validation set and used beam search decoding with varying beam sizes to derive the results. We also performed hyperparameter tuning to further fine-tune the architecture to get the optimal results on the test set. During the translation generation via the beam search decoding, we set the beam size to be 5, removed duplicated trigrams from beam search output, put a constraint on the min-length and max-length of the output sequence, and provided length penalty hyperparameter.

\textbf{Hyperparameter Tuning}: We varied the batch size from 8 to 64 and found that the model resulted in best test scores for the batch size of 8. We used a gradient accumulation strategy to increase the batch size virtually due to limited memory resources. We also varied the number of self-attention layers, the number of layers in the encoder, and the decoder of the Seq2Seq-BART architecture. 

We varied both the beam size in [3, 5, 7, 10, 15] for beam search decoding of the output and the learning rate in [1e-05, 2e-05] to optimize the loss. We froze the entire architecture, trained the topmost layer, unfroze all the weights until the last two layers, and unfroze all the layers for training and fine-tuning purposes. We had observed that by fine-tuning the entire unfrozen architecture on the training dataset, the model achieved the best test metric scores and code translations.

\begin{figure}[h!]
	\centering
	\includegraphics[width=0.8\linewidth]{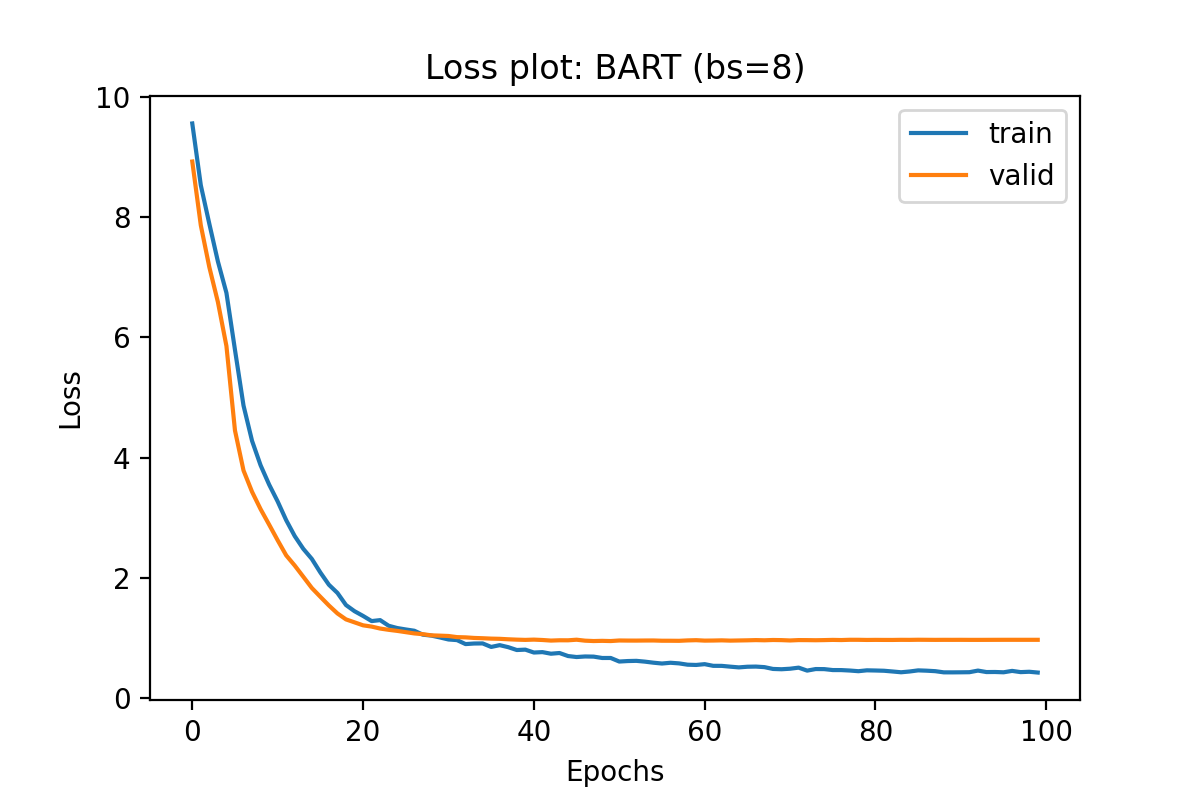}
	\caption[Loss plot for the non pretrained Seq2Seq-BART model.]{Loss plot for the Seq2Seq BART architecture without pretraining(bs=8).}
	\label{fig:loss-bart-ep100-b8}
\end{figure}

\begin{figure}
	\centering
	\includegraphics[width=0.9\linewidth]{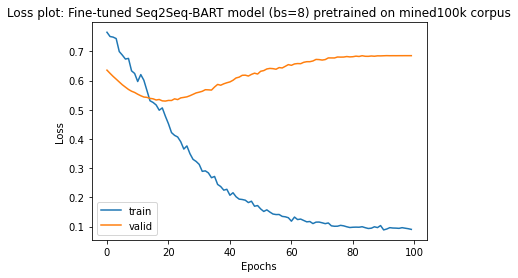}
	\caption[Loss plot for the fine-tuned Seq2Seq-BART architecture.]{Loss plot for the fine-tuned Seq2Seq-BART architecture after pretraining on mined100k corpus(bs=8).}
	\label{fig:loss-bart-finetuned100k-ep20-b8}
\end{figure}

\begin{figure}[h!]
	\centering
	\includegraphics[width=0.9\linewidth]{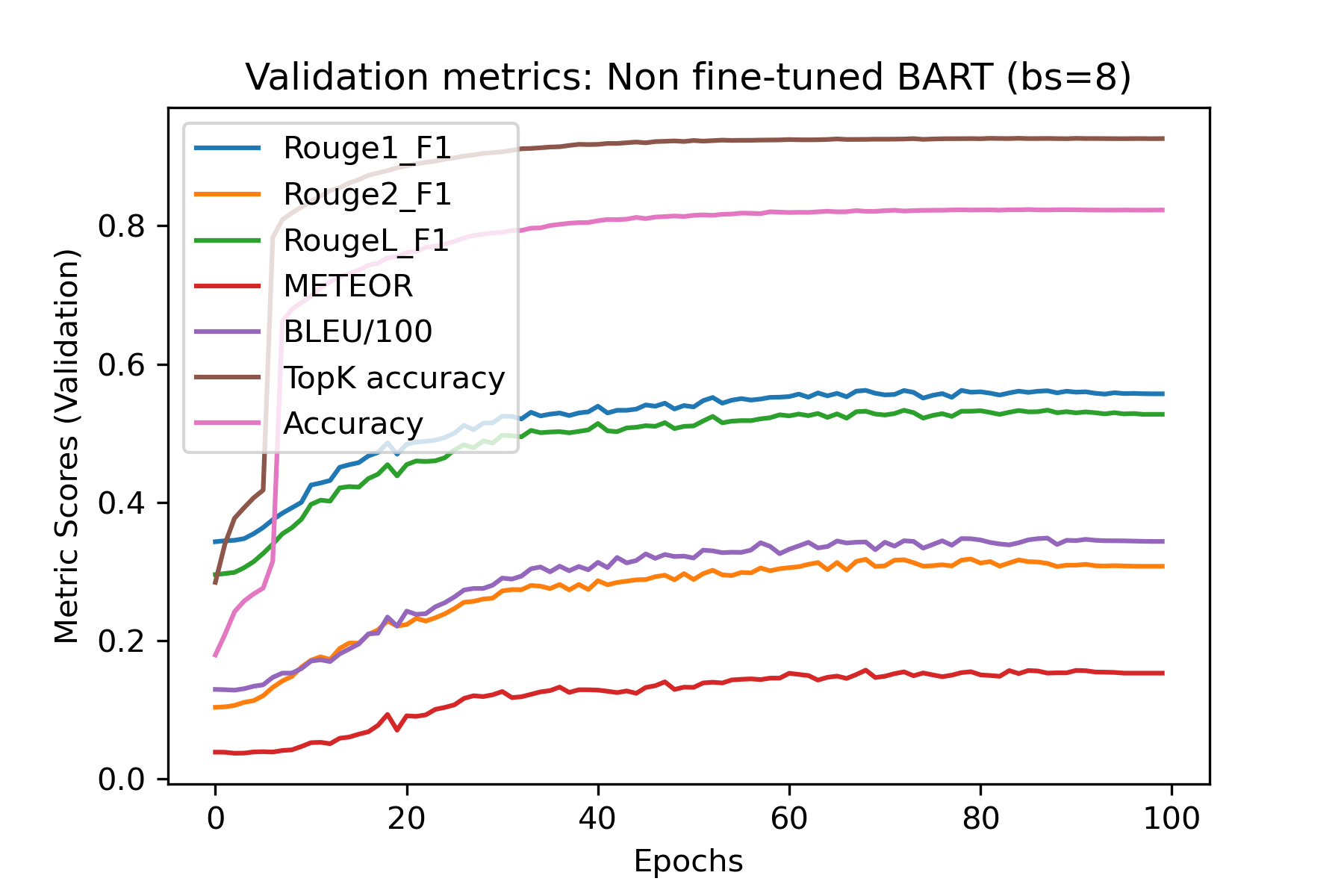}
	\caption[Validation metrics plot for the non-pretrained Seq2Seq-BART.]{Validation metrics for the Seq2Seq-BART without pretraining(bs=8).}
	\label{fig:valid-scores-bart-ep100-b8}
\end{figure}

\begin{figure}[h!]
	\centering
	\includegraphics[width=\linewidth]{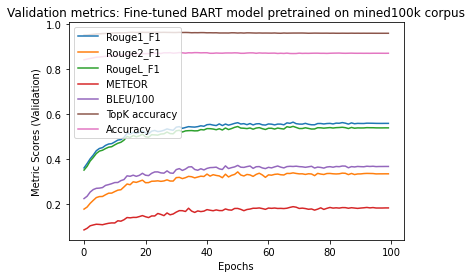}
	\caption[Validation metrics plot for the fine-tuned Seq2Seq-BART.]{Validation metrics of the fine-tuned Seq2Seq-BART after pretraining on mined100k corpus.}
	\label{fig:valid-scores-bart-finetuned100k-ep20-b8}
\end{figure}

\textbf{Results.}
\autoref{fig:loss-bart-ep100-b8} shows the plot of the training and the validation loss across epochs for the Seq2Seq-BART model without pretraining on the mined corpus.
\autoref{fig:loss-bart-finetuned100k-ep20-b8} shows the training and the validation loss across epochs for the fine-tuned Seq2Seq-BART model with pretraining on mined 100k corpus. 
\autoref{fig:valid-scores-bart-ep100-b8} shows the validation metric scores for the Seq2Seq-BART without pretraining. 
\autoref{fig:valid-scores-bart-finetuned100k-ep20-b8} shows the validation metric scores for the fine-tuned Seq2Seq-BART with pretraining on mined100k corpus. 

\autoref{tab:bleu-bart} displays the histogram of test code predictions of Seq2Seq-BART without pretraining in the range of Sentence-BLEU metric test scores.
\autoref{tab:bleu-bart-finetuned} shows the histogram of test code predictions of the fine-tuned Seq2Seq-BART pretrained on mined100k corpus in the range of Sentence-BLEU test scores.
\autoref{tab:metric-transf-roberta-bart} compares the test set metrics of Transformer, Seq2Seq-RoBERTa, and Seq2Seq-BART quantitatively.
\autoref{tab:pred-finetuned-bart-side1} and \autoref{tab:pred-finetuned-bart-side2} list down the generated code translations (beam size=15) from the test set and compare the code translations of Seq2Seq-BART without pretraining with the fine-tuned Seq2Seq-BART model, pretrained on mined100k corpus qualitatively.

\textbf{Finding 1.} From \autoref{fig:loss-bart-ep100-b8}, we saw that both the training and the validation loss decreased rapidly till 20th epochs; thereafter, both the training and the validation loss converged slowly. As the loss was converging, the model became less prone to overfit over the dataset as compared to the training of the transformer, the vanilla Seq2Seq, and the Seq2Seq-RoBERTa architectures, respectively. 

From \autoref{fig:loss-bart-finetuned100k-ep20-b8}, we observed that for the fine-tuning process, the pretrained model started training with a smaller initial loss compared to the non-pretrained model then after 20 epochs of training, the validation loss kept oscillating between 0.63584 and 0.685394 and became flat whereas, the training loss kept converging. Thus, the pretrained model did not require training more than 20 epochs for fine-tuning. 


\textbf{Finding 2.}
From \autoref{fig:valid-scores-bart-finetuned100k-ep20-b8}, we observed that both the validation accuracy and the top-k accuracy steadily increased till 20 epochs for the pretrained model which was later fine-tuned on the training set. The validation accuracy increased from 84\% to 87\% whereas, for the non fine-tuned Seq2Seq-BART model, the validation accuracy reached to 82.22\% from 17\% after the end of 100 epochs of training. The initial high score in validation accuracy while training for the pretrained model accounted for the leverage of the pretrained knowledge gained from the pretraining from the mined corpus.

\textbf{Finding 3.} 
From \autoref{fig:valid-scores-bart-ep100-b8}, we reported that both the BLEU and the ROUGE metric scores for the validation set increased as the number of training epochs increased. From \autoref{fig:valid-scores-bart-finetuned100k-ep20-b8}, the validation scores of the BLEU and the ROUGE metrics kept rising steadily till $40^{th}$ epoch for the fine-tuned model that was pretrained on the mined corpus, then it became almost flat. The validation score of BLEU metric increased from 22.30\% to 31.32\% till 40 epochs and then much slowly to 36.61\% after 100 epochs of training.

\textbf{Finding 4.}
From \autoref{tab:bleu-bart} and \autoref{tab:bleu-bart-finetuned}, we saw that there were three times more the number of accurate translations(Sentence-BLEU score between 0.8 and 0.9) and 50\% gain in the generation of the number of highest quality translations (Sentence-BLEU > 0.9) for the fine-tuned Seq2Seq-BART model over the non fine-tuned without pretraining. The higher the sentence BLEU, the better is the quality of each translation and closer to the ground truth. 

We reported a total of 142 test code predictions in the fine-tuned model for Sentence-BLEU > 0.5, whereas for the non pretrained Seq2Seq-BART model there were only 111 code predictions. We also observed that the number of poorest code translations decreased in the fine-tuned model from 88 to 70. In total, there were 23 exact matches, 66 mostly correct, 101 marginally correct, and 142 semantically equivalent code translations from the fine-tuned Seq2Seq-BART model, which was indeed the highest amongst all the architectures developed in this work.

\textbf{Finding 5.}
There were 326 valid compilable test code snippet translations and 916 valid compilable code snippets considering the top-3 predictions for each test example. After gaining the extra knowledge from the pretraining process and transferring the knowledge to the fine-tuned Seq2Seq model, there was a significant improvement for the overall corpus BLEU metric score and also for the number of accurate code predictions.

\begin{table}[!htbp]
	\centering
	
	\begin{tabular}{|c|c|}
		\hline
		\textbf{Non fine-tuned model(Sentence-BLEU)} & \textbf{\#prediction} \\ \hline
		\textgreater{}0.9                           & 16                     \\ \hline
		\textgreater{}=0.8 and \textless{}=0.9      & 6                     \\ \hline
		\textgreater{}=0.7 and \textless{}=0.8      & 18                     \\ \hline
		\textgreater{}=0.6 an \textless{}=0.7       & 30                    \\ \hline
		\textgreater{}=0.5 and \textless{}=0.6      & 41                    \\ \hline
		\textgreater{}=0.4 and \textless{}=0.5      & 69                    \\ \hline
		\textgreater{}=0.3 and \textless{}=0.4      & 64                    \\ \hline
		\textgreater{}=0.2 and \textless{}=0.3      & 77                   \\ \hline
		\textgreater{}=0.1 and \textless{}=0.2      & 68                    \\ \hline
		\textless 0.1                               & 88                   \\ \hline
		\textgreater{}0.5                           & 111                    \\ \hline
	\end{tabular}
\captionof{table}[Sentence-BLEU metric for non fine-tuned Seq2Seq-BART]{Histogram of test predictions of the non fine-tuned Seq2Seq-BART in the interval range of Sentence-BLEU scores.\label{tab:bleu-bart}}
\end{table}

\begin{table}[!htbp]
	\centering
	
	\begin{tabular}{|c|c|}
		\hline
		\textbf{Fine-tuned model(Sentence-BLEU)} & \textbf{\#prediction} \\ \hline
		\textgreater{}0.9                           & 23                    \\ \hline
		\textgreater{}=0.8 and \textless{}=0.9      & 19                     \\ \hline
		\textgreater{}=0.7 and \textless{}=0.8      & 18                     \\ \hline
		\textgreater{}=0.6 an \textless{}=0.7       & 29                   \\ \hline
		\textgreater{}=0.5 and \textless{}=0.6      & 42                    \\ \hline
		\textgreater{}=0.4 and \textless{}=0.5      & 59                    \\ \hline
		\textgreater{}=0.3 and \textless{}=0.4      & 60                    \\ \hline
		\textgreater{}=0.2 and \textless{}=0.3      & 82                   \\ \hline
		\textgreater{}=0.1 and \textless{}=0.2      & 75                    \\ \hline
		\textless 0.1                               & 70                   \\ \hline
		\textgreater{}0.5                           & 142                    \\ \hline
	\end{tabular}
	\captionof{table}[Sentence-BLEU metric for fine-tuned Seq2Seq-RoBERTa]{Histogram of test predictions of the fine-tuned Seq2Seq-BART after pretraining iin the interval range of Sentence-BLEU scores.\label{tab:bleu-bart-finetuned}}
\end{table}

\textbf{Finding 6.} From  \autoref{tab:metric-transf-roberta-bart}, we found that the Seq2Seq-BART without pretraining resulted into higher test BLEU score by 28.66\% than the fine-tuned Seq2Seq-RoBERTa, which was pretrained on mined30k corpus. It reflects that the proposed Seq2Seq-BART has better text generation capability due to the presence of a GPT decoder over the Seq2Seq-RoBERTa without even considering the pretraining technique. The Seq2Seq-BART pretrained on the mined100k and on the mined30k corpora had surpassed the fine-tuned pretrained Seq2Seq-RoBERTa model by 40.52\% and 36.7\% respectively on the BLEU-4 metric test score.

\begin{table}[!htbp]
	\centering
	\begin{tabular}{|c|c|c|c|c|c|c|}
		\hline
		\textbf{Model}    & \textbf{BLEU}    \\ \hline
		Seq2Seq-BART w/o pretraining & 24.2994 \\ \hline
		Seq2Seq-BART with pretrained BART on mined30k corpus & 25.8153 \\ \hline
		Seq2Seq-BART with pretrained BART on mined100k corpus & \textbf{26.5379} \\ \hline
		Seq2Seq with pretrained RoBERTa & 18.8853   \\ \hline
		Seq2Seq-RoBERTa w/o pretraining & 17.0032    \\ \hline
		Transformer with CoNaLa code tokenizer & 15.3834        \\ \hline
		Transformer with WordPiece tokenizer & 17.3237       \\ \hline
		Transformer with Unigram model tokenizer      & 20.9678      \\ \hline
		Transformer with BPE tokenizer & 19.3402   \\ \hline
		TranX &  25.1050 \\ \hline
	\end{tabular}
	\captionof{table}[Transformer v/s Seq2Seq-RoBERTa v/s Seq2Seq-BART]{Test set metrics comparison: Transformer, Seq2Seq-RoBERTa, and Seq2Seq-BART.\label{tab:metric-transf-roberta-bart}}
\end{table}

\textbf{Finding 7.} From \autoref{tab:pred-finetuned-bart-side1} and \autoref{tab:pred-finetuned-bart-side2}, we could observe for which cases the pretraining of the architecture on the mined corpus helped to make more accurate translations, by enriching the pretrained weights of the original BART model with the pretrained knowledge derived from the mined corpus, while fine-tuning on the training set. Therefore, the fine-tuned model could understand the context of NL intent and surrounding code better than the non-pretrained one to generate more accurate mapping of the source NL intent into the target source code generation.

\textbf{Finding 8.} In \autoref{tab:pred-finetuned-bart-side1}, we listed down some code translations obtained from the non-pretrained SeqSeq-BART model for the test examples.

For the given intent: ``run script `hello.py' with argument `htmlfilename.htm' on terminal using Python executable''. The non-pretrained Seq2Seq-BART predicted correctly `subprocess.call([`hello.py', `htmlfilename.htm'])'.
In the above example, the model was able to understand the filename `hello.py' and the argument `htmlfilename.htm' to the function call.

For the intent: ``replacing nan in the dataframe `df' with row average'', the non-pretrained model predicted `df.fillna(axis=1).mean()', whereas the correct snippet was `df.fillna(df.mean(axis=1), axis=1)'.
We saw that both the `mean()' and the `fillna()' with axis argument function call to be invoked were inferred correctly from the intent by the model but the translation could not be parsed as a valid pandas code.

For the given intent: ``Retrieve each line from a file `File.txt' as a list'', the non-pretrained model predicted `[line for line in open(`File.txt')]', whereas the correct code snippet was `[line.split() for line in open(`File.txt')]'.
In the above case, the model was able to infer the pattern and the syntax except missing to infer the split() function call.

\textbf{Finding 9.} The Seq2Seq-BART model without pretraining worked well for the NL intents like the splitting and concatenating string operation, splitting using regex, copying a file in a directory, converting a tuple to list, converting a list into a dictionary, removing an element from a dictionary, finding an absolute path of a directory, converting a string list into an integer, system time calls, finding the length of data structures like list, tuple, array, string, etc. 

\textbf{Finding 10.} From \autoref{tab:pred-finetuned-bart-side1}, we found that the code generated was of higher quality from the fine-tuned Seq2Seq-BART, which was pretrained on the mined corpus than the Seq2Seq-BART without pretraining. This table also pointed out for which type of intents, the fine-tuned model worked better than the non fine-tuned Seq2Seq-BART. The fine-tuned model worked quite well for intents like complicated regex operations of matching, splitting, searching, complicated dictionary operations such as reversing keys, advanced vector operations, advanced list slicing operations, complicated Pandas dataframe operations, object serialization, etc. 

The Seq2Seq-BART model could extract the names of variables, constants, function from the NL intent and translate into the code snippet correctly. The fine-tuned model could even infer the change in the variable name or the arguments to a data structure/method expressed in the intent and learned to apply the changes in the respective code output.

\textbf{Finding 11.}
The quantitative and the qualitative analysis of the test predictions from both the Seq2Seq-BART and Seq2Seq-RoBERTa models led us to infer that the Seq2Seq-BART had generated more diverse translated code because of auto-regressive left-to-right GPT decoder, which was trained with a causal language modeling (CLM) objective. Therefore, it seemed to be more powerful in predicting the next token in a sequence than just a BERT-based decoder, which had been ineffective for the text generation task.

\textbf{Finding 12.} We finally observed that our proposed fine-tuned Seq2Seq-BART model had surpassed the test BLEU score of the state-of-the-art neural semantic parser, TranX, which is heavily featured engineered. Additionally, the TranX had produced only 206 valid parsable code snippets via AST parsing from the test set, whereas the proposed Seq2Seq-BART generated 326 valid parsable code snippets.

\begin{sidewaystable}[!htbp]
	\captionof{table}[Test predictions 1: Fine-tuned Seq2Seq-BART v/s non-pretrained BART]{Comparison of test predictions: fine-tuned Seq2Seq-BART v/s Seq2Seq-BART.	\label{tab:pred-finetuned-bart-side1}}
	
	\footnotesize
		\resizebox{\columnwidth}{!}{\begin{tabularx}{\linewidth}{|*{4}{>{\RaggedRight\arraybackslash}X|}}
		\hline
		\textbf{Intent}                                                                                        & \textbf{BART prediction}                                                                                                                                   & \textbf{Fine-tuned BART prediction}                                                                                                                                                                                                   & \textbf{Snippet}                                                                     \\ \hline
		Download a file `url' over HTTP.                                                                      & {[}`wget.download(url)', `urllib.request.url'," wget.download(`url')"{]}                                                                               & {[} `r =   requests.get(url)'," r = requests.get(url, headers=\{`Content-Type':   `application",                                                                  & r =   requests.get(url)                                                              \\ \hline
		Enable debug mode on Flask application `app'. & {[}`app.debug()', `app.disabledebug()', `app.debug\_mode()'{]}                                                                                        & {[}`app.run(debug=True)', `app.run(debug=False)', `app.debug = True'{]}                                                                                                                                                          & app.run(debug=True)                                                                  \\ \hline
		Split string `QH QD JC KD JS' into a list on white spaces. & {[}"{[}x for x in re.split(`QH QD JC KD JS', " {[}x for x in re.findall(`QH QD JC   KD JS"," {[}x.strip(`QH QD JC KD JS') for x in re"{]} & {[}`"""QH QD JC KD JS""".split()', "re.split('\textbackslash{}\textbackslash{}\textbackslash{}\textbackslash{}W+', `QH QD JC KD JS')", ' """QH QD JC KD   JS""".split(\textbackslash{}',\textbackslash{}')'{]}                   & """QH QD JC KD JS""".split()                                                       \\ \hline
		Get reverse of list items from list `b' using extended slicing.                                       & {[}`{[}x for x in a   if x not in b{]}', `{[}x for x in b if x not in b{]}', `{[}x{[}1{]} for x in b{]}'{]}                                             & {[}`{[}x{[}::-1{]} for x   in a{]}', `{[}x{[}::(-1){]} for x in a{]}', `{[}x{[}::-1{]} for x in b{]}'{]}                                                                                                                           & {[}x{[}::-1{]} for x in   b{]}                                                       \\ \hline
		Cast an int `i' to a string and concat to string `ME'.                                                & {[}`""ME""".join(i.split())', `""ME""".join(i)',"`ME' = str(i)"{]}                                                                                  & {[}"`ME' +   str(i)"," `ME'.join(str(i) for i in i)"," `ME'.join('\%s   \%s' \% (i,))"{]}                                                                                                                                            & `ME' + str(i)                                                                        \\ \hline
		Removing duplicates in list `t'.                                                                      & {[}` {[}x for x in t   if x != 0{]}', `{[}x for x in list(t){]}', `{[}x for x in t if x not in x{]}'{]}                                                  & {[}`list(set(t))',   `new\_list = list(set(t))', `t = list(set(t))'{]}                                                                                                                                                             & list(set(t))                                                                         \\ \hline
		Removing duplicates in list `source\_list'.                                                           & {[}`{[}x for x in   source\_list{]}', `new\_list = {[}x for x in source\_list{]}', `closed(source\_list)'{]}                                          & {[}`new\_list =   list(source\_list)', `final\_list = list(source\_list)', `map(list,   set(source\_list))'{]}                                                                                                                     & list(set(source\_list))                                                              \\ \hline
		Get attribute `attr' from object `obj'.                                                               & {[}`obj.attr()', `obj.attr()'," obj.attribute(`attr')"{]}                                                                                               & {[}"   getattr(obj, `attr')","getattr(obj, `attr')", `getattr(obj,   attr)'{]}                                                                                                                                                      & getattr(obj,   `attr')                                                               \\ \hline
		Encode `Decimal(`3.9')' to a JSON string. & {[}"json.decimal(`3.9')", "json.decimal(`3.9'){[}0{]}", "json.encode('Decimal(`3.9')')"{]}"                                                 & {[}"   json.dumps(Decimal(`3.9'))"," json.dumps(Decimal(`3.9'),   ensure\_ascii","json.dumps(decimal(`3.9'))"{]}                                                                                                                     & json.dumps(Decimal(`3.9'))                                                           \\ \hline
		Check if object `obj` is a string. & {[}`isinstance(obj,   string)', `isinstance(obj, str)', `isinstance(obj, bool)'{]}                                                                      & {[}`isinstance(obj,   str)', `if isinstance(obj, str):\textbackslash{}n pass', `if hasattr(obj, string):\textbackslash{}n   pass'{]}                                                                                               & isinstance(obj,   str)                                                               \\ \hline
		Finding the index of elements containing substring `how' and `what' in a list of strings   `myList'. & {[}' myList{[}i:i + j   for i, j in enumerate(my'," myList.index({[}`how', `what'{]})", ` myList{[}i:i + j for i, j in zip(myList'{]}                    & {[}" {[}item for item in myList if 'how' in item{]}"," {[}x for x in `how' if `what' in x{]}"," {[}item for item in `how' if `what' in item{]}"{]}                                                                                & {[}i for i, j in   enumerate(myList) if `how' in j.lower() or `what' in j.lower(){]} \\ \hline
		Reverse the keys and values in a dictionary `myDictionary'. & {[}`dict((k, v) for   k, v in myDictionary.items())', `dict((k, v) for k, v in   myDictionary.items()', `dict((k, v) for k, v in list(myDictionary.'{]} & {[}`dict((k, v) for k, v in myDict.items()', `dict((v, k) for k, v in myDict.items()', `dict((v, k) for k, v in myDict.items())'{]}                                                                                            & \{i{[}1{]}: i{[}0{]} for i in list(myDictionary.items())\}                         \\ \hline

	\end{tabularx}}
\end{sidewaystable}

\begin{sidewaystable}[!htbp]
	\captionof{table}[Test predictions 2: Fine-tuned Seq2Seq-BART v/s non-pretrained BART]{Comparison of test predictions: fine-tuned Seq2Seq-BART v/s non fine-tuned Seq2Seq-BART.	\label{tab:pred-finetuned-bart-side2}}
	
	\footnotesize
	\resizebox{\columnwidth}{!}{\begin{tabularx}{\linewidth}{|*{4}{>{\RaggedRight\arraybackslash}X|}}
			\hline
			\textbf{Intent}                                                                                        & \textbf{BART prediction}                                                                                                                                   & \textbf{Fine-tuned BART prediction}                                                                                                                                                                                                   & \textbf{Snippet}                                                                     \\ \hline
			Reload a module   `module'. & {[}`module.reload()', `module.reshape()', `module.load()'{]}                                                                                         & {[}`import   imp\textbackslash{}nimp.reload(module)', `import imp\textbackslash{}nimport imp\textbackslash{}nimp.reload(module)',  & import imp   imp.reload(module)                                                      \\ \hline
			Calculate mean across dimension in a 2d array `a'.                                                    & {[}`np.mean(np.arange(a), axis=0)', `np.mean(np.arange(a, axis=0)', `np.mean(np.arange(a), axis=1)'{]}                                              & {[}`np.mean(a, axis=0)', `np.mean(a, axis=1)', `np.mean(a, axis=(0,1))'{]}                                                                                                                                                      & np.mean(a, axis=1)                                                                   \\ \hline
			Get the value associated with unicode key `from\_user' of first dictionary in list `result'. & {[}"result{[}`from\_user'{]} = True", "return result{[}`from\_user'{]}", "print(result{[}`from\_user'{]} for item in result))"{]} & {[}"result{[} `from\_user'{]}","{[}x{[}`from\_user'{]} for x in result{]}", " {[}x{[}`from\_user'{]} for x in result{]}"{]} & result{[}0{]}{[}`from\_user'{]}                                                      \\ \hline
			Open a file `path/to/
			FILE\_NAME.ext' in write mode. & {[}"shutil.write(`path/to/
			FILE\_NAME.ext')", "shutil.open(`path/to/
			FILE\_NAME.ext', `write", "os.path.write(`path/to/
			FILE\_NAME.ext')"{]} & {[}"f = open(`path/to/
			FILE\_NAME.ext', `w')", "shutil.rmtree(`path/to/
			FILE\_NAME.ext')", "file = open(`path/to/FILE\_NAME.ext', `w')"{]} & new\_file = open(`path/to/FILE\_NAME.ext', `w') \\ \hline
			Add key "mynewkey" to dictionary `d' with value "mynewvalue".                                         & {[}"d{[}`mynewkey'{]} = `mynewvalue'"," d = \{`mynewkey':   `mynewvalue'\}","d{[}`mynewkey'{]} = d.get(`mynewvalue')"{]} & {[}"d =   \{`mynewkey': `mynewvalue'\}", "d.update(\{
			`mynewkey':`mynewvalue'\})", "d.setdefault(key, {[}{]}).append(`mynewvalue')"{]}                                                                                             & d{[}`mynewkey'{]} = `mynewvalue'                                                   \\ \hline
			Enable warnings using action `always'. & {[}"driver.find\_element\_by\_id(
			`always')", "drivers.find\_element\_by\_id(
			`always')", "os.system(`always')"{]} & {[}"warnings.warn(`always')", "   warnings.filterwarnings(`always')", "warnings.warn(`always',   False)"{]} & warnings.simplefilter(`always')                                                      \\ \hline
			Multiply a matrix `P' with a 3d tensor `T' in scipy. & {[}`np.multiply(P,   T).reshape(axis=0).', `np.multiply(P, T).reshape(axis=1).', `P =   np.multiply(P, T).reshape(axis='{]}                             & {[}`scipy.constant({[}{[}1, 2{]}, {[}3, 4{]}{]}', `scipy.constant({[}{[}1, 2, 3{]}, {[}4,', `A{[}(np.arange(M.shape{[}0{]}) != 0),'{]}                                                                                         & scipy.tensordot(P,   T, axes={[}1, 1{]}).swapaxes(0, 1)                              \\ \hline
			Print a list `l' and move first 3 elements to the end of the list. & {[}`   print(sorted(list(items=1))', `print(itertools.product(*1))', `print(sorted(list(range(1)))'{]}                                                 & {[}"   print(`\textbackslash{}\textbackslash{}n'.join(l))"," print(`\textbackslash{}\textbackslash{}n'.join(str(p) for p in   l))"," print(`\textbackslash{}\textbackslash{}n'.join(str(i) for i in l))"{]}                           & print(l{[}3:{]} +   l{[}:3{]})                                                       \\ \hline
			Get elements from list `myList', that have a field `n' value 30. & {[}` myList{[}i:i + j for i in range(30){]}', `myList = {[}x for x in range(30){]}', `myList{[}i{]}:i + j   for i in range(30){]}'{]} & {[}`{[}x for x in myList if x.field == 30{]}', "{[}item for item in myList if `n' in item{]}", `{[}x for x in myList if x == 30{]}'{]} & {[}x for x in myList if x.n == 30{]}                                               \\ \hline

	\end{tabularx}}
\end{sidewaystable}

\subsection{RQ2. What did we find from the ablation studies of the developed architectures?}\label{RQ2}
\subsubsection{RQ2a. Ablation Study: What would happen if we varied the number of self-attention heads in Transformer for training?}
\textbf{Motivation.}
Ablation is the removal of a component of an AI system. An ablation study reports the performance of an AI system by removing certain components, to understand the contribution of each component to the overall system. The motivation is to see what components of the transformer architecture are important for the overall test metrics. Here, we tried to analyze how important is the number of self-attention heads of the transformer for our target objective. 

\textbf{Approach.}
We set the depth of the transformer model to 3. We varied the number of self-attention heads in the range of [2,32], whereas the transformer had 12 self-attention heads by default.  We used the BLEU code tokenizer as provided in the CoNaLa challenge to tokenize the source code. We trained the architecture with a batch size of 64 for a maximum of 50 epochs with an early stopping mechanism. The hidden dimension of each block of the transformer and the dimension of the model were set to 512, 256, respectively. The default learning rate of 1e-03 and Adam optimizer were used for model weight updates as mentioned in \citep{vaswani2017attention}.

\textbf{Results.}
\autoref{tab:ablation-transf-heads} displays the results of the ablation study of the transformer regarding the number of self-attention heads in the model architecture.

\begin{table}[!htbp]
	\centering
	\centering
	\Large
	\begin{adjustbox}{width=\textwidth}
		\begin{tabular}{|c|c|c|c|c|c|c|}
			\hline
			\textbf{Attention heads} & \textbf{Validation BLEU} & \textbf{Test BLEU} & \textbf{Validation RougeL F1} & \textbf{Test RougeL F1} & \textbf{Validation token accuracy} & \textbf{Test token accuracy} \\ \hline
			2                                  & 20.6100             & 16.7480           & 0.5730                   & 0.5497                 & 16.5211                       & 13.5680                      \\ 
			4                                  & 21.6802            & 17.3580            & 0.5871                  & 0.5598                & \textbf{18.4361}                      & 14.3770                       \\ 
			8                                  & \textbf{23.5436}           & 17.6857           & \textbf{0.6074}                 & 0.5707                 & 17.2164                       & 14.1223                      \\ 
			16                                 & 22.8340             & 17.2390            & 0.5957                  & 0.5594                 & 16.8025                       & 13.4464                     \\ 
			32                                 & 22.3587           & \textbf{17.7187}           & 0.5977                  & \textbf{0.5725}                 & 18.3544                       & \textbf{15.0941}                     \\ \hline
		\end{tabular}
	\end{adjustbox}
	\captionof{table}[Ablation study: Self-attention heads in Transformer]{Ablation study of the self-attention heads of Transformer while keeping the number of layers fixed at 3.\label{tab:ablation-transf-heads}}
\end{table}


\textbf{Finding 1.} From \autoref{tab:ablation-transf-heads}, we observed that as the number of heads was increased, the validation BLEU score increased by 14.25\% and the test BLEU score increased by 5.5\% until it plateaued at eight self-attention heads. From 16 to 32 self-attention heads, the validation BLEU score dropped further by 2.13\%, whereas the test metric increased by 2.7\%.

\textbf{Finding 2.} \autoref{tab:ablation-transf-heads} showed that the validation token accuracy increased as the number of self-attention heads was increased to 8, then the token accuracy decreased up to 16 self-attention heads. Furthermore, the test token accuracy increased and was reported the highest with 32 self-attention heads. Therefore, in general, the test token accuracy score was increased as the number of self-attention heads was increased.

\textbf{Finding 3.} From \autoref{tab:ablation-transf-heads}, we observed that the validation and the test RougeL F1-score metrics increased by 6\% and 3.82\% respectively as the self-attention heads increased to 8 from 2,  then kept decreasing by 1.9\% and 1.98\% respectively till 16 self-attention heads and then increased by 2.35\%, and reporting to be the highest test RougeL F1-score with 32 self-attention heads.

\textbf{Finding 4.} Multi-head attention seemed to be useful to attend over the different aspects of the source context. The multiple heads gave the network more freedom to learn better features. Thus, the results improved as we increased the number of self-attention units but start getting saturated at 8 heads because the complexity of the model became larger compared to the availability of enough training data. 

\subsubsection{RQ2b. Ablation Study: What would happen if we varied the depth of the Transformer architecture for training?}
\textbf{Motivation.}
As discussed previously, our motivation is to see what components of the Transformer architecture are important for the overall test metrics. Here, we tried to analyze how relevant is the depth, i.e.,  the number of layers in the encoder and  the decoder of Transformer architecture for our target objective.

\textbf{Approach.}
We set the number of self-attention heads in the Transformer model to eight. We varied the number of layers from one to six instead of the default depth of 6 for both the encoder and the decoder of the base Transformer model. We considered that both the encoder and the decoder had same number of layers for each experiment performed for the ablation study. We used the BLEU code tokenizer as provided in the CoNaLa challenge to tokenize the source code. 

We varied the depth of the Transformer and trained the architecture with a batch size of 64 for a maximum of 50 epochs with an early stopping mechanism. The hidden dimension of each block of the transformer and the dimension of the model were set to 512, 256 respectively. The default learning rate of 1e-03 and Adam optimizer were used as in \citep{vaswani2017attention} for training and updating the model weights for learning.

\textbf{Results.}
\autoref{tab:ablation-transf-layers} displays the results of the ablation study of the transformer about the depth of the transformer model.

\textbf{Finding 1.} From \autoref{tab:ablation-transf-layers}, we observed that the performance of the transformer increased when the number of layers was increased to three and thereafter, we saw a sudden drop in both token accuracy and test BLEU scores. We also observed a decrease in the validation BLEU metric when the number of layers was higher than five. 

\textbf{Finding 2.} From \autoref{tab:ablation-transf-layers}, we observed that both the validation and test scores of the BLEU metric increased when the number of layers was increased from one to three, by 16.73\% and 6.85\% respectively, then decreased after the third layer. There was a significant drop of validation BLEU score by 13.46\% from the fifth layer to the sixth layer.

\textbf{Finding 3.} From \autoref{tab:ablation-transf-layers}, we found that both the validation and the test F1 scores of RougeL metric increased when the number of layers got increased from one to three, by 5.3\% and 5.85\% respectively, and then decreased after the third layer. Thus, the three layers of transformer architecture were sufficient to exceed the performance of an LSTM/GRU-based encoder-decoder for this task. 

\begin{table}[!htbp]
	\centering
	\Large
	\begin{adjustbox}{width=\textwidth}
		\begin{tabular}{|c|c|c|c|c|c|c|}
			\hline
			\multicolumn{1}{|c|}{\textbf{Number of layers}} & \textbf{Validation BLEU} & \textbf{Test BLEU} & \textbf{Validation RougeL F1} & \textbf{Test RougeL F1} & \textbf{Validation token accuracy} & \textbf{Test token accuracy} \\ \hline
			1                                                                          & 18.7609            & 15.2000             & 0.5673                  & 0.5400                  & 15.7220                       & 14.3319                      \\ 
			2                                                                          & 21.4045            & 16.1740           & 0.5779                  & 0.5469                 & 18.0871                      & 15.5930                      \\
			3                                                                          & 21.9000    & \textbf{16.2420}  & \textbf{0.5976}         & \textbf{0.5716}       & \textbf{20.1941}             & \textbf{16.6751}            \\ 
			4                                                                          & 21.1420             & 15.0200            & 0.5892                 & 0.5539                & 18.9563                      & 15.5323                      \\ 
			5                                                                          & \textbf{22.1571}   & 15.0414           & 0.5899                  & 0.5493               & 19.7900                       & 16.4548                      \\ 
			6                                                                          & 19.1737           & 16.0100           & 0.5743                  & 0.5642                & 15.6427                       & 14.2967                      \\ \hline
		\end{tabular}
	\end{adjustbox}
\captionof{table}[Ablation study: Depth of Transformer]{Ablation study on the number of layers of Transformer while keeping the attention heads fixed at 8.\label{tab:ablation-transf-layers}}
\end{table}

\subsubsection{RQ2c. Ablation Study: What would happen if we varied the number of self-attention heads in Seq2Seq-RoBERTa?}

\textbf{Motivation.}
Our motivation is to see what components of the Seq2Seq-RoBERTa architecture are relevant for the overall test metrics. Here, we tried to analyze how crucial the number of self-attention heads for the Seq2Seq-RoBERTa is. 

\textbf{Approach.}
We used the lighter RoBERTa model, i.e, DistilRoBERTa for our experiments. We fixed the depth of the DistilRoBERTa model at 1. We varied the number of self-attention heads in the range of [2, 12], whereas there were 12 self-attention heads in the DistilRoBERTa. We trained the architecture with a batch size of 8 for a maximum of 50 epochs with an early stopping mechanism. The hidden dimension of each block of the RoBERTa was set to 768. The learning rate of 1e-03 with adam optimizer and default hyperparameters were used for training as mentioned in \citep{devlin-etal-2019-bert}. 

\textbf{Results.}
After an ablation study on the number of self-attention heads in Seq2Seq-RoBERTa, we found out that by increasing the number of self-attention heads, there was an increase in the test metrics scores.

\textbf{Finding 1.} From  \autoref{tab:roberta-ablation-heads}, the test BLEU metric increased by 9.84\% when the number of self-attention heads was increased to 12 from 2, considering beam search decoding with a beam size of 15.

\textbf{Finding 2.} 
From \autoref{tab:roberta-ablation-heads}, the test Rouge1 F1 metric score increased by 5.8\% when the number of self-attention heads was raised to 8 from 6, and by 2.8\% from 8 to 12 attention heads, considering greedy decoding of the output.
\begin{table}[!htbp]
	\centering
	\Large
	\resizebox{\linewidth}{!}{
		\begin{tabular}{|c|c|c|c|c|c|}
			\hline
			\multicolumn{1}{|c|}{\textbf{Attention heads}} & \textbf{Decoding method} & \textbf{Test BLEU} & \textbf{Test Rouge1 Precision} & \textbf{Test Rouge1 Recall} & \textbf{Test Rouge1 F1-score} \\ \hline
			\multirow{5}{*}{2}                                                                       & Greedy          & 12.8621  & 0.3469               & 0.3259            & 0.3139        \\ 
			& Beam size = 4   & 12.8621  & 0.3469               & 0.3259           & 0.3139        \\ 
			& Beam size = 7   & 12.6930  & 0.3371              & 0.3221           & 0.3066        \\ 
			& Beam size = 10  & 12.8602  & 0.3375              & 0.3227           & 0.3072      \\ 
			& Beam size = 15  & 12.7773   & 0.3380              & 0.3243           & 0.3076        \\ \hline
			\multirow{5}{*}{4}                                                                       & Greedy          & 13.2852  & 0.3357               & 0.3054           & 0.2988        \\ 
			& Beam size = 4   & 13.9182   & 0.3273              & 0.3144           & 0.3000        \\ 
			& Beam size = 7   & 14.0908  & 0.3291              & 0.3216         & 0.3050        \\ 
			& Beam size = 10  & 13.9419   & 0.3279               & 0.3211           & 0.3045       \\ 
			& Beam size = 15  & 13.8604   & 0.3275              & 0.3202           & 0.3039       \\ \hline
			\multirow{5}{*}{6}                                                                       & Greedy          & 13.3928  & 0.3197              & 0.3110           & 0.2938        \\ 
			& Beam size = 4   & 13.3928 & 0.3197              & 0.3110           & 0.2938        \\ 
			& Beam size = 7   & 13.3976  & 0.3085              & 0.3106           & 0.2871       \\ 
			& Beam size = 10  & 13.5370  & 0.3110             & 0.3128           & 0.2897       \\ 
			& Beam size = 15  & 13.4991 & 0.3088               & 0.3111           & 0.2878        \\ \hline
			\multirow{5}{*}{8}                                                                       & Greedy          & 13.4887  & \textbf{0.3544}             & 0.3075           & 0.3109       \\ 
			& Beam size = 4   & 13.6894  & 0.3366             & 0.3103           & 0.3048       \\ 
			& Beam size = 7   & 13.8962 & 0.3365             & 0.3125           & 0.3050       \\ 
			& Beam size = 10  & 13.8202  & 0.3365              & 0.3114          & 0.3050        \\ 
			& Beam size = 15  & 13.8844  & 0.3364              & 0.3143           & 0.3062       \\ \hline
			\multirow{5}{*}{12}                                                                      & Greedy          & 13.3493  & 0.3474              & 0.3179          & 0.3117       \\ 
			& Beam size = 4   & 13.8000    & 0.3399              & 0.3261           & 0.3126      \\ 
			& Beam size = 7   & 13.9197  & 0.3340               & 0.3253           & 0.3095      \\ 
			& Beam size = 10  & \textbf{14.0733}  & 0.3337              & 0.3260           & 0.3096       \\ 
			& Beam size = 15  & 14.0307 & 0.3350              & \textbf{0.3265}           & \textbf{0.3150}       \\ \hline
	\end{tabular}}
	\captionof{table}[Ablation study: Number of self-attention heads for Seq2Seq-RoBERTa]{Ablation study on the number of self-attention heads for DistilRoBERTa while keeping layers fixed at 1. (batch\_size=8, 50 epochs)\label{tab:roberta-ablation-heads}}
\end{table}

\subsubsection{RQ2d. Ablation Study: What would happen if we varied the depth of the hybrid Seq2Seq-RoBERTa  architecture?}
\textbf{Motivation.}
Our motivation is to see which components of the Seq2Seq-RoBERTa architecture are crucial for the overall test metrics. Here, we tried to analyze how relevant is the depth, i.e., the number of layers of the Seq2Seq-RoBERTa for our target objective. 

\textbf{Approach.}
As discussed before, we used DistilRoBERTa model for our experiments. We fixed the number of self-attention heads in the DistilRoBERTa model to 8. We varied the number of layers from 1 to 5, whereas the highest depth in the DistilRoBERTa was 6. We considered that both the encoder and the decoder had same number of layers for each experiment performed for the ablation study. We trained the architecture with a batch size of 8 and for a maximum of 50 epochs with an early stopping mechanism. The learning rate of 1e-03 with adam optimizer, default hyperparameters, and hidden dimension of each block of 768 were used for training as mentioned in \citep{devlin-etal-2019-bert}. 

\textbf{Results.}
Here, we discussed the results of the ablation study about the number of layers in the model where we kept increasing the number of layers in the model while keeping the number of self-attention heads fixed. We found out that with the increase in the number of layers in the encoder and the decoder, there was an improvement in the overall test scores of BLEU, and ROUGE metrics respectively. \autoref{tab:roberta-ablation-layers} displays the ablation study results for the depth of the model architecture. 

\textbf{Finding 1.} From  \autoref{tab:roberta-ablation-layers}, we found that the test BLEU metric increased by 6.6\% when the number of layers was increased to 5 from 3, considering greedy decoding of the output. 

\textbf{Finding 2.} 
From \autoref{tab:roberta-ablation-layers}, we observed that the test Rouge1-F1 metric score increased by 7.87\% when the depth of the architecture was raised to 5 from 3, considering beam search decoding of the output.

\begin{table}[!htbp]
	\centering
	\Large
	\resizebox{\linewidth}{!}{
		\begin{tabular}{|c|c|c|c|c|c|}
			\hline
			\multicolumn{1}{|l|}{\textbf{Number of layers}} & \textbf{Decoding Method} & \textbf{Test BLEU} & \textbf{Rouge1 F1} & \textbf{Rouge2 F1} & \textbf{RougeL F1} \\ \hline
			\multirow{4}{*}{3}                              & Greedy                   & 14.2945          & 0.3238           & 0.1148           & 0.3106           \\ 
			& Beam size= 4             & 15.1319          & 0.3258            & 0.1240            & 0.3127          \\ 
			& Beam size = 10           & 15.3172          & 0.3268           & 0.1250           & 0.3139          \\ 
			& Beam Size = 15           & 15.2510           & 0.3254            & 0.1246           & 0.3125           \\ \hline
			\multirow{5}{*}{5}                              & Greedy                   & 15.2415           & 0.3506           & 0.1248          & \textbf{0.3349}  \\ 
			& Beam size= 4             & 15.9160           & 0.3520           & 0.1267           & 0.3334           \\ 
			& Beam size= 7             & 15.9965            & \textbf{0.3525}            & \textbf{0.1276}  & 0.3340           \\ 
			& Beam Size = 10           & 16.0659           & \textbf{0.3525}  & 0.1256            & 0.3337          \\ 
			& Beam Size = 15           & \textbf{16.0700}  & 0.3508            & 0.1274           & 0.3330            \\ \hline
	\end{tabular}}
	\captionof{table}[Ablation study: Depth of the Seq2Seq-RoBERTa]{Ablation study on the depth of RoBERTa layers in the hybrid Se2Seq while keeping the attention heads to be 8. (batch\_size=8, 50 epochs)\label{tab:roberta-ablation-layers}}
\end{table}

\subsection{RQ3. Do pretraining knowledge and transfer learning help to improve the results? Are the improvements to the fine-tuned models statistically significant?}
\textbf{Motivation.}
Transfer learning is a means of extracting knowledge from a source setting and applying it to a different target setting. A pretrained model is a saved network that was previously trained on a large unannotated dataset, typically on large-scale (NL, code) pairs for our task. This pretraining technique saves the weights trained from the large corpus. These weights can be reused to fine-tune the pretrained model on the respective downstream, relatively, smaller training annotated dataset. The intuition behind transfer learning for NL2Code translation is that a model, trained on a large and general enough dataset, will effectively serve as a generic model of the (NL, code) pairs. We can then take advantage of these learned feature maps without having to start from scratch by training a large model on a large dataset.

\textbf{Approach.}
Both BART and RoBERTa model architectures were pretrained on large unlabeled English corpora. The proposed hybrid Seq2Seq-RoBERTa and Seq2Seq-BART architectures were pretrained on the large CoNaLa mined100k corpus where the pretraining objective was to learn the mappings of natural language intent to their respective code snippets. The pretrained weights were later fine-tuned by training on the smaller labeled train set for the Code2NL objective. 

\textbf{Results.}
As we observed in both Seq2Seq-RoBERTa and Seq2Seq-BART results, the fine-tuning of the pretrained model performed better than the non fine-tuned models. The pretraining process enriched these architectures to gain knowledge from the mappings of natural language intent to code snippet from the mined corpus. The latter technique of fine-tuning these pretrained hybrid Seq2Seq architectures on the training set improved the test BLEU score metric by 9.2\% over non-pretrained Seq2Seq-BART architecture and by 11.2\% over non-pretrained Seq2Seq-RoBERTa.

\subsubsection{Are the improvements to the fine-tuned models statistically significant?}
\textbf{Motivation.}
Are the experimental results coincidental?  To answer the question, we require to use a standard statistical tool such as statistical significance testing \citep{dror-etal-2018-hitchhikers} thereby, ensuring that experimental results are not coincidental. The statistical significance testing in NLP research helps us take an inferred decision whether we should rely on empirical evaluation to validate our hypotheses and, whether the obtained results and improvements from the experiments are just occurences by random choice or coincidental.

The way to verify hypotheses from experiments is by applying pairwise algorithms to the same datasets such as, the same test set, to compare the performance of two algorithms/models in consideration. This is based on the reasoning that if one algorithm is consistently better than the other with a sufficiently large marginal difference on the same test set, then it should also be better on future unknown datasets. Statistical significance testing makes sure that the probability of falsely concluding that one algorithm is better than the other is very small. Thus, the significance testing ensures that the difference between the two algorithms, as observed in an individual comparison, is not coincidental.

\textbf{Approach}
The distribution of the test statistic was unknown. Hence, we preferred non-parametric significance testing such as sampling-based testing, i.e., paired bootstrap test. The test took the values of the evaluation measures under consideration to estimate the $p$-value based on the test statistic values in the samples t-tests by repeatedly sampling from the test data.  

However, the unique structured nature of natural language data is reflected in specialized evaluation measures such as BLEU and ROUGE. The BLEU measure is used for the pairwise bootstrap test to figure out the distribution of the BLEU measure from the algorithms in comparison. 

The distribution of the evaluation metric measures is of great importance to statistical significance testing. Statistical significance tests quantify the likelihood of the samples of evaluation metrics such as the observable BLEU score, given the assumption of being drawn from the same distribution. If this assumption, or null hypothesis, is rejected, it suggests that the difference in evaluation metric score is statistically significant.

We performed the statistically significant pairwise bootstrap test to compare different systems in pairs to judge whether their difference in magnitude in the test BLEU scores are statistically significant or not.

\textbf{Results.}
We used the BLEU metric as an evaluation metric for comparison between the two models. We used 10000 samples for the sampling-based pairwise bootstrap test.

\textbf{System 1}: Seq2Seq-RoBERTa model trained for 50 epochs with batch size of 8 and beam size set to 15. \\
\textbf{System 2}: Fine-tuned Seq2Seq-RoBERTa model pretrained on mined30k corpus.

\begin{table}[!htbp]
	\centering
	\centering
	\begin{tabular}{|c|c|c|}
		\hline
	\textbf{Statistic}	& \textbf{System 1}           & \textbf{System 2}           \\ \hline
		\textbf{Win ratio}                 & 0.046              & 0.954              \\ 
		\textbf{95 \% Confidence Interval} & {[}0.000, 0.011{]} & {[}0.000, 0.014{]} \\ 
		\textbf{Mean}                      & 0.007              & 0.010              \\ 
		\textbf{Median}                    & 0.008             & 0.011              \\ \hline
	\end{tabular}
	\captionof{table}[Significance Testing of non-pretrained Seq2Seq-RoBERTa v/s fine-tuned Seq2Seq-RoBERTa]{Signifiance Test comparison: non-pretrained Seq2Seq-RoBERTa v/s fine-tuned Seq2Seq-RoBERTa with pretraining. \label{tab:sig-roberta}}
\end{table}
From \autoref{tab:sig-roberta}, we saw that the system 2 was superior with $p$-value= 0.046 ($p$-value < 0.05). This meant that the results obtained from the fine-tuned Seq2Seq-RoBERTa model with pretraining were statistically significant than the non-pretrained Seq2Seq-RoBERTa model.

\textbf{System 1}: Seq2Seq-BART model trained for 100 epochs with batch size of 8. \\
\textbf{System 2}: Fine-tuned Seq2Seq-BART model pretrained on mined30k corpus.

\begin{table}[!htbp]
	\centering
	\centering
	\begin{tabular}{|c|c|c|}
		\hline
		\textbf{Statistic} & \textbf{System 1}           & \textbf{System 2}           \\ \hline
		\textbf{Win ratio}                 & 0.038              & 0.962              \\ 
		\textbf{95 \% Confidence Interval} & {[}0.018, 0.031{]} & {[}0.021, 0.036{]} \\ 
		\textbf{Mean}                      & 0.024              & 0.029              \\ 
		\textbf{Median}                    & 0.025             & 0.029             \\ \hline
	\end{tabular}
\captionof{table}[Significance Testing of Seq2Seq-BART v/s fine-tuned  Seq2Seq-BART]{Significance Test comparison: Seq2Seq-BART (non-pretrained) v/s fine-tuned Seq2Seq-BART. \label{tab:sig-bart}}
\end{table}

From \autoref{tab:sig-bart}, we found that the system 2 was superior with $p$-value = 0.038 ($p$-value < 0.05). This meant that the results obtained from the fine-tuned Seq2Seq-BART with pretraining were statistically significant than the non-pretrained Seq2Seq-BART model.

\textbf{System 1}: Seq2Seq-RoBERTa model trained for 50 epochs with batch size of 8 and beam size set to 15. \\
\textbf{System 2:} Seq2Seq-BART model trained for 100 epochs with batch size of 8.

\begin{table}[!htbp]
	\centering
	\centering
	\begin{tabular}{|c|c|c|}
		\hline
	\textbf{Statistic}	& \textbf{System 1}           & \textbf{System 2}           \\ \hline
		\textbf{Win ratio}                 & 0                  & 1.0                \\ 
		\textbf{95 \% Confidence Interval} & {[}0.000, 0.011{]} & {[}0.018, 0.031{]} \\ 
		\textbf{Mean}                      & 0.007              & 0.025              \\ 
		\textbf{Median}                    & 0.008              & 0.025              \\ \hline
	\end{tabular}
	\captionof{table}[Significance Testing of Seq2Seq-RoBERTa v/s Seq2Seq-BART]{Significance Test comparison: Seq2Seq-RoBERTa v/s Seq2Seq-BART. \label{tab:sig-roberta-bart}}
\end{table}

From \autoref{tab:sig-roberta-bart}, we observed that the system 2 was superior with $p$-value < 0.05. This meant that the results obtained from the Seq2Seq-BART model were statistically significant than the non-pretrained Seq2Seq-RoBERTa model.

\textbf{System 1}: Vanilla Seq2Seq model with batch size of 8 and dropout probability of 0.3. \\
\textbf{System 2:}: Transformer model (3 layers, 8 attention heads) model with dropout probability set to 0.25.

\begin{table}[!htbp]
	\centering
	\centering
	\begin{tabular}{|c|c|c|}
		\hline
	\textbf{Statistic}	& \textbf{System 1}           & \textbf{System 2}           \\ \hline
		\textbf{Win ratio}                 & 0.000              & 1.000              \\ 
		\textbf{95 \% Confidence Interval} & {[}0.000, 0.000{]} & {[}0.146, 0.181{]} \\ 
		\textbf{Mean}                      & 0.000              & 0.163              \\ 
		\textbf{Median}                    & 0.000              & 0.163              \\ \hline
	\end{tabular}
\captionof{table}[Significance Testing of Seq2Seq v/s Transformer]{Significance Test comparison: vanilla Seq2Seq v/s Transformer. \label{tab:sig-transf-seq}}
\end{table}

From \autoref{tab:sig-transf-seq}, we found that the system 2 was superior with $p$-value < 0.05. This meant that the results obtained from the vanilla transformer with the CoNaLa code tokenizer model were statistically significant than the vanilla Seq2Seq model.

\subsection{RQ4. Does the SentencePiece tokenizer improve the performance of the Transformer model?}
\textbf{Motivation.}
Traditional language models (LMs) in NLP involve RNNs that predict a single token at a time. Hence, they operate at the token level.  This strategy often leads to large vocabulary sizes for source code, because identifiers in programming languages often correspond to entire phrases in natural language. As the number of unique identifiers increases with the size of the corpus, this makes it difficult to train code LMs on large corpora. 

Thus, we used subword tokenizers by letting the model predict subword units rather than full tokens at each time step of the RNN. A subword unit is a n-gram of characters that appear as a subsequence of some token in the corpus. In this experiment, we used SentencePiece, which is a language-independent subword tokenizer and detokenizer designed for Neural-based text processing.

\textbf{Approach.}
Instead of using the standard code tokenizer provided in the CoNaLa challenge, we used different SentencePiece subword tokenizer models such as Unigram, WordPiece, and BPE to tokenize the source code. Initially, we trained these models respectively on the CoNaLa training dataset to learn the subword vocabulary. Then these trained models were loaded to generate the respective SentencePiece tokens that tokenize the source code as subword tokenization. To conduct experiments using different SentencePiece subword tokenizers, the configuration of the transformer model to train were used as follows:
D\_MODEL = 256,
N\_LAYERS = 3,
N\_HEADS = 8,
HIDDEN\_SIZE = 512,
MAX\_LEN = 50,
DROPOUT = 0.25,
LR = 1e-3,
N\_EPOCHS = 50,
GRAD\_CLIP = 1.0, with an early stopping mechanism.

\textbf{Results.}

\textbf{Finding 1.} From \autoref{tab:metric-transf-tok}, we observed that the Google's SentencePiece\footnote{\href{https://github.com/google/sentencepiece}{https://github.com/google/sentencepiece}} Unigram trained tokenizer resulted in the highest test corpus BLEU score followed by Byte-Pair encoding, and WordPiece subword tokenizers models respectively, for the transformer architecture. 

\textbf{Finding 2.} We found that the SentencePiece's Unigram model tokenizer model performed better than the CoNaLa code \href{https://github.com/conala-corpus/conala-baseline/blob/d109d90da57103f858539fc40110c41e2e90685e/eval/conala_eval.py#L94}{tokenizer} provided in the CoNaLa challenge by 36.30\%. The transformer-WordPiece performed better by 12.6\% and the transformer-BPE performed better by 25.72\% over the CoNaLa code tokenizer. These results showed that with better preprocessing of the source code, the Transformer-Unigram and the Transformer-BPE architectures performed better than the proposed Seq2Seq-RoBERTa architecture on the BLEU metric.
\begin{table}[!htbp]
	\centering
	\Large
	\begin{adjustbox}{width=\textwidth}
		\begin{tabular}{|c|c|c|c|c|c|c|}
			\hline
			\Large
			\textbf{Tokenizer}    & \textbf{BLEU}     & \textbf{Rouge1 F1} & \textbf{Rouge2 F1} & \textbf{Rouge4 F1} & \textbf{RougeL F1} & \textbf{Token Accuracy} \\ \hline
			WordPiece & 17.3237 & 0.6786       &0.5339      & 0.2448        & 0.6786     & 9.1647       \\ 
			Unigram       & \textbf{20.9678} &  0.6718       & 0.5214      & 0.2376       &  0.6718     & 12.5242       \\ 
			BPE & 19.3402 & \textbf{0.6937} & \textbf{0.5495} & \textbf{0.2567} & \textbf{0.6937} & 10.1486 \\ 
			CoNaLa code tokenizer & 15.3834 &  0.5449 &  0.2628 & 0.1311 & 0.5347 & \textbf{14.7041} \\ \hline
		\end{tabular}
	\end{adjustbox}
	\captionof{table}[Test metrics for Transformer using SentencePiece tokenizers]{Test metrics for Transformer using SentencePiece tokenizers. (batchsize=64, 3 layers, 8 heads, LR=1e-03, D\_MODEL=256, DROPOUT=0.25)	\label{tab:metric-transf-tok}}
\end{table}

\textbf{Finding 3.} The use of a subword unit LM for code had two potential advantages. Firstly, as the model had a smaller vocabulary size due to reduced level of data sparsity, it might have better performance over the non-subword tokenizers. Secondly, the model could handle the $OoV$ problem by synthesizing the missing $OoV$ tokens that were seen in the training data using the smaller subtoken units.

\subsection{RQ5. Does byte-pair encoding of code tokens capture more information than the AST and the AMT?}
\textbf{Motivation.}
There has been a lot of research on modeling source code as Abstract Syntax Tree (AST) representation. \citep{10.1145/2983990.2984041} and \citep{Li2018CodeCW} researched upon code completion task by predicting the next AST node in the flattened sequence of ASTs. \citep{wang2020fullline} found out that syntax-based approaches do not outperform token sequence-based ones, and even perform worse on the Py150 dataset. It has been observed that the action sequence for a Python program is often longer than its source code token sequence, which brings an extra burden to language models to model the source code of programming languages.

The semantic parser, TranX, suffers to convert AST into action sequence for function declaration or for/while iterators code lines, as their syntax is not complete, therefore, the code lines cannot be parsed into ASTs.  Thus, the above-discussed disadvantages and limitations of semantic parsing motivated us to model source code modeling with the sequence token-based modeling neural tokenization models such as Unigram, Byte-Pair Encoding(BPE), abd WordPiece to tokenize the source code.

\textbf{Approach.}
Our approach modeled the source code as token sequence-based language modeling instead of an abstract meaning representation of source code representation. We then compared our proposed architectures with AST-based modeling neural semantic parser, TranX, to see whether the AST modeling of code tokens can be substituted by token sequence-based modeling of source code of programming languages (lexical representation) to generate code snippets. 

\textbf{Results.}
In our work, we didn't model the source code as AST. Hence, we could not say that AST/AMT results in better modeling of source code representation and contextual understanding. However, our proposed approach had surpassed the performance of the neural semantic parser, TranX, that uses the underlying AST representation and action sequences of source code.  

\textbf{Finding 1.} We inferred that modeling source code as neural language modeling with BPE tokenizer instead of AST/AMT could also result in better code representation, especially for the NL2Code objective. Thus, it did not outrightly weigh out the importance of possibly combing AST with Seq2Seq-BART in future for better-structured code predictions from NL. 

\textbf{Finding 2.} The main advantage of ASDL action sequences over source code tokens is that an action sequence gets converted to a syntactically correct code snippet. Maybe in the future, we would like to use AST modeling in our hybrid Seq2Seq architecture to see whether AST modeling outperforms the sequence token-based modeling of source code.

\subsection{RQ6. Does the proposed hybrid Seq2Seq-BART architecture work bidirectionally to reverse the hypothesis?}
\textbf{Motivation.}
The idea was to see whether reversing the proposed objective, i.e., generating natural language intents from the code snippet (Code2NL) from the hybrid architecture, works well as the proposed objective in mapping NL intent to code snippets. If it works, the model architecture will be versatile and can be used to generate comments, docstrings, and documentation with the source code as an input. We further use the reverse hypothesis for data augmentation technique to expand the training dataset for the NL2Code objective.

\textbf{Approach.}
We reversed the input and the output for the Seq2Seq-BART architecture. We fed the code snippets into the Encoder and mapped them to the natural language intent in the Decoder of the Seq2Seq-BART architecture. We used the same values for the hyperparameters used for the Nl2Code-BART model.

\textbf{Results.}
Here, we discuss the result of the experiment to test whether the proposed hybrid Seq2Seq-BART model has bidirectionality ability, if we reverse the target and inputs, we saw whether it works to generate natural language comments/docstrings given a code snippet. We evaluated the generated natural language intent with the original intent in the test set to measure the BLEU, ROUGE, and METEOR test metrics.

\begin{figure}[!h]
	\centering
	\includegraphics[width=0.6\linewidth]{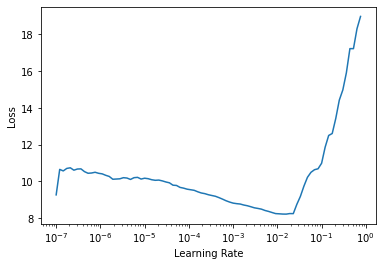}
	\caption[Optimal learning rate plot for Code2NL]{Optimal learning rate for Code2NL task.}
	\label{fig:optimal-lr-code2nl}
\end{figure}

\textbf{Finding 1.} In \autoref{fig:optimal-lr-code2nl}, we found the optimal learning rate for the same training dataset but for the Code2NL objective. 

\textbf{Finding 2.} In ~\autoref{tab:code2nl}, we showed the test metrics for the single experiment conducted for the Code2NL objective using the Seq2Seq-BART architecture. We observed that the proposed architecture worked quite well for the reversed hypothesis. Therefore, the hybrid model architecture developed in this thesis works bidirectionally, i.e., for both NL2Code and Code2NL tasks respectively.

\begin{table}[!htbp]
	\centering
	\large
	\begin{tabular}{|c|c|c|c|c|}
		\hline
		\textbf{BLEU} & \textbf{Rouge1 F1} & \textbf{Rouge2 F1} & \textbf{RougeL F1} & \textbf{METEOR} \\ \hline
		21.80291      & 0.45516            & 0.21644            & 0.407032           & 0.3050          \\ \hline
	\end{tabular}
	\captionof{table}[Test metrics for the Code2NL.]{Test metrics for the Code2NL\label{tab:code2nl}.}
\end{table}     

\subsection{RQ7. Do data augmentation and pretraining techniques improve the results of the proposed hybrid Seq2Seq-BART architecture?}
\textbf{Motivation.}
The unavailability of a large training dataset has limited us in exploring the full potential of the proposed architecture. Thus, we proposed to augment the training and validation datasets with the generated data from the reverse hypothesis of generating code snippets from natural language. The back-translation approach augments the original dataset for further training the proposed architecture with more training data to see whether the performance of a model improves.

\textbf{Approach.}
We trained the Seq2Seq-BART architecture on the augmented CoNaLa datasets of $3x$ and $5x$ sized of the original dataset. We used the pretrained Seq2Seq-BART model, thereafter, fine-tuned the model on the augmented dataset.

\textbf{Results.}
Here, we discuss the result of the experiments to see whether, with more training data, the Seq2Seq-BART architecture performs better than the previously conducted experiments on the original dataset. \autoref{tab:metrics-bart-augmented} compares the performance of the Seq2Seq-BART on the different sets of augmented datasets. 

\textbf{Finding 1.} We observed that when the Seq2Seq-BART trained on the augmented $3x$ size of the original dataset, it had a 6\% improvement of BLEU-4 metric test score than on the model trained on the original dataset.

\textbf{Finding 2.}
We found that the test scores decreased for the model trained on the augmented 5$x$ dataset compared to the $3x$ dataset. The drop in the test metrics score was particularly due to the added noise of adding 4$x$ sized more of the newly generated training data. However, the test metric scores were still higher than the model trained on the original test set by 3.5\%.

\textbf{Finding 3.} Similarly, we observed that when the Seq2Seq-BART pretrained on the mined 100k corpus, thereafter, fine-tuned on the $3x$ augmented training dataset, there was an improvement of the original test set score by 4.84\%. This framework had reported the highest test BLEU score among all the architectures developed and implemented in this work.

\textbf{Finding 4.}
We observed that the model fine-tuned on $5x$ dataset after pretraining on mined100k corpus had relatively performed worse, compared to the models fine-tuned on the normal training set by 4.6\%, and on the $3x$ training set by 9\%. This drop in the test score might be due to added noise in the augmented dataset by adding newly generated NL intents via a generative process from the Code2NL system. 

\textbf{Finding 5.}
Therefore, we inferred that the $3x$ augmented dataset that we had exclusively created in this thesis work made the developed hybrid Seq2Seq-BART architecture to achieve state-of-the-art results. 

\begin{table}[!htbp]
	\centering
	\Large
		\begin{adjustbox}{width=\textwidth}
	\begin{tabular}{|c|c|}
		\hline
		\textbf{Seq2Seq-BART on Augmented Datasets}             & \textbf{Test BLEU} \\ \hline
		Seq2Seq-BART on $3x$ size of original dataset                                      & 25.7710             \\ \hline
		Seq2Seq-BART on $5x$ size of original dataset                                     & 25.1601            \\ \hline
		Seq2Seq-BART on CoNaLa dataset                       & 24.2990             \\ \hline
		Fine-tuned Seq2Seq-BART on CoNaLa, pretrained on mined100k corpus & 26.5379           \\ \hline
		Fine-tuned Seq2Seq-BART on $3x$ CoNaLa, pretrained on mined100k corpus & \textbf{27.8235}            \\ \hline
		Fine-tuned Seq2Seq-BART on $5x$ CoNaLa, pretrained on mined100k corpus & 25.3153            \\ \hline
		TranX on CoNaLa train dataset & 25.1050 \\ \hline
	\end{tabular}
\end{adjustbox}
	\captionof{table}{Results of Seq2Seq-BART models on Augmented Datasets.\label{tab:metrics-bart-augmented}}
	
\end{table}

\subsection{RQ8. How well does the code completion task work with the use of neural language model?}
\textbf{Motivation.}
What if a system suggests code elements to developers given a partially complete code snippet as an input by the developer? A code completion system suggests pieces of code by understanding the context of the incomplete code snippet written by the developer. Ideally, a code completion system depends on rule-based systems where various rules are hard-coded to achieve certain code completion.

For code completion, the aim is to develop a machine learning approach that learns to predict the next code tokens given an incomplete code line. Alternatively, the model gives the predicted probabilities of sentences in the form of source code tokens.

\textbf{Approach.}
After the success of pretrained contextual embeddings approaches for natural languages, we present the attempt to apply the underlying techniques to model the source code of Python programming language by tokenizing source code as sequence token-based split of source code, instead of representing source code as an abstract semantic representation or complicated abstract syntax tree.  

In particular, BERT \citep{devlin-etal-2019-bert} produced a bidirectional Transformer encoder \citep{vaswani2017attention} by training it to predict values of masked tokens, and whether two sentences follow each other in natural discourse. The pre-trained model can be fine-tuned for downstream supervised tasks and to produce state-of-the-art results on several natural-language understanding benchmarks. 

We derived a contextual embedding and language modeling of source code by training a RoBERTa model on source code. We called our language model CuRoBERTa-LM, short for Code Understanding RoBERTa language model. We trained the RoBERTa masked language model from scratch on the training set of Algorithms Python code repositories. Before training, we had removed the comments, docstring from the Python source code files. We  had split the dataset into an 80-20\% split for training and validation purposes. 

We used the subword tokenizer, ByteLevel BPE, to model the source code then trained the RoBERTa tokenizer for source code by setting $min\_frequency=1$ to consider the rare words in the corpus as the given corpus is not very big. The training of the tokenizer was not memory intensive for the smaller dataset. We set the $vocab\_size=52000$, $max\_position\_embeddings=514$, $num\_attention\_heads=12$, $num\_hidden\_layers=6$, $mlm\_probability=0.15$ in the RoBERTa model configuration for training. It took 6hrs to train the RoBERTa language model on the Algorithm Python source code repository. 

\textbf{Results.}
Here, we discuss the results of the fill-mask task of code completion. We used the Huggingface fill-mask pipeline\footnote{\href{https://huggingface.co/transformers/main_classes/pipelines.html}{https://huggingface.co/transformers/main\_classes/pipelines.html}} to mask a code token and to check whether the trained language model can predict the masked token given the surrounding tokens. 

\textbf{Finding 1.} From \autoref{tab:fill-mask}, we observed that the CuRoBERTa-LM for Python source code model correctly predicted the masked token and understood the complex declarative and conditional logic statements such as, if-else, while constructs, generators, and iterators.

However, the model gives higher score to the $sum=a*b$ logical statement over the correct $sum=a+b$ expression. The intelligent part was that the system had generated a different logical expression for the $sum = a <mask> b$ including summation, multiplication, difference, and modular division between two logical operands for the given masked code expression. Thus, the feature could assist developers to pick up the correct logical expression from the model that would fit in the live coding setting.

\begin{figure}[]
	\centering
	\includegraphics[width=0.65\linewidth]{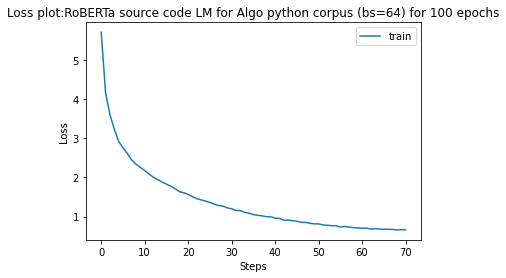}
	\caption[Loss plot for the CuRoBERTa-LM]{Loss plot for the CuRoBERTa-LM trained on Algorithm Python corpus.}
	\label{fig:loss-roberta-lm-algo-py-ep100-b64}
\end{figure}

\begin{table}[!htbp]
	\centering
	\begin{tabular}{|c|c|c|}
		\hline
		\textbf{Fill Mask task}                                                                & \textbf{Prediction/Output}                                        & \textbf{Score} \\ \hline
		"\textless{}mask\textgreater os"                                                       & `import os'                                             & \textbf{0.9979}       \\ \hline
		\multirow{4}{*}{if (x is not None) \textless{}mask\textgreater  ( x \textgreater{} 1)}     & 'if (x is not None) and (x\textgreater{}1)             & \textbf{0.6732}      \\ 
		& 'if (x is not None) / (x\textgreater{}1)               & 0.0770        \\ 
		& 'if (x is not None) | (x\textgreater{}1)               & 0.0769      \\ 
		& if (x is not None) or (x\textgreater{}1)               & 0.0291      \\ \hline
		\multirow{3}{*}{if self.graph{[}u{]}.count({[}w, v{]}) \textless{}mask\textgreater 0:} & if self.graph{[}u{]}.count({[}w, v{]}) == 0:           & \textbf{0.5888}       \\ 
		& 'if self.graph{[}u{]}.count({[}w, v{]})!= 0:'          & 0.3329        \\ 
		& if self.graph{[}u{]}.count({[}w, v{]}) \textgreater 0: & 0.0360        \\ \hline
		\multirow{4}{*}{sum = a \textless{}mask\textgreater b}                                 & sum = a * b                                            & \textbf{0.7289}       \\ 
		& sum = a + b                                            & 0.1288       \\ 
		& 'sum = a - b                                           & 0.0186      \\ 
		& sum = a // b'                                          & 0.0118      \\ \hline
		\multirow{5}{*}{mdist = {[}\textless{}mask\textgreater for i in range(V){]}}           & mdist = {[}0 for i in range(V){]}                      & \textbf{0.7845}        \\ 
		& mdist = {[}i for i in range(V){]}                      & 0.0978        \\ 
		& mdist = {[}True for i in range(V){]}                   & 0.0320        \\ 
		& mdist = {[}False for i in range(V){]}                  & 0.0235        \\ 
		& mdist = {[}1 for i in range(V){]}                      & 0.0088        \\ \hline
	\end{tabular}
	\captionof{table}{Fill-mask task for code completion.\label{tab:fill-mask}}
\end{table}

\chapter{Conclusion}\label{conclusion}

In this thesis, we have shown that parsable Python source code snippets can be directly generated from natural language intent using deep learning architectures, without necessarily using heavily feature engineered semantic parsers, to parse input structured programming source code to produce Abstract Syntax Trees representation of the source code and to convert them back to the source code representation. 

Firstly, we have proposed a novel hybrid Sequence-to-Sequence architecture with BERT as an encoder and GPT as a decoder(Seq2Seq-BART) that pre-processes source code input using byte-level Byte Pair encoding tokenizer and produces parsable Python code directly from the decoder without requiring any AMT. This proposed Seq2Seq-BART architecture has resulted in generating better code translations from NL and has exceeded the performance of the vanilla Seq2Seq, Transformer, and the proposed Seq2Seq-RoBERTa architectures on BLEU-4 metric score for natural language to code translation, i.e., generating source code from natural language intent. The developed architecture has also surpassed the test BLEU metric score of the state-of-the-art neural semantic parser, TranX, on the CoNaLa dataset.

The proposed architecture also works well for the reverse hypothesis of translating source code to natural language. The reversed architecture can be used to automatically generate code documentation such as comments and Python docstrings from source code input. We used a pre-training strategy along with data augmentation technique in the proposed NL2Code system framework to enhance the quality of the code translation. The pre-training strategy facilitated the proposed Seq2Seq architecture to learn the mapping of the natural language intent to the source code from the CoNaLa mined corpus. Then, the pretrained model was fine-tuned on the training set and validated on the validation set, thereafter, was used on the test dataset to evaluate the translation quality of the model architecture. We have used BLEU, ROUGE metrics to measure the quality of the translation of the code generated from the natural language intent. 

Secondly, we have also introduced a valid compilable code snippet metric for the CoNaLa challenge where we have counted the number of generated source code translations with a valid parsable Python source code. We have also introduced augmented training and validation datasets for the CoNaLa challenge using the proposed Code2NL back-translation system of generating new natural language intent from source code snippet using the reverse hypothesis of NL2Code, i.e., Code2NL. Furthermore, we have shown that the code generation of our algorithm can be improved by using the pretraining approach and the data augmentation technique.

Thirdly, we have also performed an empirical study about various deep learning architectures that translate the natural language to source code. We undertook ablation studies about the Transformer-based and the BERT-based architectures to find out which components of the system are more correlated to the target objective. We have also shown that the output from the Transformer architecture can be improved for the Nl2Code task by using subword tokenizers such as the Byte-Pair Encoding (BPE). 

Lastly, we have also built a RoBERTa based language model, namely CuRoBERTa-LM, for Python source code and used the trained source code language model for code completion task, as a fill-in-mask token task to be specific. The trained language model is quite effective in predicting the masked token given the surrounding unmasked tokens of a logical expression. The proposed CuRoBERTa-LM can understand the context of the complex logical deductions from Python expressions/ statements. This shows that the powerful neural language models which were exclusively built for NLP tasks can also be trained for programming source code, and can be used for downstream tasks like code search, code summarization, variable mismatch, function naming, etc. 

\section{Future Work}
Firstly, we would like to explore our novel hybrid Seq2Seq architecture on more datasets such as Django \citep{oda2015ase:pseudogen1}, WikiSQL \citep{zhongSeq2SQL2017} to see how well the proposed architecture generalizes to other programming languages and language-specific domains. We would also like to incorporate ASTs into our proposed architecture to see whether the incorporation of AST into our proposed architecture results in better structured semantic code translations from natural language intent. 

Secondly, we would also like to use the recently introduced metric, BERTSCORE \citep{zhang2020bertscore} as an evaluation metric for the translated code snippets. In contrast to string matching (e.g., in BLEU) or matching heuristics (e.g., in METEOR), this metric uses contextual embeddings to represent the tokens, and compute matching using cosine similarity, which is optionally weighted with inverse document frequency scores. Thus, the BERTSCORE addresses the common pitfalls in n-gram-based metrics that often fail to capture distant dependencies, match paraphrases, and penalize semantically critical ordering changes.

Lastly, we would create a dataset that maps natural language documentation/ docstrings from functions, with the source code of the entire function spanning multiple lines of code, and to use this dataset to train our proposed architecture to generate multi-line source code snippets as generating an entire function code declaration. This will allow our proposed system to explore contextual awareness over multiple lines of source code and enable it to learn more about the surrounding context of the source code, thus, making the system capable of injecting code snippets from natural language intent into any point of insertion in a live code writing set by the developer.

	
	\bibliographystyle{abbrvnat}
	\bibliography{references}

\begin{thebibliography}{55}
\providecommand{\natexlab}[1]{#1}
\providecommand{\url}[1]{\texttt{#1}}
\expandafter\ifx\csname urlstyle\endcsname\relax
  \providecommand{\doi}[1]{doi: #1}\else
  \providecommand{\doi}{doi: \begingroup \urlstyle{rm}\Url}\fi

\bibitem[Con()]{Conalacode}
{CoNaLa} code tokenizer implementation.
\newblock
  \url{https://github.com/conala-corpus/conala-baseline/blob/d109d90da57103f858539fc40110c41e2e90685e/eval/conala_eval.py#L94}.
\newblock Accessed: 2021-01-05.

\bibitem[Alon et~al.(2019)Alon, Zilberstein, Levy, and Yahav]{10.1145/3290353}
U.~Alon, M.~Zilberstein, O.~Levy, and E.~Yahav.
\newblock Code2vec: Learning distributed representations of code.
\newblock \emph{Proc. ACM Program. Lang.}, 3\penalty0 (POPL), Jan. 2019.
\newblock \doi{10.1145/3290353}.
\newblock URL \url{https://doi.org/10.1145/3290353}.

\bibitem[Ba et~al.(2016)Ba, Kiros, and Hinton]{ba2016layer}
J.~L. Ba, J.~R. Kiros, and G.~E. Hinton.
\newblock Layer normalization, 2016.

\bibitem[Bahdanau et~al.(2016)Bahdanau, Cho, and Bengio]{bahdanau2016neural}
D.~Bahdanau, K.~Cho, and Y.~Bengio.
\newblock Neural machine translation by jointly learning to align and
  translate, 2016.

\bibitem[Banerjee and Lavie(2005)]{banarjee2005}
S.~Banerjee and A.~Lavie.
\newblock {METEOR}: An automatic metric for {MT} evaluation with improved
  correlation with human judgments.
\newblock In \emph{Proceedings of the {ACL} Workshop on Intrinsic and Extrinsic
  Evaluation Measures for Machine Translation and/or Summarization}, pages
  65--72, Ann Arbor, Michigan, June 2005. Association for Computational
  Linguistics.
\newblock URL \url{https://www.aclweb.org/anthology/W05-0909}.

\bibitem[Bhoopchand et~al.(2016)Bhoopchand, Rocktäschel, Barr, and
  Riedel]{bhoopchand2016learning}
A.~Bhoopchand, T.~Rocktäschel, E.~Barr, and S.~Riedel.
\newblock Learning python code suggestion with a sparse pointer network, 2016.

\bibitem[Chung et~al.(2014)Chung, Gulcehre, Cho, and
  Bengio]{chung2014empirical}
J.~Chung, C.~Gulcehre, K.~Cho, and Y.~Bengio.
\newblock Empirical evaluation of gated recurrent neural networks on sequence
  modeling, 2014.

\bibitem[Dai et~al.(2019)Dai, Yang, Yang, Carbonell, Le, and
  Salakhutdinov]{dai-etal-2019-transformer}
Z.~Dai, Z.~Yang, Y.~Yang, J.~Carbonell, Q.~Le, and R.~Salakhutdinov.
\newblock Transformer-{XL}: Attentive language models beyond a fixed-length
  context.
\newblock In \emph{Proceedings of the 57th Annual Meeting of the Association
  for Computational Linguistics}, pages 2978--2988, Florence, Italy, July 2019.
  Association for Computational Linguistics.
\newblock \doi{10.18653/v1/P19-1285}.
\newblock URL \url{https://www.aclweb.org/anthology/P19-1285}.

\bibitem[Devlin et~al.(2019)Devlin, Chang, Lee, and
  Toutanova]{devlin-etal-2019-bert}
J.~Devlin, M.-W. Chang, K.~Lee, and K.~Toutanova.
\newblock {BERT}: Pre-training of deep bidirectional transformers for language
  understanding.
\newblock In \emph{Proceedings of the 2019 Conference of the North {A}merican
  Chapter of the Association for Computational Linguistics: Human Language
  Technologies, Volume 1 (Long and Short Papers)}, pages 4171--4186,
  Minneapolis, Minnesota, June 2019. Association for Computational Linguistics.
\newblock \doi{10.18653/v1/N19-1423}.
\newblock URL \url{https://www.aclweb.org/anthology/N19-1423}.

\bibitem[Dong and Lapata(2016)]{dong-lapata-2016-language}
L.~Dong and M.~Lapata.
\newblock Language to logical form with neural attention.
\newblock In \emph{Proceedings of the 54th Annual Meeting of the Association
  for Computational Linguistics (Volume 1: Long Papers)}, pages 33--43, Berlin,
  Germany, Aug. 2016. Association for Computational Linguistics.
\newblock \doi{10.18653/v1/P16-1004}.
\newblock URL \url{https://www.aclweb.org/anthology/P16-1004}.

\bibitem[Dror et~al.(2018)Dror, Baumer, Shlomov, and
  Reichart]{dror-etal-2018-hitchhikers}
R.~Dror, G.~Baumer, S.~Shlomov, and R.~Reichart.
\newblock The hitchhiker{'}s guide to testing statistical significance in
  natural language processing.
\newblock In \emph{Proceedings of the 56th Annual Meeting of the Association
  for Computational Linguistics (Volume 1: Long Papers)}, pages 1383--1392,
  Melbourne, Australia, July 2018. Association for Computational Linguistics.
\newblock \doi{10.18653/v1/P18-1128}.
\newblock URL \url{https://www.aclweb.org/anthology/P18-1128}.

\bibitem[Gage(1994)]{10.5555/177910.177914}
P.~Gage.
\newblock A new algorithm for data compression.
\newblock \emph{C Users J.}, 12\penalty0 (2):\penalty0 23–38, Feb. 1994.
\newblock ISSN 0898-9788.

\bibitem[Gehring et~al.(2017)Gehring, Auli, Grangier, Yarats, and
  Dauphin]{pmlr-v70-gehring17a}
J.~Gehring, M.~Auli, D.~Grangier, D.~Yarats, and Y.~N. Dauphin.
\newblock Convolutional sequence to sequence learning.
\newblock In D.~Precup and Y.~W. Teh, editors, \emph{Proceedings of the 34th
  International Conference on Machine Learning}, volume~70 of \emph{Proceedings
  of Machine Learning Research}, pages 1243--1252. PMLR, 06--11 Aug 2017.
\newblock URL \url{http://proceedings.mlr.press/v70/gehring17a.html}.

\bibitem[He et~al.(2015)He, Zhang, Ren, and Sun]{he2015deep}
K.~He, X.~Zhang, S.~Ren, and J.~Sun.
\newblock Deep residual learning for image recognition, 2015.

\bibitem[Hellendoorn and Devanbu(2017)]{10.1145/3106237.3106290}
V.~J. Hellendoorn and P.~Devanbu.
\newblock Are deep neural networks the best choice for modeling source code?
\newblock In \emph{Proceedings of the 2017 11th Joint Meeting on Foundations of
  Software Engineering}, ESEC/FSE 2017, page 763–773, New York, NY, USA,
  2017. Association for Computing Machinery.
\newblock ISBN 9781450351058.
\newblock \doi{10.1145/3106237.3106290}.
\newblock URL \url{https://doi.org/10.1145/3106237.3106290}.

\bibitem[Hendrycks and Gimpel(2020)]{hendrycks2020gaussian}
D.~Hendrycks and K.~Gimpel.
\newblock Gaussian error linear units (gelus), 2020.

\bibitem[Hindle et~al.(2012)Hindle, Barr, Su, Gabel, and
  Devanbu]{10.5555/2337223.2337322}
A.~Hindle, E.~T. Barr, Z.~Su, M.~Gabel, and P.~Devanbu.
\newblock On the naturalness of software.
\newblock In \emph{Proceedings of the 34th International Conference on Software
  Engineering}, ICSE '12, page 837–847. IEEE Press, 2012.
\newblock ISBN 9781467310673.

\bibitem[Iyer et~al.(2018)Iyer, Konstas, Cheung, and
  Zettlemoyer]{iyer2018mapping}
S.~Iyer, I.~Konstas, A.~Cheung, and L.~Zettlemoyer.
\newblock Mapping language to code in programmatic context, 2018.

\bibitem[Joshi et~al.(2020)Joshi, Chen, Liu, Weld, Zettlemoyer, and
  Levy]{joshi2020spanbert}
M.~Joshi, D.~Chen, Y.~Liu, D.~S. Weld, L.~Zettlemoyer, and O.~Levy.
\newblock Spanbert: Improving pre-training by representing and predicting
  spans, 2020.

\bibitem[Karampatsis et~al.(2020)Karampatsis, Babii, Robbes, Sutton, and
  Janes]{Karampatsis2020BigC}
R.-M. Karampatsis, H.~Babii, R.~Robbes, C.~Sutton, and A.~Janes.
\newblock Big code != big vocabulary: Open-vocabulary models for source code.
\newblock \emph{2020 IEEE/ACM 42nd International Conference on Software
  Engineering (ICSE)}, pages 1073--1085, 2020.

\bibitem[Khandelwal et~al.(2018)Khandelwal, He, Qi, and
  Jurafsky]{khandelwal-etal-2018-sharp}
U.~Khandelwal, H.~He, P.~Qi, and D.~Jurafsky.
\newblock Sharp nearby, fuzzy far away: How neural language models use context.
\newblock In \emph{Proceedings of the 56th Annual Meeting of the Association
  for Computational Linguistics (Volume 1: Long Papers)}, pages 284--294,
  Melbourne, Australia, July 2018. Association for Computational Linguistics.
\newblock \doi{10.18653/v1/P18-1027}.
\newblock URL \url{https://www.aclweb.org/anthology/P18-1027}.

\bibitem[Kudo(2018)]{kudo2018subword}
T.~Kudo.
\newblock Subword regularization: Improving neural network translation models
  with multiple subword candidates, 2018.

\bibitem[Kudo and Richardson(2018)]{kudo2018sentencepiece}
T.~Kudo and J.~Richardson.
\newblock Sentencepiece: A simple and language independent subword tokenizer
  and detokenizer for neural text processing, 2018.

\bibitem[Lewis et~al.(2019)Lewis, Liu, Goyal, Ghazvininejad, Mohamed, Levy,
  Stoyanov, and Zettlemoyer]{lewis2019bart}
M.~Lewis, Y.~Liu, N.~Goyal, M.~Ghazvininejad, A.~Mohamed, O.~Levy, V.~Stoyanov,
  and L.~Zettlemoyer.
\newblock Bart: Denoising sequence-to-sequence pre-training for natural
  language generation, translation, and comprehension, 2019.

\bibitem[Li et~al.(2018)Li, Wang, Lyu, and King]{Li2018CodeCW}
J.~Li, Y.~Wang, M.~Lyu, and I.~King.
\newblock Code completion with neural attention and pointer networks.
\newblock \emph{ArXiv}, abs/1711.09573, 2018.

\bibitem[Lin(2004)]{lin-2004-rouge}
C.-Y. Lin.
\newblock {ROUGE}: A package for automatic evaluation of summaries.
\newblock In \emph{Text Summarization Branches Out}, pages 74--81, Barcelona,
  Spain, July 2004. Association for Computational Linguistics.
\newblock URL \url{https://www.aclweb.org/anthology/W04-1013}.

\bibitem[Ling et~al.(2016)Ling, Blunsom, Grefenstette, Hermann,
  Ko{\v{c}}isk{\'y}, Wang, and Senior]{ling-etal-2016-latent}
W.~Ling, P.~Blunsom, E.~Grefenstette, K.~M. Hermann, T.~Ko{\v{c}}isk{\'y},
  F.~Wang, and A.~Senior.
\newblock Latent predictor networks for code generation.
\newblock In \emph{Proceedings of the 54th Annual Meeting of the Association
  for Computational Linguistics (Volume 1: Long Papers)}, pages 599--609,
  Berlin, Germany, Aug. 2016. Association for Computational Linguistics.
\newblock \doi{10.18653/v1/P16-1057}.
\newblock URL \url{https://aclanthology.org/P16-1057}.

\bibitem[Liu et~al.(2016)Liu, Wang, Shin, Gonzalez, and Song]{liu2016neural}
C.~Liu, X.~Wang, R.~Shin, J.~E. Gonzalez, and D.~Song.
\newblock Neural code completion.
\newblock 2016.

\bibitem[Liu et~al.(2020{\natexlab{a}})Liu, Li, Wei, Xia, Fu, and
  Jin]{10.1145/3387904.3389261}
F.~Liu, G.~Li, B.~Wei, X.~Xia, Z.~Fu, and Z.~Jin.
\newblock A self-attentional neural architecture for code completion with
  multi-task learning.
\newblock In \emph{Proceedings of the 28th International Conference on Program
  Comprehension}, ICPC '20, page 37–47, New York, NY, USA,
  2020{\natexlab{a}}. Association for Computing Machinery.
\newblock ISBN 9781450379588.
\newblock \doi{10.1145/3387904.3389261}.
\newblock URL \url{https://doi.org/10.1145/3387904.3389261}.

\bibitem[Liu et~al.(2020{\natexlab{b}})Liu, Li, Wei, Xia, Fu, and
  Jin]{liu2020selfattentional}
F.~Liu, G.~Li, B.~Wei, X.~Xia, Z.~Fu, and Z.~Jin.
\newblock A self-attentional neural architecture for code completion with
  multi-task learning, 2020{\natexlab{b}}.

\bibitem[Liu et~al.(2019)Liu, Ott, Goyal, Du, Joshi, Chen, Levy, Lewis,
  Zettlemoyer, and Stoyanov]{liu2019roberta}
Y.~Liu, M.~Ott, N.~Goyal, J.~Du, M.~Joshi, D.~Chen, O.~Levy, M.~Lewis,
  L.~Zettlemoyer, and V.~Stoyanov.
\newblock Roberta: A robustly optimized bert pretraining approach, 2019.

\bibitem[Luong et~al.(2015)Luong, Pham, and Manning]{luong2015effective}
M.-T. Luong, H.~Pham, and C.~D. Manning.
\newblock Effective approaches to attention-based neural machine translation,
  2015.

\bibitem[Oda et~al.(2015)Oda, Fudaba, Neubig, Hata, Sakti, Toda, and
  Nakamura]{oda2015ase:pseudogen1}
Y.~Oda, H.~Fudaba, G.~Neubig, H.~Hata, S.~Sakti, T.~Toda, and S.~Nakamura.
\newblock Learning to generate pseudo-code from source code using statistical
  machine translation.
\newblock In \emph{Proceedings of the 2015 30th IEEE/ACM International
  Conference on Automated Software Engineering (ASE)}, ASE '15, pages 574--584,
  Lincoln, Nebraska, USA, November 2015. IEEE Computer Society.
\newblock ISBN 978-1-5090-0025-8.
\newblock \doi{10.1109/ASE.2015.36}.
\newblock URL \url{https://doi.org/10.1109/ASE.2015.36}.

\bibitem[Papineni et~al.(2002)Papineni, Roukos, Ward, and
  Zhu]{10.3115/1073083.1073135}
K.~Papineni, S.~Roukos, T.~Ward, and W.-J. Zhu.
\newblock Bleu: A method for automatic evaluation of machine translation.
\newblock In \emph{Proceedings of the 40th Annual Meeting on Association for
  Computational Linguistics}, ACL '02, page 311–318, USA, 2002. Association
  for Computational Linguistics.
\newblock \doi{10.3115/1073083.1073135}.
\newblock URL \url{https://doi.org/10.3115/1073083.1073135}.

\bibitem[Post(2018)]{post-2018-call}
M.~Post.
\newblock A call for clarity in reporting {BLEU} scores.
\newblock In \emph{Proceedings of the Third Conference on Machine Translation:
  Research Papers}, pages 186--191, Belgium, Brussels, Oct. 2018. Association
  for Computational Linguistics.
\newblock URL \url{https://www.aclweb.org/anthology/W18-6319}.

\bibitem[Rabinovich et~al.(2017)Rabinovich, Stern, and
  Klein]{rabinovich-etal-2017-abstract}
M.~Rabinovich, M.~Stern, and D.~Klein.
\newblock Abstract syntax networks for code generation and semantic parsing.
\newblock In \emph{Proceedings of the 55th Annual Meeting of the Association
  for Computational Linguistics (Volume 1: Long Papers)}, pages 1139--1149,
  Vancouver, Canada, July 2017. Association for Computational Linguistics.
\newblock \doi{10.18653/v1/P17-1105}.
\newblock URL \url{https://www.aclweb.org/anthology/P17-1105}.

\bibitem[Radford et~al.(2019)Radford, Wu, Child, Luan, Amodei, and
  Sutskever]{Radford2019LanguageMA}
A.~Radford, J.~Wu, R.~Child, D.~Luan, D.~Amodei, and I.~Sutskever.
\newblock Language models are unsupervised multitask learners.
\newblock 2019.

\bibitem[Raychev et~al.(2016{\natexlab{a}})Raychev, Bielik, and
  Vechev]{10.1145/2983990.2984041}
V.~Raychev, P.~Bielik, and M.~Vechev.
\newblock Probabilistic model for code with decision trees.
\newblock In \emph{Proceedings of the 2016 ACM SIGPLAN International Conference
  on Object-Oriented Programming, Systems, Languages, and Applications}, OOPSLA
  2016, page 731–747, New York, NY, USA, 2016{\natexlab{a}}. Association for
  Computing Machinery.
\newblock ISBN 9781450344449.
\newblock \doi{10.1145/2983990.2984041}.
\newblock URL \url{https://doi.org/10.1145/2983990.2984041}.

\bibitem[Raychev et~al.(2016{\natexlab{b}})Raychev, Bielik, and
  Vechev]{10.1145/3022671.2984041}
V.~Raychev, P.~Bielik, and M.~Vechev.
\newblock Probabilistic model for code with decision trees.
\newblock \emph{SIGPLAN Not.}, 51\penalty0 (10):\penalty0 731–747, Oct.
  2016{\natexlab{b}}.
\newblock ISSN 0362-1340.
\newblock \doi{10.1145/3022671.2984041}.
\newblock URL \url{https://doi.org/10.1145/3022671.2984041}.

\bibitem[Schuster and Nakajima(2012)]{6289079}
M.~Schuster and K.~Nakajima.
\newblock Japanese and korean voice search.
\newblock In \emph{2012 IEEE International Conference on Acoustics, Speech and
  Signal Processing (ICASSP)}, pages 5149--5152, 2012.
\newblock \doi{10.1109/ICASSP.2012.6289079}.

\bibitem[Sennrich et~al.(2016{\natexlab{a}})Sennrich, Haddow, and
  Birch]{sennrich-etal-2016-neural}
R.~Sennrich, B.~Haddow, and A.~Birch.
\newblock Neural machine translation of rare words with subword units.
\newblock In \emph{Proceedings of the 54th Annual Meeting of the Association
  for Computational Linguistics (Volume 1: Long Papers)}, pages 1715--1725,
  Berlin, Germany, Aug. 2016{\natexlab{a}}. Association for Computational
  Linguistics.
\newblock \doi{10.18653/v1/P16-1162}.
\newblock URL \url{https://www.aclweb.org/anthology/P16-1162}.

\bibitem[Sennrich et~al.(2016{\natexlab{b}})Sennrich, Haddow, and
  Birch]{sennrich2016neural}
R.~Sennrich, B.~Haddow, and A.~Birch.
\newblock Neural machine translation of rare words with subword units,
  2016{\natexlab{b}}.

\bibitem[Smith and Topin(2018)]{smith2018superconvergence}
L.~N. Smith and N.~Topin.
\newblock Super-convergence: Very fast training of neural networks using large
  learning rates, 2018.

\bibitem[Sutskever et~al.(2014)Sutskever, Vinyals, and
  Le]{sutskever2014sequence}
I.~Sutskever, O.~Vinyals, and Q.~V. Le.
\newblock Sequence to sequence learning with neural networks, 2014.

\bibitem[Tu et~al.(2014)Tu, Su, and Devanbu]{10.1145/2635868.2635875}
Z.~Tu, Z.~Su, and P.~Devanbu.
\newblock On the localness of software.
\newblock In \emph{Proceedings of the 22nd ACM SIGSOFT International Symposium
  on Foundations of Software Engineering}, FSE 2014, page 269–280, New York,
  NY, USA, 2014. Association for Computing Machinery.
\newblock ISBN 9781450330565.
\newblock \doi{10.1145/2635868.2635875}.
\newblock URL \url{https://doi.org/10.1145/2635868.2635875}.

\bibitem[Vaswani et~al.(2017)Vaswani, Shazeer, Parmar, Uszkoreit, Jones, Gomez,
  Kaiser, and Polosukhin]{vaswani2017attention}
A.~Vaswani, N.~Shazeer, N.~Parmar, J.~Uszkoreit, L.~Jones, A.~N. Gomez,
  L.~Kaiser, and I.~Polosukhin.
\newblock Attention is all you need, 2017.

\bibitem[Wang et~al.(1997)Wang, Appel, Korn, and Serra]{Wang1997TheZA}
D.~C. Wang, A.~Appel, J.~L. Korn, and C.~S. Serra.
\newblock The zephyr abstract syntax description language.
\newblock In \emph{DSL}, 1997.

\bibitem[Wang et~al.(2020)Wang, Shen, Li, and Jin]{wang2020fullline}
W.~Wang, S.~Shen, G.~Li, and Z.~Jin.
\newblock Towards full-line code completion with neural language models.
\newblock \emph{CoRR}, abs/2009.08603, 2020.
\newblock URL \url{https://arxiv.org/abs/2009.08603}.

\bibitem[Wilcox(2003)]{WILCOX2003237}
R.~R. Wilcox.
\newblock 8 - comparing two independent groups.
\newblock In R.~R. Wilcox, editor, \emph{Applying Contemporary Statistical
  Techniques}, pages 237--284. Academic Press, Burlington, 2003.
\newblock ISBN 978-0-12-751541-0.
\newblock \doi{https://doi.org/10.1016/B978-012751541-0/50029-8}.
\newblock URL
  \url{https://www.sciencedirect.com/science/article/pii/B9780127515410500298}.

\bibitem[Yin and Neubig(2017)]{yin-neubig-2017-syntactic}
P.~Yin and G.~Neubig.
\newblock A syntactic neural model for general-purpose code generation.
\newblock In \emph{Proceedings of the 55th Annual Meeting of the Association
  for Computational Linguistics (Volume 1: Long Papers)}, pages 440--450,
  Vancouver, Canada, July 2017. Association for Computational Linguistics.
\newblock \doi{10.18653/v1/P17-1041}.
\newblock URL \url{https://www.aclweb.org/anthology/P17-1041}.

\bibitem[Yin and Neubig(2018)]{yin-neubig-2018-tranx}
P.~Yin and G.~Neubig.
\newblock {TRANX}: A transition-based neural abstract syntax parser for
  semantic parsing and code generation.
\newblock In \emph{Proceedings of the 2018 Conference on Empirical Methods in
  Natural Language Processing: System Demonstrations}, pages 7--12, Brussels,
  Belgium, Nov. 2018. Association for Computational Linguistics.
\newblock \doi{10.18653/v1/D18-2002}.
\newblock URL \url{https://www.aclweb.org/anthology/D18-2002}.

\bibitem[Yin et~al.(2018)Yin, Deng, Chen, Vasilescu, and Neubig]{yin2018mining}
P.~Yin, B.~Deng, E.~Chen, B.~Vasilescu, and G.~Neubig.
\newblock Learning to mine aligned code and natural language pairs from stack
  overflow.
\newblock In \emph{International Conference on Mining Software Repositories},
  MSR, pages 476--486. ACM, 2018.
\newblock \doi{https://doi.org/10.1145/3196398.3196408}.

\bibitem[Zettlemoyer and Collins(2005)]{Zettlemoyer05learningto}
L.~S. Zettlemoyer and M.~Collins.
\newblock Learning to map sentences to logical form: Structured classification
  with probabilistic categorial grammars.
\newblock In \emph{In Proceedings of the 21st Conference on Uncertainty in AI},
  pages 658--666, 2005.

\bibitem[Zhang et~al.(2020)Zhang, Kishore, Wu, Weinberger, and
  Artzi]{zhang2020bertscore}
T.~Zhang, V.~Kishore, F.~Wu, K.~Q. Weinberger, and Y.~Artzi.
\newblock Bertscore: Evaluating text generation with bert, 2020.

\bibitem[Zhong et~al.(2017)Zhong, Xiong, and Socher]{zhongSeq2SQL2017}
V.~Zhong, C.~Xiong, and R.~Socher.
\newblock Seq2sql: Generating structured queries from natural language using
  reinforcement learning.
\newblock \emph{CoRR}, abs/1709.00103, 2017.

\end{thebibliography}
	
\end{document}